\documentclass{aastex701}
\usepackage{amsmath}
\usepackage{makecell}   
\usepackage{seqsplit}   

\begin{document}

\title{AstroSkyFlow: an astronomical sky image flow simulator for time domain survey validation and machine learning}

\author[0009-0003-9882-8542]{Kexin Li}
\affiliation{State Key Laboratory of Dark Matter Physics, Tsung-Dao Lee Institute \& School of Physics and Astronomy, Shanghai Jiao Tong University, Shanghai 201210, China}
\email{kexin_li@sjtu.edu.cn}  

\author[0000-0001-5295-1682]{Yicheng Rui} 
\affiliation{State Key Laboratory of Dark Matter Physics, Tsung-Dao Lee Institute \& School of Physics and Astronomy, Shanghai Jiao Tong University, Shanghai 201210, China}
\email[show]{ruiyicheng@sjtu.edu.cn}  

\author[0000-0001-6039-0555]{Fabo Feng}
\affiliation{State Key Laboratory of Dark Matter Physics, Tsung-Dao Lee Institute \& School of Physics and Astronomy, Shanghai Jiao Tong University, Shanghai 201210, China}
\email[show]{ffeng@sjtu.edu.cn}

\author[0009-0001-9616-1823]{Shuyue Zheng}
\affiliation{State Key Laboratory of Dark Matter Physics, Tsung-Dao Lee Institute \& School of Physics and Astronomy, Shanghai Jiao Tong University, Shanghai 201210, China}
\email{zhengshuyue@sjtu.edu.cn}

\author[0000-0001-5426-2397]{Anton Pomazan}
\affiliation{Shanghai Astronomical Observatory, Chinese Academy of Sciences, Shanghai 200030, People’s Republic of China}
\email{antpomaz@shao.ac.cn}

\author[0009-0009-3434-4796]{Yiyang Guo}
\affiliation{State Key Laboratory of Dark Matter Physics, Tsung-Dao Lee Institute \& School of Physics and Astronomy, Shanghai Jiao Tong University, Shanghai 201210, China}
\email{yiyangguo@sjtu.edu.cn}

\author[0000-0001-6637-6973]{Jie Zheng}
\affiliation{National Astronomical Observatories, Chinese Academy of Sciences, People’s Republic of China}
\email{jiezheng@nao.cas.cn}

\author{Lin-Qiao Jiang}
\affiliation{College of Physics and Optoelectronic Engineering, Leshan Normal University, Leshan 614000, People’s Republic of China}
\affiliation{Key Laboratory of Detection and Application of Space Effect in Southwest Sichuan, Leshan 614000, People’s Republic of China}
\affiliation{Center for Applied Optics Research, Leshan Normal University, Leshan 614000, People’s Republic of China}
\email{jianglinqiao11@163.com}






\begin{abstract}
Modern time-domain optical surveys produce massive data volumes that require robust, high-fidelity simulated datasets for developing and validating automated pipelines and machine-learning models. We present AstroSkyFlow, a modular sky-image simulator that generates on-demand, time-dependent flux variations and models the full observing stack — from celestial sources and atmospheric effects to sensor response. Given a simulated observing schedule, AstroSkyFlow produces multi-epoch, time-series images with realistic noise and variability. Compared to real observational data, AstroSkyFlow reproduces noise characteristics and point spread function properties more accurately than the widely used SkyMaker simulator. In addition, AstroSkyFlow successfully recovers injected photometric and motion signals, such as exoplanet transits and asteroid trails. AstroSkyFlow enables the generation of labeled, high-fidelity datasets essential for training machine learning pipelines and conducting rigorous injection-recovery tests for analysis pipelines for next-generation time-domain surveys. 

\end{abstract}



\section{Introduction} 

Time-domain optical surveys and robotic telescopes produce continuous, high-cadence image streams that exemplify the Fourth Paradigm of data-intensive science \citep{Hey12}. These surveys deliver massive volumes of imaging data whose scientific exploitation increasingly depends on automated pipelines and machine-learning (ML) methods to detect, classify and measure transiting exoplanets \citep{Borucki10}, transients \citep{Masci19} and other variable sources. High-fidelity simulated training data are crucial for modern pipelines and ML approaches. Simulated image datasets provide abundant, controllable training sets with ground-truth labels, observation-based fluxes, parameterized morphologies and physics-informed noise realizations, which are essential for supervised training, stress-testing, domain-adaptation studies and for quantifying ML model biases and failure modes (e.g., \citealt{2019MNRAS.489.3582D, 2020AJ....159..212J}). Simulated datasets reproduce the full physical process of observation under controlled and repeatable conditions, enabling direct tests of detection efficiency, photometric bias, astrometric performance and systematic error that would be difficult or impossible to isolate from real data alone. 

Among current optical surveys, Tianyu is a one-meter telescope designed to detect the second solar system and explore the dynamic universe \citep{2024AcASn..65...34F}. It will be built in Lenghu, Qinghai, China in 2026. With its multi-cadence observation modes, Tianyu is able to observe time-domain objects with different magnitude ranging from transiting long-period exoplanets ($V>11$) to supernovae ($V<21$). With the survey strategy illustrated in the white paper, Tianyu will generate about 500 megabytes of raw data per second \citep{2024AcASn..65...34F}, which is about 16 times the raw data generated by Zwicky Transient Facility (ZTF)\citep{Masci19}. \cite{2025PASP..137f4501R} provide the architecture of the Tianyu software system and present a preliminary relative photometry pipeline developed based on this architecture. Such a telescope both motivates and demands the testing of analysis pipelines and ML methods: to build robust algorithms for Tianyu we need high-fidelity simulated datasets that reproduce the instrumental, atmospheric, and astrophysical variability. Meanwhile, simulators enable survey strategy optimization by allowing rapid exploration of observing cadences and pipeline configurations to maximize scientific yield. 

Modern image simulators make different trade-offs in modular, realistic stellar and galaxy populations, optical system and sensor effects. Photon Monte-Carlo engines such as PhoSim \citep{2015ApJS..218...14P} provide detailed, first-principles (used in LSST telescope simulations) but are computationally intensive; SkyMaker \citep{2010ascl.soft10066B} is lightweight and user friendly but omits certain instrument-specific, atmospheric effects and time-domain astrophysical variability. Other tools focus on the morphology of galaxies and cosmological signals (e.g., \citealt{2012MNRAS.423.3163K, 2015A&C....10..121R}). As a result, many existing simulators are either overly time-consuming, or are tailored to galaxy and cosmology science, or are highly project-specific. These tools are not optimized to produce realistic, labeled multi-epoch image sequences movies with the throughput and flexibility required for practical use, especially for validating pipeline development on robotic telescopes, time-domain and ML applications.

To address these needs we present an astronomical sky image flow simulator (AstroSkyFlow) which is specifically designed to generate high-fidelity, multi-epoch image sequences for survey pipeline validation and ML training. AstroSkyFlow is a real-time schedule-driven simulator that accepts user schedules including transits, flares, binaries, occultations, supernovae, satellite passes and asteroids. Furthermore, it produces per-exposure FITS images and per-target products. A key advantage of the AstroSkyFlow is its ability to generate real-time photometric and motion variations on demand. Meanwhile, the scheduler input is considered therefore can be easily integrated into the real observational control system (OCS) and autonomous data processing pipeline by replacing the image simulation with the image capture command. Our framework aims to maximize configurability while preserving the observational realism that matters for photometry and transient detection. Although the AstroSkyFlow was originally designed for Tianyu, it is not exclusively applicable to Tianyu. By setting a series of straightforward input parameters, the simulator can be re-configured for any observation system without rewriting physics modules. This design makes it especially suited for algorithm testing, end-to-end pipeline benchmarking, data-challenge generation and machine-learning applications.

The structure of the paper is as follows. Section \ref{sec:arc} summarizes the top-level design of the AstroSkyFlow, including the overall architecture and main module components; Section \ref{sec:image} introduces realization of the core module image simulation including two sequential parts: an external scene photon simulator that models celestial sources and astrophysical fields, followed by a hardware simulator that incorporates the optics system and sensor response; Section \ref{sec:cite} shows the image results and the comparison with real observation and SkyMaker. Section \ref{sec: con} gives a summary and discussion on this work.



\section{Design of the simulator\label{sec:arc}} 

\subsection{Overall Architecture}
\begin{figure*}[ht!]
\includegraphics[width=1\textwidth]{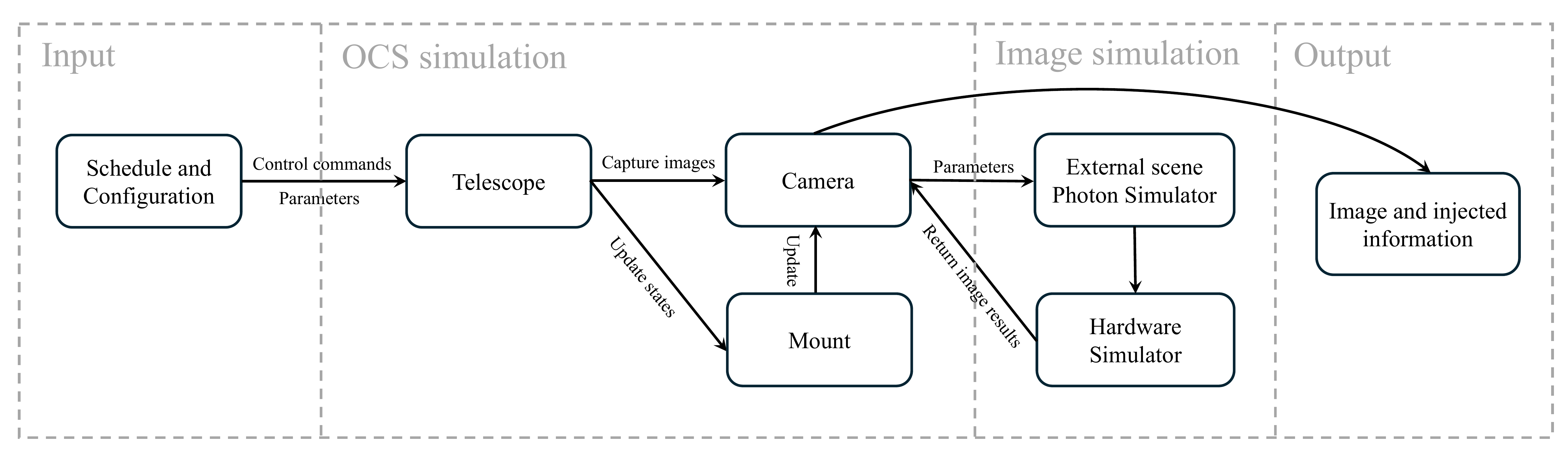}
\caption{Flow chart of the simulation pipeline's main processes.}
\label{fig:flow}
\end{figure*}

The simulator models the whole-chain process that a telescope acquires images from the observing schedule. The architecture of AstroSkyFlow is shown in Fig. \ref{fig:flow}. Excluding input and output modules, AstroSkyFlow consists of two components: an OCS simulator and an image simulator. Upon receiving instructions from the scheduler, the OCS simulator slews the virtual telescope to the target sky region. 
After the telescope has slewed, the camera initiates image simulation to produce images based on the telescope state and observing-site conditions. The image simulation consists of two parts: an external scene photon simulator which models celestial sources and astrophysical fields, and a hardware simulator that incorporates the optics system and sensor response.
Read-out time is specified in the configuration file and is used both by the OCS simulator and by the image simulator. During read-out procedure the camera is considered busy and the OCS will not issue new exposure commands. The read-out time directly affects the duty cycle and scheduling. 
In this architecture AstroSkyFlow effectively maps an observing schedule to synthetic images, making it directly useful for autonomous telescope development and for integration into an observatory control or data-processing pipeline.

\subsection{OCS simulation}
The OCS simulation emulates the operational behavior of a real observatory control system. Given a chronological observing schedule, it updates the mount status, simulates pointing to specific regions of the sky which involves calculating the required slewing time by taking into account angular distance, stabilization time and tracking speed, and incorporates pointing errors into the simulation. The module also evaluates visibility constraints including the target altitude and safety limits. Subsequently, it issues exposure commands at the scheduled times to initiate image capture, rejects requests when the camera is occupied and tracks the camera's idle status. 

\subsection{Image simulation}
\begin{figure*}[ht!]
\centering
\includegraphics[width=0.8\textwidth]{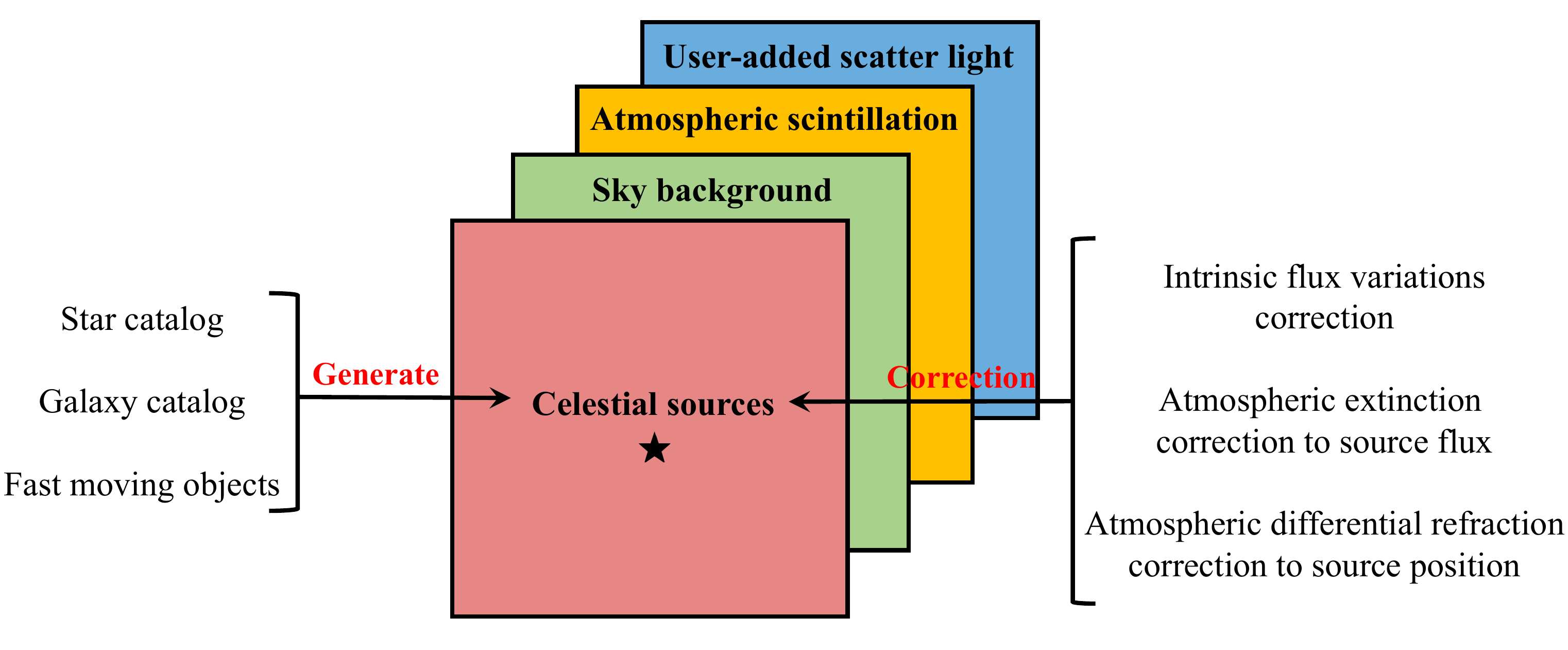}
\caption{Structure and specific implementation process of external scene photon simulator.}
\label{fig:image_photon}
\end{figure*}

Once the OCS confirms the telescope is pointing at the target coordinates steadily, the image simulation is triggered which serves as the core functionality of the entire simulator for generating astronomical images. This simulation includes two parts: an external scene photon simulator and a hardware simulator. External scene photon simulator synthesizes a comprehensive distribution of light reaching a certain plane from all possible sources and various effects. It begins by generating celestial sources including the point spread function (PSF) of point sources, the morphological structures of galaxies, and characteristic streaks of fast moving objects. These source models are then corrected for intrinsic and observational effects: intrinsic photometric variability, atmospheric extinction which attenuates source flux, and atmospheric differential refraction (ADR) which shifts source coordinates\footnote{The ADR correction is implemented as a configurable option. If the simulated telescope is equipped with an atmospheric dispersion corrector (ADC), this effect can be disabled by the user.}. Then we add the diffuse sky background field and spatially correlated scintillation field. To further enhance flexibility, user can add scatter light field, supporting both FITS and NumPy formats. If the user-added scatter light maps do not align with the target frame dimensions, AstroSkyFlow automatically rescales the data to the required resolution via OpenCV-based interpolation. This stage results in a raw photon count map per pixel that includes source signal and realistic atmospheric noise. The structure and specific implementation process of external scene photon simulator are shown in Fig. \ref{fig:image_photon}. Hardware simulator incorporates the optics system and sensor response. For optics system, geometric obscuration and pupil effects are taken into account through a vignetting model. The sensor response is simulated by modeling the complete chain—converting photons to electrons, then to voltage, and finally to Analog-to-Digital Units (ADUs), while incorporating key CCD or CMOS noise thereby producing realistic signatures in the final images.

\section{Realization of the image simulation} \label{sec:image}

\subsection{External scene photon simulator}
\subsubsection{Source synthesis and projection} \label{sec:source_synthesis}
Accurate rendering of celestial sources is the foundation of realistic image simulation: catalogs must be projected to the focal plane, point sources must be rendered with a realistic PSF, and extended sources must preserve morphological parameters. 
The sources in our simulation are categorized into three types: static stars, static galaxies, and fast-moving objects. 
We label stars and galaxies as ``static", because typical stellar proper motions are a few to a few tens of milliarcseconds per year, their apparent motions are negligible compared to pixel sizes for typical survey exposures. In the current version of AstroSkyFlow, we do not simulate their proper motion and parallax. Adding precise astrometric motion is a goal of a future release and can be enabled when required for long-baseline or astrometric studies. 
A second aspect of the ``static" is photometrically static at this stage. The fluxes are derived from magnitudes through the standard magnitude–flux conversion. Subsequent sections introduce corrections to time-dependent photometric variations.
The stars and galaxies are loaded from user-provided catalogs that follow a standardized format, including essential parameters: the object designation – Gaia DR3 ID, right ascension (R.A.) and declination (Decl.), magnitude and error in the specific passband. To ensure computational efficiency, we first filter these catalogs to select only the entries that fall within the field of view (FOV). This process limits the subsequent calculations to relevant sources.
In contrast, positions for fast-moving objects including satellites and asteroids are not from static catalogs but are dynamically obtained via public ephemerides and propagate to exposure times - internally subdivides each exposure into short sub-intervals to produce streaks when needed. The celestial coordinates are transformed to image pixel coordinates using a world coordinate system (WCS) derived from telescope parameters and camera geometry. 

Stars appear as point sources in images, with precise photometry and astrometry being highly sensitive to the PSF shape. When the exposure time $t_{\rm exp}$ is much shorter than the atmospheric coherence time $t_0$, the image point is determined by the instantaneous wavefront and exhibits a speckle pattern. Under long exposure averaging, these speckles are smoothed over time, yielding an approximately regular Moffat profile \citep{1969A&A.....3..455M}. According to \cite{1990JOSAA...7.1224F}, the atmospheric coherence time is much smaller:
\begin{equation}\label{eq:scint}
t_0 = 0.314\frac{r_0}{\overline{V}},\quad \mathrm{FWHM} \approx \mathrm{seeing} \approx 0.98\lambda/r_0,
\end{equation}
where $r_0$ is Fried parameter and $\overline{V}$ is the layer-weighted average representative wind speed. FWHM denotes the full width at half maximum of the profile. Consequently, under the normal exposure conditions the instantaneous speckle pattern induced by atmospheric turbulence averages to a smooth core add wing profile well represented by a Moffat function. We render each star using a two-dimensional Moffat PSF
\begin{equation}\label{eq:scint}
I(x,y) = I_0 \left[1 + \frac{(x-x_{\mathrm{center}})^2+(y-y_{\mathrm{center}})^2}{\alpha^2}\right]^{-\beta}, 
\end{equation}
where $\alpha$ and $\beta$ are shape parameters chosen from the seeing and telescope-optics statistics. The Moffat FWHM is related to $\alpha$ and $\beta$ by
\begin{equation}\label{eq:alpha_from_fwhm}
\mathrm{FWHM}/2 \;=\;  2\,\alpha\sqrt{2^{1/\beta}-1} .
\end{equation}
By requiring that the integrated profile equals the catalog total photon count $F_{\mathrm{tot}}$, we obtain the analytic normalization
\begin{equation}\label{eq:total}
F_{\mathrm{tot}}= \pi I_0 \frac{\alpha^2}{\beta-1}.
\end{equation}
Hence the peak intensity amplitude $I_0$ is
\begin{equation}\label{eq:I0_from_Ftot}
I_0 \;=\; \frac{F_{\mathrm{tot}}(\beta-1)}{\pi\,\alpha^2}.
\end{equation}
In practice, the Moffat convolution is computationally truncated beyond a radial distance where $I(x,y)$ becomes negligible compared to various noises. Typically, the convolution region is limited to 5 – 15 times the FWHM, which balances computational efficiency with photometric accuracy. 

\begin{figure*}[ht!]
\centering
\includegraphics[width=0.9\textwidth]{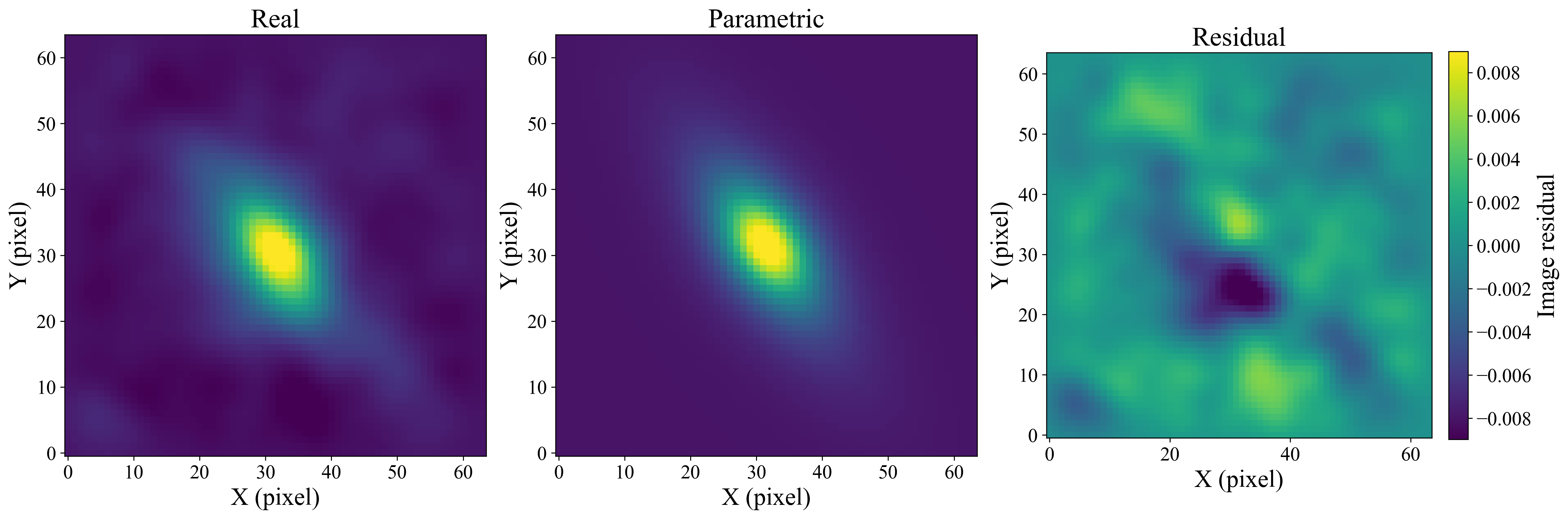}
\caption{The real galaxy, parametric reconstruction and their residual.}
\label{fig:galaxy}
\end{figure*}

Galaxies are modeled using parameterized surface-brightness profiles to ensure that their simulated morphology, axis ratio, and position angle align with the catalog inputs. Specifically, we employ Sérsic profiles \citep{1963BAAA....6...41S} based on the catalog-supplied axis-ratio and position angle. 
Sérsic model is appropriate for morphologically simple, effectively single-component galaxies, but it is not intended to capture the full structural diversity of more complex systems.
To map these distributions onto each pixel grid, the total flux is first determined using the catalog photometry and the instrumental passband. Subsequently, we utilize the GalSim \citep{2015A&C....10..121R} to represent the surface brightness and perform accurate image convolution and rendering. 
To validate the parametric implementation of GalSim, we use real galaxy images, including Hubble Space Telescope observations, provided by GalSim\footnote{\url{https://galsim-developers.github.io/GalSim/_build/html/real_gal.html}}. Following the catalog definitions of \cite{2010ApJ...709...97L} and parametric models derived by fitting procedure of \cite{2012MNRAS.421.2277L}, we select 100 galaxies that are a prior considered viable for a single-Sérsic representation. The agreement between the real galaxies and our parametric reconstructions is quantified using four complementary metrics: pixel residual, radial-profile relative error, adaptive-moment size relative error $|\Delta \sigma|$, and adaptive-moment shape difference $|\Delta e|$. Our results summarized as mean, median, and $90^{\mathrm{th}}$ percentile of four metrics demonstrate high fidelity: pixel residuals ($0.132, 0.123, 0.200$), radial-profile errors ($0.031, 0.026, 0.051$), size relative errors $|\Delta \sigma|$ ($0.023, 0.017, 0.046$), and shape differences $|\Delta e|$ ($0.042, 0.033, 0.089$). Fig. \ref{fig:galaxy} shows the real galaxy, parametric reconstruction and their residual for one example. We have validated the Sérsic-based simulation for simple galaxies, and modeling complex morphologies remains a target for future improvements.

\begin{figure*}[ht!]
\centering
\includegraphics[width=0.8\textwidth]{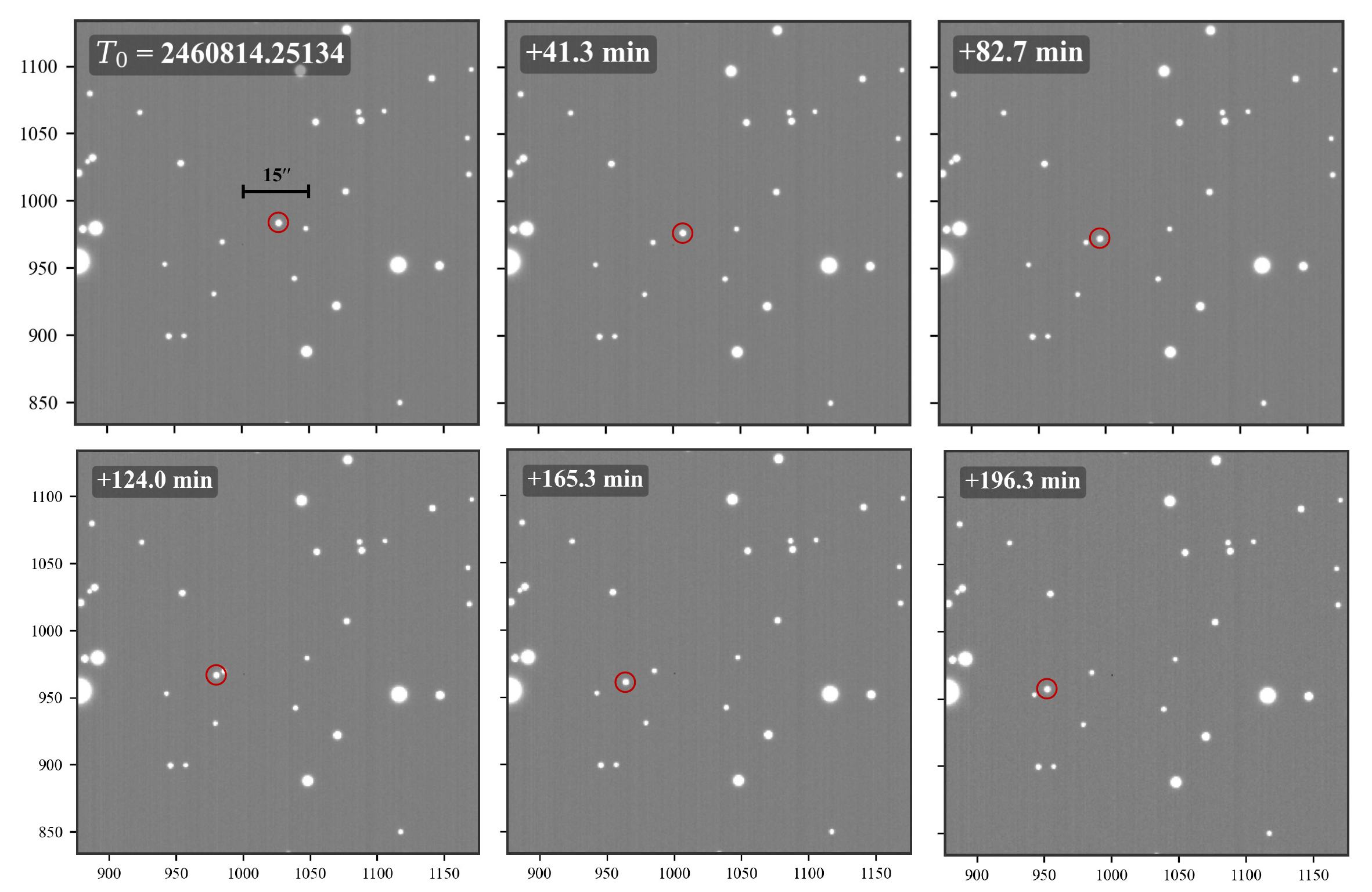}
\caption{The slow motion of asteroid  582 Olymias (A906 BN) in continuous observation. The pixel scale is 0.303 arcsecond per pixel. The movement is below the threshold. We regard it as a stationary point source.}
\label{fig:aster}
\end{figure*}

For the satellite, we get the Two Line Element from the celestrak\footnote{\url{https://celestrak.org}}. Only some public satellites are available and their data are time-sensitive. During the simulation, the latest Two Line Element data are needed as the input file. We use the skyfield package \citep{2019ascl.soft07024R} to determine whether the satellite is within the field of view and in the sunlight for each exposure. After determining the start and end pixel coordinates, 
we simulate satellite tracks by propagating the object position through the exposure and representing each sampled point with a Moffat PSF. AstroSkyFlow supports temporal oversampling within a single exposure to model trailing from fast-moving objects: we can control the number of sub-sample positions per pixel.
For asteroid data, NASA provides an application programming interface (API) via the Jet Propulsion Laboratory’s Solar System Dynamics (SSD) and Center for Near-Earth Object Studies (CNEOS) services. Using the SB identification API, we can identify small bodies within a specified field of view.
To determine how these objects appear in the final image, AstroSkyFlow compares their total displacement during the exposure to a user-defined pixel threshold. If the movement is below the threshold, the object is treated as a stationary point source. An example of this is 582 Olympias (A906 BN), as shown in Fig. \ref{fig:aster}. If the motion exceeds the threshold, the object forms a streak within the single exposure. We employ the same temporal oversampling method used for satellite simulations to produce the streaks of asteroids. Examples of trails caused by the rapid motion of asteroids are provided in Fig. \ref{fig:muguang_ios} and \ref{fig:tianyu_ios} in Section \ref{r and i}.

\subsubsection{Intrinsic flux variations}
High-fidelity astronomical images must go beyond static sources, because many astrophysical sources exhibit measurable brightness variations over the course of an observing sequence. These include periodic phenomena such as exoplanet transits \citep{2010exop.book...55W} and eclipsing binaries \citep{KallrathMilone2009}, stochastic variability such as stellar flares \citep{2024LRSP...21....1K}, obscuration events such as occultations \citep{2022MNRAS.511.1167G}, and transients such as supernovae \citep{1997ARA&A..35..309F}. Accurate photometric variability is essential for science and pipeline validation: it enables faithful recovery of light curves and transient parameters, prevents systematic biases in photometry, provides realistic negative or positive examples for machine-learning classifiers, and permits robust tests of detection efficiency and false-positive rates. In our pipeline, we correct intrinsic flux variations in two complementary ways: First, for user-specified variable sources of interest we generate the light curves using the physical models. The light curves for different variable sources are generated using distinct models. Second, for variable sources in the field that are not included in the user-specified source list, we automatically identify known variables by cross-matching the field catalog with the Variable Star Index (VSX) database\footnote{\url{https://www.aavso.org/vsx/index.php?view=search.top}} and model their variability patterns based on archival TESS and Kepler light curves. The former approach takes precedence over the latter when both are applicable. The resulting relative flux modulation is used to replace the corresponding entry in the master star catalog. Through these two methods, images containing real-time photometric information can be obtained.

 For transiting systems, we use Batman \citep{2015PASP..127.1161K} to generate normalized relative flux change of host stars according to the user-provided transit catalog. This flux replaces the same target's flux in the total star catalog. It is worth noting that Batman itself only does occlusion in geometric positions. Limb-darkening coefficients which are relevant to the filters must be supplied correspond to the system bandpass. 
 
 For eclipsing binaries, we employ Phoebe \citep{2016ApJS..227...29P} to generate normalized relative flux variations according to the user-provided binary catalog, and replace the corresponding flux. Before using Phoebe, we follow the tutorial\footnote{\url{https://phoebe-project.org/docs/2.4/tutorials/passbands}} to add the wavelength and transmission curves of telescope system to Phoebe for subsequent use. It is crucial that the user-defined band's name in the configuration of AstroSkyFlow matches the corresponding bandpass name registered in the Phoebe. Alternatively, one may directly use the bands provided within Phoebe. Considering the constraints between different physical parameters in Phoebe, different binary morphologies  including Algol-type (EA, detached), $\beta$ Lyrae-type (EB, semi-contact), and W Ursae Majoris-type (EW, contact) require distinct parameters. We adopt Phoebe’s default atmospheric model \citep{2003IAUS..210P.A20C} and limb-darkening interpolation. We fall back to a blackbody approximation by setting limb-darkening coefficients manually in case of modeling failure.

For stellar flares, we adopt the analytic white-light flare template of \citet{2022AJ....164...17T}, consisting of a Gaussian heating pulse convolved with a biexponential cooling profile. Each flare is parameterized by an amplitude, full timing width at half maximum ($t_{1/2}$), and central time, and the template is scaled to match fitted cadence-dependent parameters. This template was primarily calibrated using Kepler data and subsequently validated with TESS observations, which utilize a broader passband with a longer effective wavelength. The results show that the white-light approximation produces only minor band-dependent differences and can be applied directly in the visible and near-infrared ranges. According to the user-provided flare catalog, we calculate the relative flux variations and replace the corresponding flux in total catalog.

For the light curve of occultation, first, we use the wavelength and transmission curves of the telescope system to calculate the wavelength-dependent coefficients for subsequent use. According to the user-provided occultation catalog, we calculate the Fresnel scale and the dimensionless parameter $\rho$ \citep{1987AJ.....93.1549R, 2007AJ....134.1596N} to determine whether diffraction effects are significant. When $\rho$ is relatively large, geometric approximation is adopted for calculation. When $\rho$is relatively small, the complete Fresnel diffraction pattern is solved using the Lommer series \citep{1987AJ.....93.1549R, 2007AJ....134.1596N}.
The Fresnel scale is
\begin{equation}
F(\lambda)\;=\;\sqrt{\frac{D_{\mathrm {obs}}\,\lambda}{2}} , 
\end{equation}
which sets the transverse length over which diffraction is important, and form the dimensionless ratio to decide the regime:
\begin{equation}
\rho \;=\; \frac{r_{\mathrm {ast}}}{F(\lambda)} ,
\end{equation}
if the ratio is larger than the threshold, we use a fast geometric approximation to calculate the normalized flux based on the overlap area of two disks and their brightness ratio. And if ratio is smaller, we compute the full Fresnel diffraction pattern using the Lommel series $U_\mathrm{n}$ 
\begin{equation}
U_\mathrm n(\mu,\nu)\;=\;\sum_{k=0}^{\infty} (-1)^k\left(\frac{\mu}{\nu}\right)^{\,n+2k}\,J_{\,n+2k}\!\bigl(\pi\,\mu\,\nu\bigr),
\end{equation}
where \(J_m\) is the Bessel function of the first kind of order \(m\),
\begin{equation}
\eta \;=\; \frac{|r_{\mathrm{position}}|}{F(\lambda)},\qquad
\beta \;=\; \frac{\pi}{2}\bigl(\eta^2+\rho^2\bigr),
\end{equation}
inside the occulting disk ($\eta <= \rho$):
\begin{equation}
\text{for }\eta \le \rho:\qquad
I(\eta,\rho) \;=\; U_0(\eta,\rho)^{2} + U_1(\eta,\rho)^{2},
\end{equation}
outside the occulting disk ($\eta > \rho$):
\begin{equation}
\text{for }\eta > \rho:\qquad
I(\eta,\rho) \;=\; 1 + U_1(\rho,\eta)^{2} + U_2(\rho,\eta)^{2}
         - 2\,U_1(\rho,\eta)\,\sin\beta + 2\,U_2(\rho,\eta)\,\cos\beta .
\end{equation}
This flux replaces the same target's flux in the total star catalog.

\begin{figure*}[ht!]
  \gridline{%
    \fig{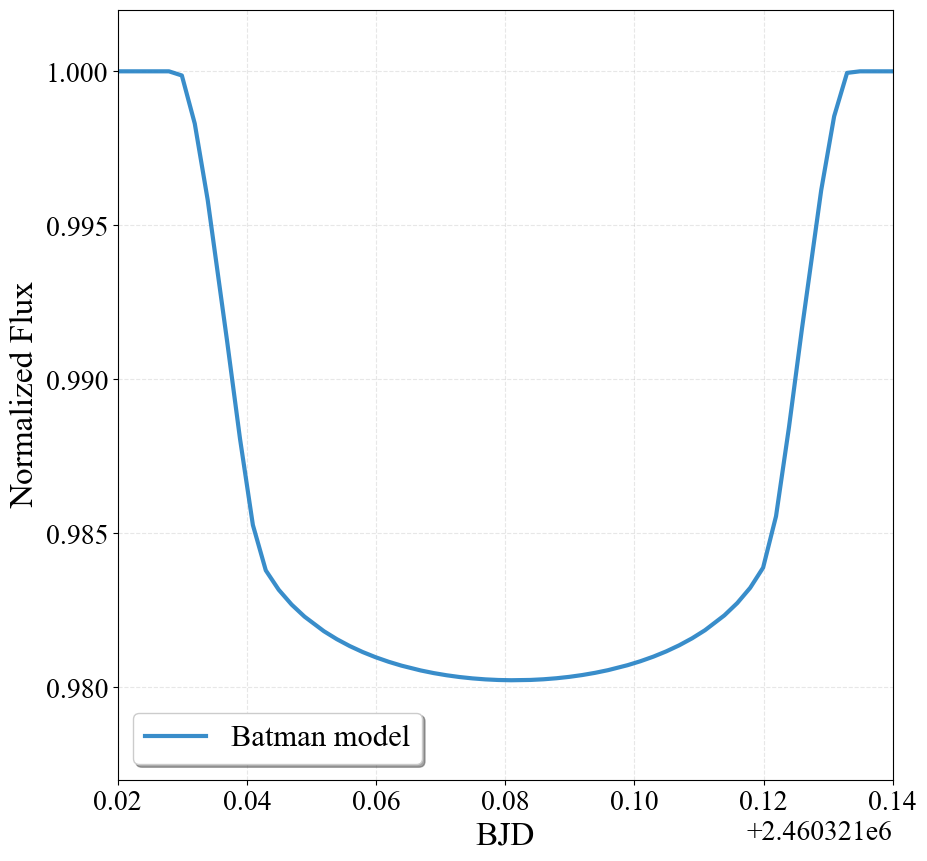}{0.28\textwidth}{(a) Transit model light curve.}%
    \fig{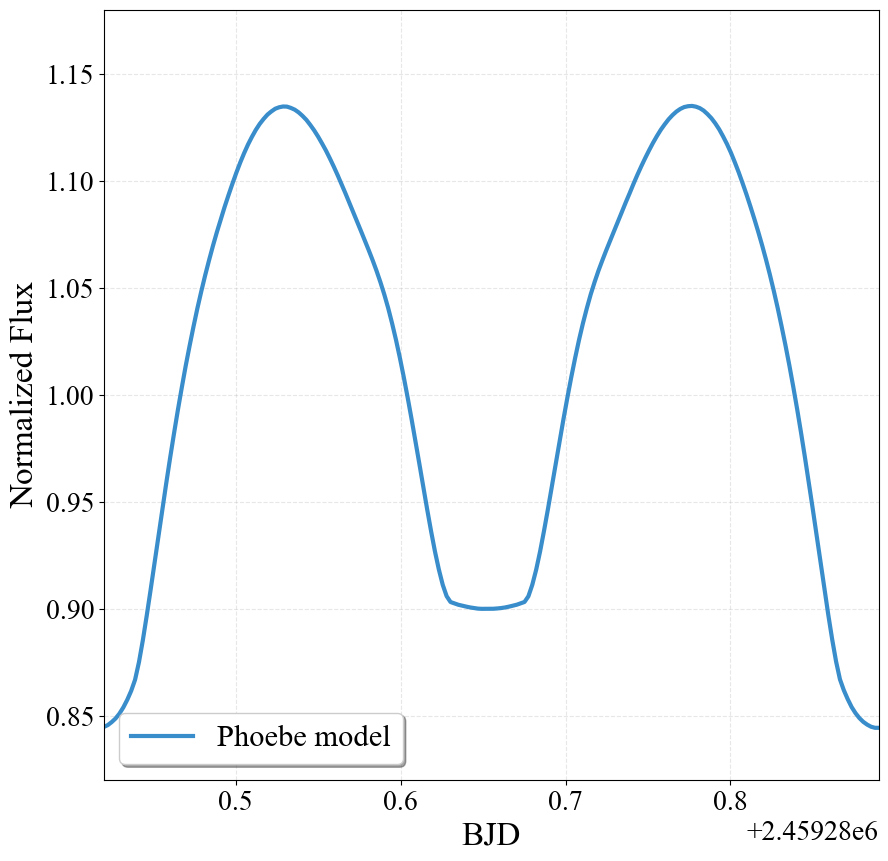}{0.28\textwidth}{(b) EW binary model light curve.}%
    \fig{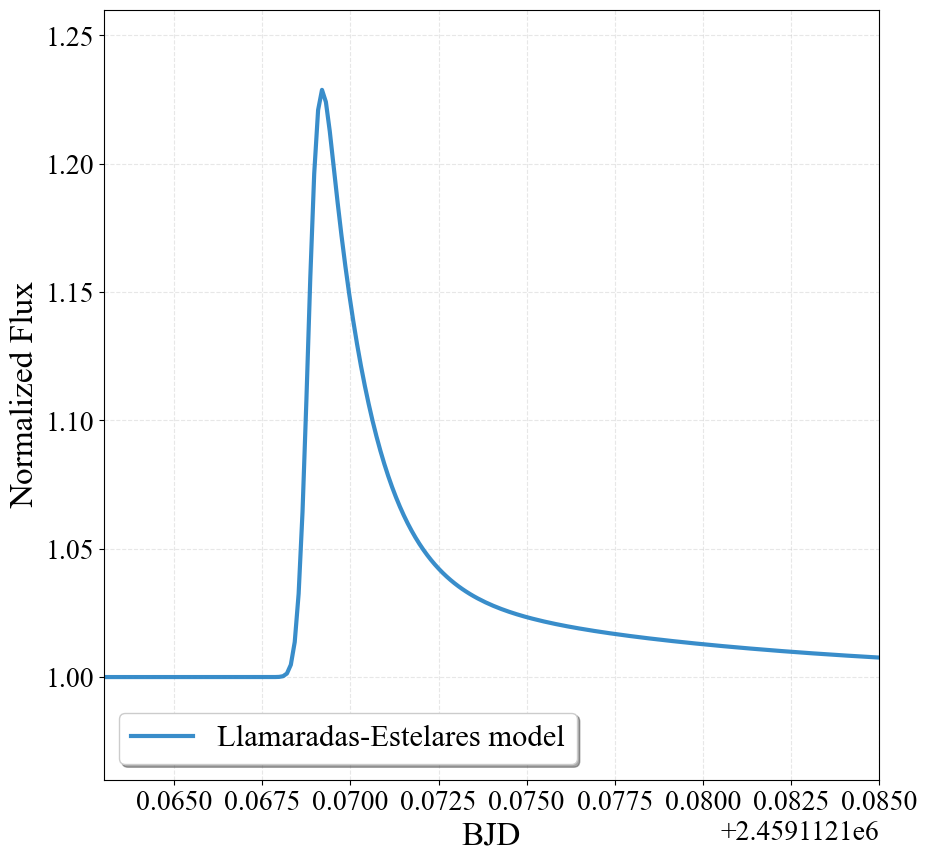}{0.3\textwidth}{(c) Flare model light curve.}%
  }
  \gridline{%
    \fig{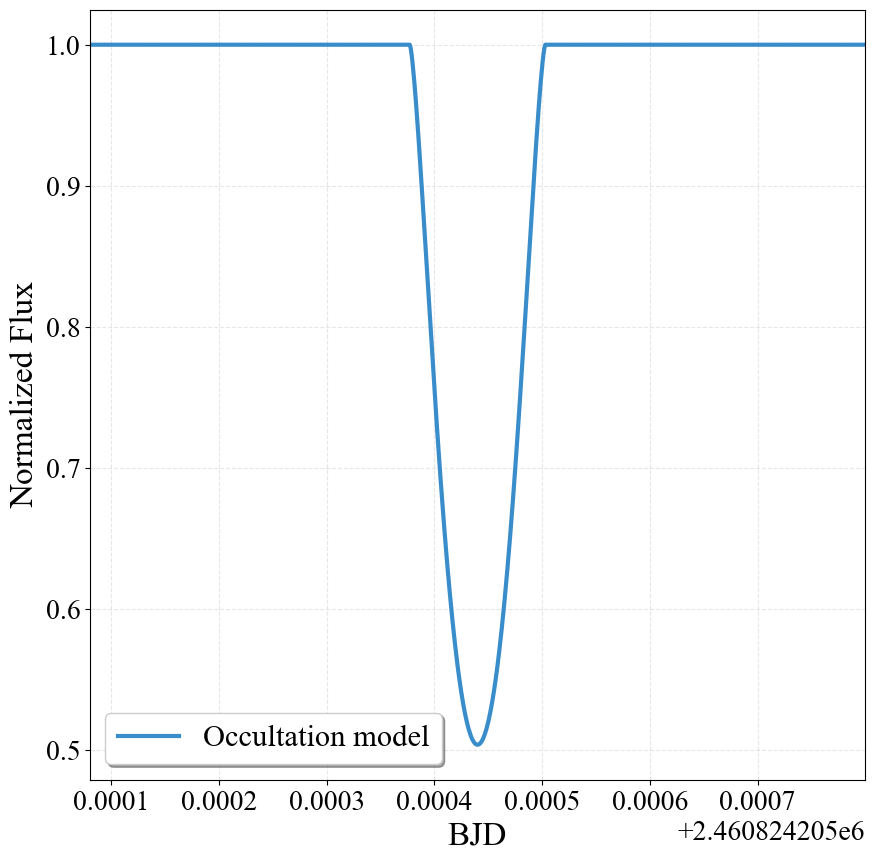}{0.28\textwidth}{(d) Occultation model light curve with geometric approximation.}
    \fig{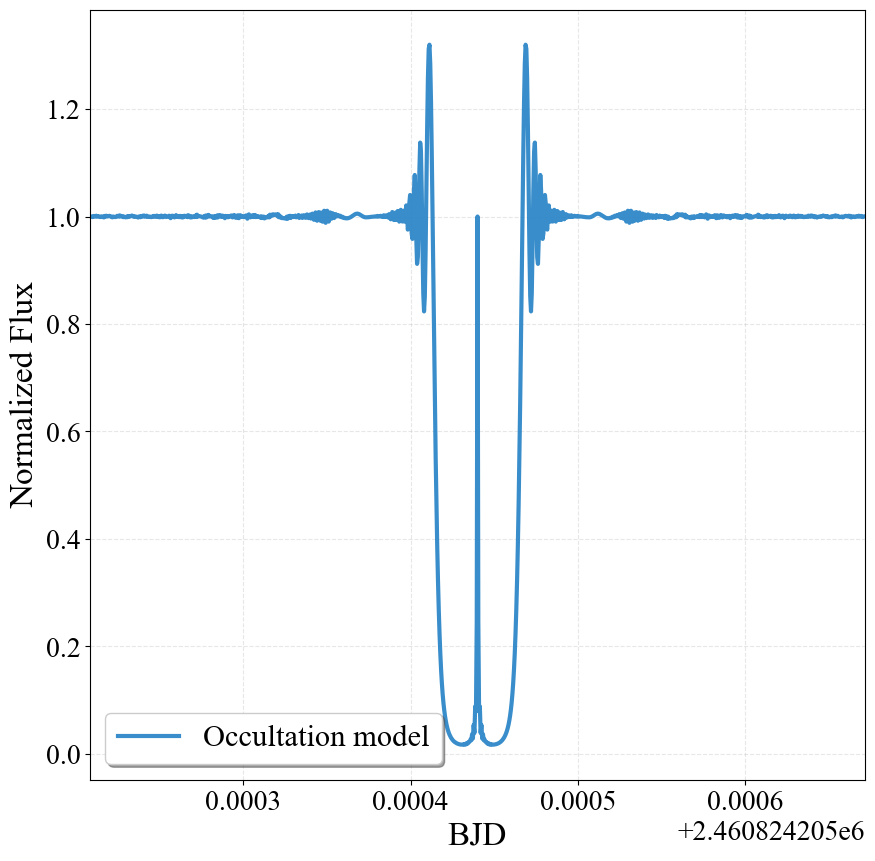}{0.28\textwidth}{(e) Occultation model light curve in Fresnel diffraction pattern.}%
    \fig{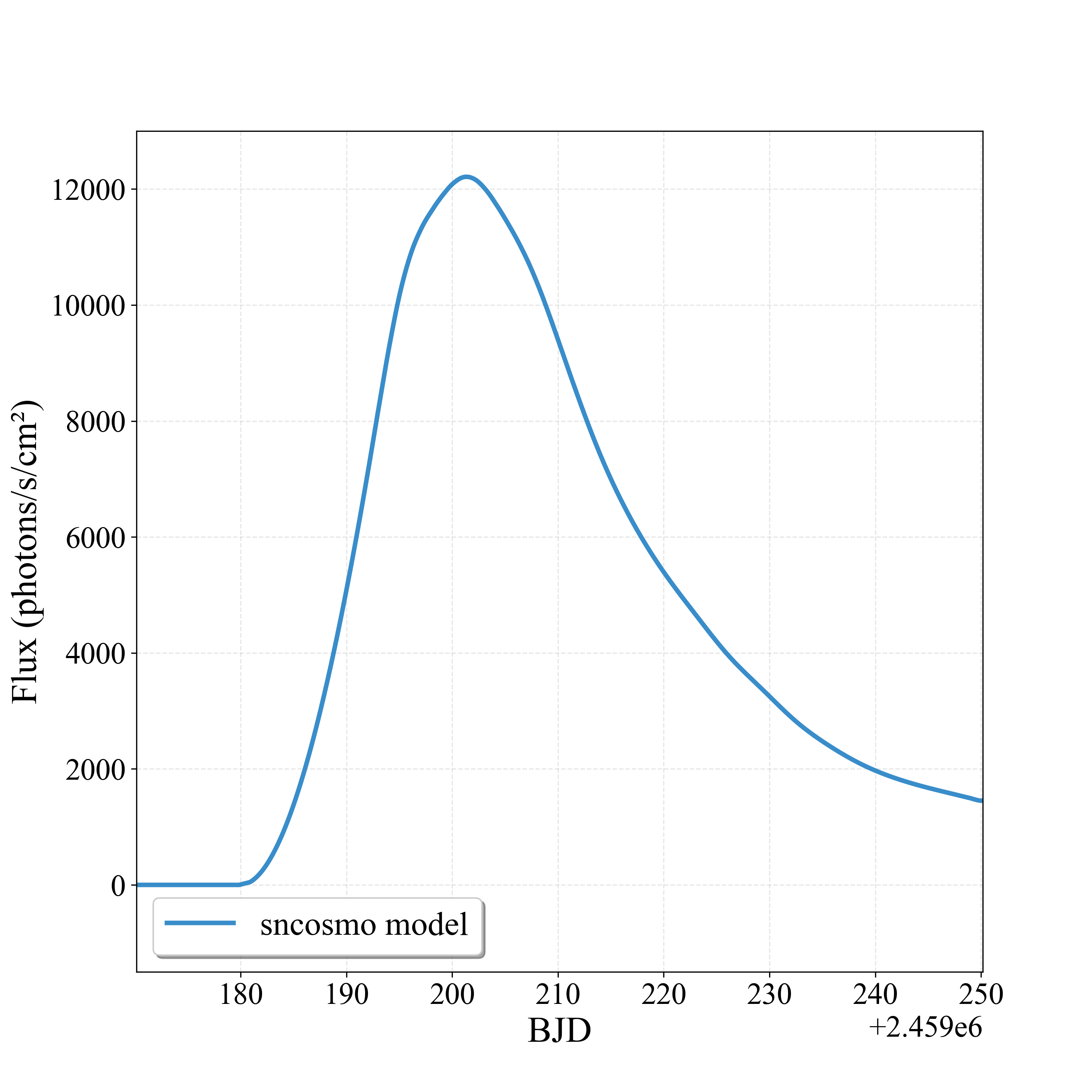}{0.3\textwidth}{(f) Supernova model light curve.}%
  }
  \caption{Simulated light curves of different types of variables using different models. And the specific parameters employed during the simulation are listed in Table \ref{tab:target_parameters}.}
  \label{fig:show}
\end{figure*}

For the supernovae, we use sncosmo \citep{barbary_2025_15019859} to generate the flux change of supernovae according to the user-provided supernova catalog. Considering that registering a new bandpass in sncosmo is a straightforward process, we have integrated this functionality directly into our pipeline following the official tutorial\footnote{\url{https://sncosmo.readthedocs.io/en/stable/bandpasses.html}}. We only need to provide the filter and transmission information. Subsequently, to select host galaxies for the supernovae, we use rejection sampling \citep{2007nras.book.....P}.
The purpose of rejection sampling is to generate supernova samples that follow the target redshift distribution. In our implementation, for a galaxy catalog of size $N$, the proposal distribution is taken to be uniform, $g(z) = 1/N$, where $z$ denotes the redshift. The function $R(z)$ represents the assumed redshift-dependent supernova rate model. To ensure valid rejection sampling, we define a scaling constant $M = \max[R(z)/g(z)]$, such that $M g(z) \geq R(z)$ is satisfied for all $z$ within the catalog's domain. For each iteration, a candidate galaxy with redshift $z_0$ is first drawn from the proposal distribution. Subsequently, a random variable $u_0$ is sampled from a uniform distribution $\mathcal{U}(0, M g(z_0))$. The candidate is accepted as a host index if $u_0 < R(z_0)$; otherwise, it is rejected. Through this iterative process, the resulting subsample converges to the target distribution $R(z)$. Supernova occurrence probability can be described using an analytical form $R(z) \propto (1+z)^\alpha$ \citep{2008ApJ...682..262D, 2015ApJ...813...93S}. In this implementation, we adopt a simplified analytical form,
\begin{equation}\label{eq:R_z}
R(z) = 3 \times 10^{-4} \times (1+z)^{2.11}.
\end{equation}
After selecting the host galaxies, we superimpose the photon distribution of the supernovae onto the host galaxies. Fig. \ref{fig:show} presents the simulated light curves of different types variables using the method described above. All parameters employed in the simulations are comprehensively listed in Table \ref{tab:target_parameters}.

\begin{table}[htbp]
\centering
\caption{Physical and Observational Parameters of Different Astronomical Targets}
\label{tab:target_parameters}
\resizebox{\textwidth}{!}{%
\begin{tabular}{lcclcll}
\hline
Variable type & Target name & Reference & Parameter & Description & Value & Unit \\
\hline
 transit & WASP-11 b & \cite{2017AA...602A.107B} &$T_0$ & reference mid-transit time &  2454759.68753& BJD(TDB) \\
& & &$P$ & orbital period & 3.7224793 & day \\
& & & $R_\mathrm p/R_\star$ & planet-to-star radius ratio & 0.132 & -- \\
& & &$a/R_\star$ & scaled semi-major axis & 12.71 & $R_\star$ \\
& & &$i$ & orbital inclination & 89.03 & degree \\
& & &$e$ & eccentricity & 0 & -- \\
& & &$u_1$ & limb darkening coefficient 1 & 0.5 & -- \\
& & &$u_2$ & limb darkening coefficient 2 & 0.1 & -- \\
& & &$u_3$ & limb darkening coefficient 3 & 0.1 & -- \\
& & &$u_4$ & limb darkening coefficient 4 & -0.1 & -- \\
\hline
binary (EW) & UCAC4 554-003513 & \cite{2023PASP..135e4201L} & $T_0$ & reference primary eclipse time & 2459162.296 & BJD (TDB) \\
& & &$P$ & orbital period & 0.4724851 & day \\
& & &$q$ & mass ratio ($M_2/M_1$) & 0.129 & -- \\
& & &$i$ & orbital inclination & 80 & degree \\
& & &$a$ & semi-major axis & 2.94 & $R_\odot$ \\
& & &$T_{\text{eff},1}$ & primary effective temperature & 6928 & K \\
& & &$R_1$ & primary radius & 1.67 & $R_\odot$ \\
& & &$T_{\text{eff},2}$ & secondary effective temperature & 6570 & K \\
& & &$R_2$ & secondary radius & 1.5 & $R_\odot$ \\
\hline
flare & TIC 197829751 & \cite{2022AJ....164...17T} & $t_{\text{peak}}$ & time of peak brightness & 2460824.22 & BJD (TDB) \\
& & &FWHM & full width at half maximum & 0.05 & day \\
& & &$\Delta F_{\text{peak}}$ & peak fractional amplitude & 0.25 & -- \\
\hline
occultation & simulated target 1 & -- & $T_{\text{ref}}$ & reference time (center of the event) & 2460824.20544 & BJD (TDB) \\
& & & Mode & calculation mode: monochromatic or filters' name & V & -- \\
& & & $\lambda$ & wavelength in \AA (if monochromatic) & -- & -- \\
& & & $b$ & impact parameter & 2.0 & km \\
& & & $R_{\text{occulter}}$ & occulter radius & 2000 & m \\
& & & $D_{\text{obs}}$ & occulted object distance & $1.496 \times 10^{11}$ & m \\
& & & $\theta_\star$ & stellar angular diameter & 0.2 & mas \\
& & & $v_{\text{rel}}$ & relative speed of occulter& 0.1 & km/s \\
& & & $f$ & brightness ratio between the occulter and the occulted object & 0.0 & -- \\
& & & $\rho_{\text{limit}}$ & geometry or fresnel diffraction threshold & 5.0 & -- \\
\hline
occultation & simulated target 2 & -- & $T_{\text{ref}}$ & reference time (center of the event) & 2460824.20544 & BJD (TDB) \\
& & & Mode & calculation mode: monochromatic or filters' name & V & -- \\
& & & $\lambda$ & wavelength in \AA (if monochromatic) & -- & -- \\
& & & $b$ & impact parameter & 2.0 & km \\
& & & $R_{\text{occulter}}$ & occulter radius & 1000 & m \\
& & & $D_{\text{obs}}$ & occulted object distance & $1.496 \times 10^{11}$ & m \\
& & & $\theta_\star$ & stellar angular diameter & 0.1 & mas \\
& & & $v_{\text{rel}}$ & relative speed of occulter & 0.5 & km/s \\
& & & $f$ & brightness ratio between the occulter and the occulted object & 0.0 & -- \\
& & & $\rho_{\mathrm {limit}}$ & geometry or fresnel diffraction threshold & 5.0 & -- \\
\hline
supernova & simulated target 3 & -- & $T_0$ & time of peak brightness & 2459200.163 & BJD (TDB)\\
& & & $z$ & redshift & 0.015 & -- \\
& & & $x_1$ & stretch parameter in SALT3 model & 0.1 & -- \\
& & & $c$ & color parameter in SALT3 model & -0.1 & -- \\
& & & filter & observation filter & V & -- \\
& & & $M_\mathrm{V}$ & peak absolute magnitude & -19.3 & mag \\
\hline
\end{tabular}
}
\end{table}

\begin{figure*}[ht!]
\centering
\includegraphics[width=0.9\textwidth]{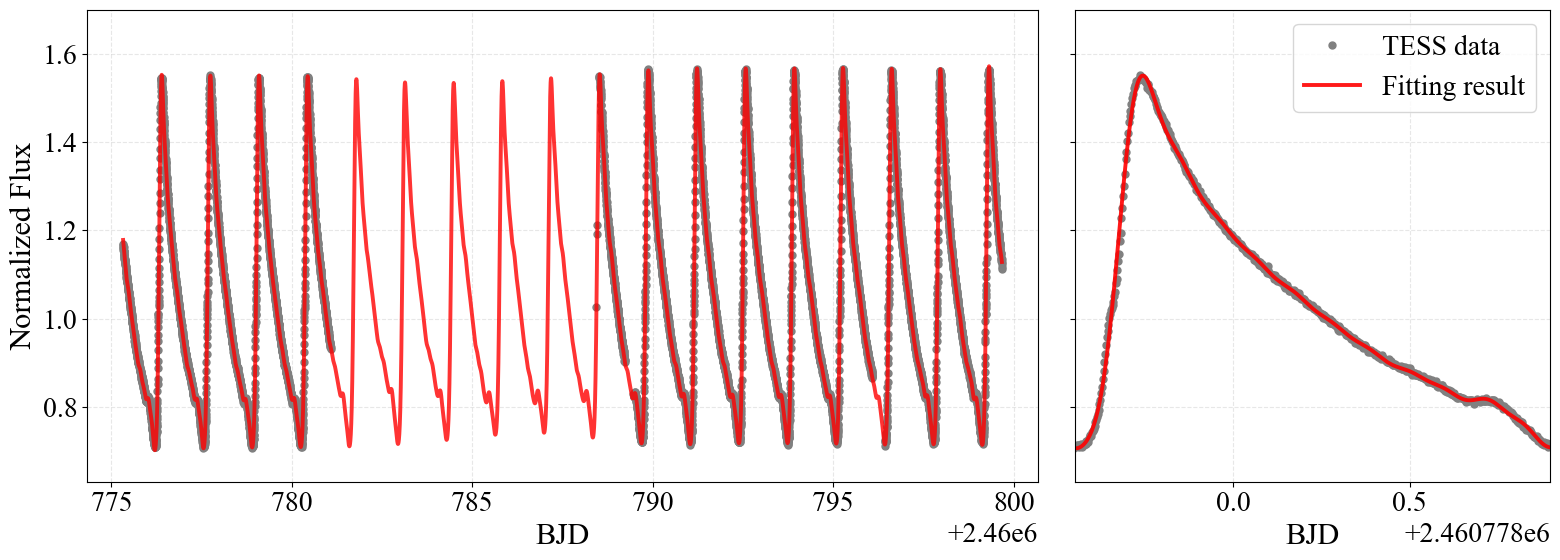}
\caption{The fitting result for a ACEP star XX Vir in TESS Sector 91. It's obtained by an iterative frequency analysis using the Lomb–Scargle periodogram. The left panel shows the TESS data points and fitting results in TESS Sector 91. The right panel displays the results when zoomed into one period.}
\label{fig:ls}
\end{figure*}

The second way is template reconstruction for the variable stars not covered by user-specified catalogs via an automated procedure. First, we select variables in star catalog of a specific FOV through cross-matching with the VSX database. For these matching results, we identify the most continuous and highest-quality light-curve data from a single TESS or Kepler sector using the lightkurve package \citep{2018ascl.soft12013L}. This also means cutting the period off at less than 27.4 days. Specifically, we prioritize data products from the Science Processing Operations Center (SPOC) pipeline over those from the Quick-Look Pipeline (QLP) due to their more rigorous preprocess. We give priority to choosing those with shorter observation cadence to increase the number of data points per light curve. To ensure the quality of these archival TESS/Kepler data and the accuracy of the reconstruction results, it is essential to calibrate these data. First, we retain only data points with a quality flag of 0, indicating that the corresponding measurements are of good quality. Measurements with nonzero quality flags are discarded, as they may be contaminated by instrumental artifacts or other systematic effects. Second, we select the most appropriate flux product available from the simple aperture photometry (SAP) flux, presearch data conditioning SAP (PDCSAP) flux, kepler-spline SAP (KSPSAP) flux, detrended flux and systematics-removed flux. If no corresponding TESS or Kepler data exists, we skip this target. For these selected data, we perform an iterative frequency analysis using the Lomb–Scargle periodogram \citep{1976Ap&SS..39..447L} which is a standard method for detecting periodic signals in time series. The specific steps are finding the peak signal, determining whether the false alarm probability (FAP) is less than the threshold value of 0.01 and using the original light curve to subtract this peak signal. We repeat until FAP is more than the 0.01. Finally, we obtain all extracted peak signals. The set of extracted frequencies, amplitudes, and phases is combined to produce the reconstructed normalized light curve and they can be used to predict flux at any epochs.
In this reconstruction we ignore possible differences in the light curve at different wavelengths and extract the shape of the normalized light curve. Fig. \ref{fig:ls} displays the filtered TESS Sector 91 data for an anomalous Cepheid (ACEP) XX Vir identified through this automated procedure, along with the fitting results.

\subsubsection{Atmospheric extinction}
\begin{figure*}[ht!]
\centering
\gridline{%
  \fig{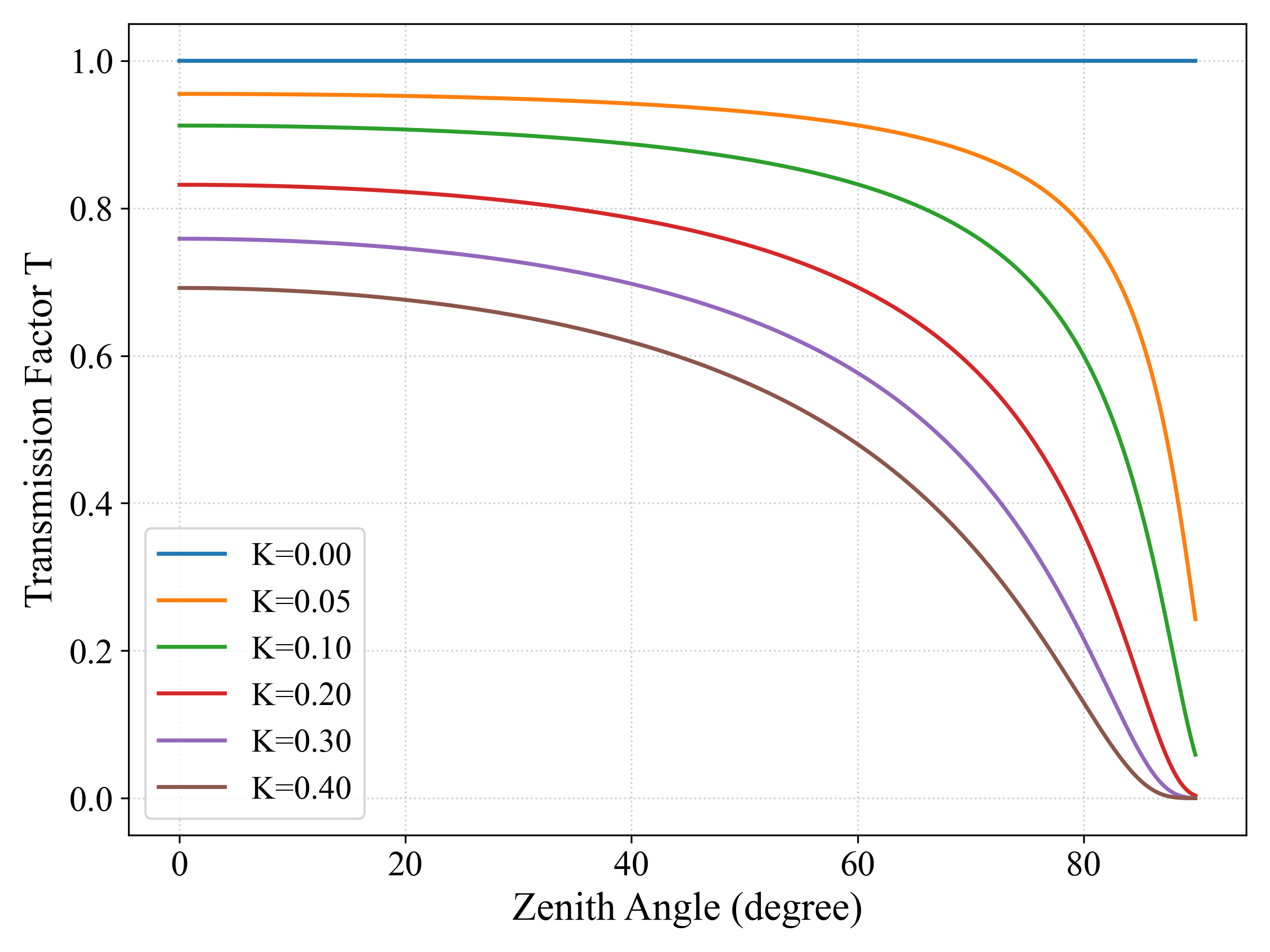}{0.43\textwidth}{(a) Transmission factor T for different zenith angle and extinction coefficient.}%
  \fig{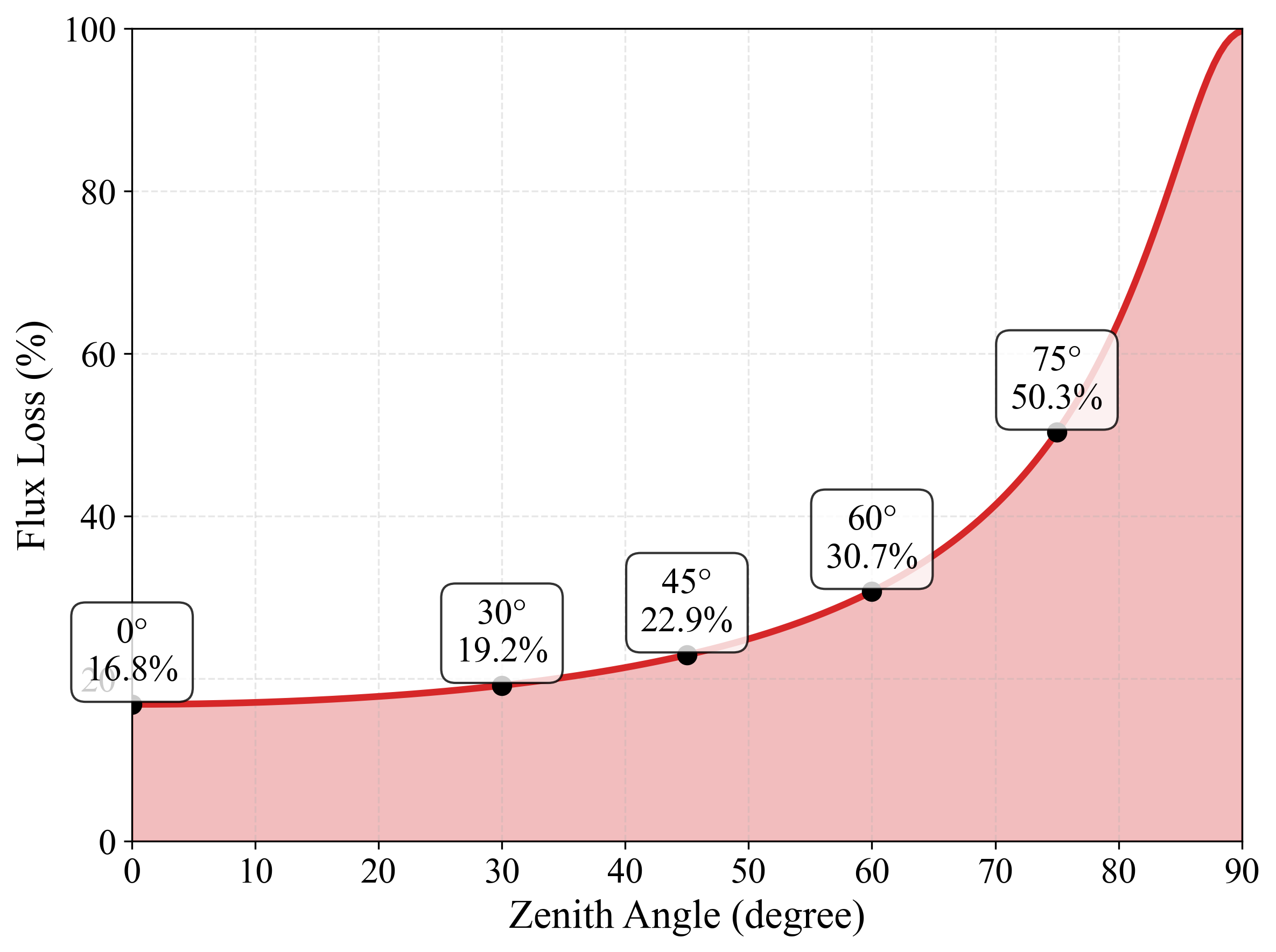}{0.43\textwidth}{(b) Flux loss due to atmospheric extinction (K = 0.20).}%
}
\caption{Atmospheric extinction at different zenith angle and extinction coefficient. The left panel illustrates how the transmission factor varies with zenith angle under different extinction coefficient. The right panel shows flux loss due to atmospheric extinction when extinction coefficient is 0.20.}
\label{fig:extinction}
\end{figure*}

Atmospheric extinction is a primary observational effect that reduces the measured flux of all celestial sources. The flux attenuation depends on the zenith angle. This correction must be applied precisely before generating the final photon map to ensure photometric precision. The airmass is evaluated using an empirical rational approximation:
\begin{equation}\label{eq:airmass}
X(z)\;=\;\frac{1.002432\,\mu^{2} + 0.148386\,\mu + 0.0096467}
{\mu^{3} + 0.149864\,\mu^{2} + 0.0102963\,\mu + 0.000303978},
\qquad \mu=\cos z,
\end{equation}
where z is the zenith angle. Clipping z to \([0,\pi/2]\) enforces physical limits near the horizon or zenith. This implementation follows the empirical fit given in the optics literature \citep{Young:94}. Given the extinction coefficient \(K\) in the unit of magnitudes per airmass, the atmospheric extinction fraction is
\begin{equation}\label{eq:transmission}
T \;=\; 10^{-0.4\,K\,X(z)}.
\end{equation}
Transmission factor \(T\) is introduced as an additional multiplier to the source flux. This accounts for the airmass-dependent dimming caused by the atmosphere. Atmospheric extinction at different zenith angle and extinction coefficient is shown in Fig. \ref{fig:extinction}. The resulting corrected flux is passed to the subsequent stages of the simulation pipeline.

\subsubsection{Atmospheric differential refraction}
ADR produces wavelength-dependent shifts in the apparent positions of celestial sources. 
Crucially, this displacement is not constant across the field of view. It varies with the sky position of each source. We adopt the method described by \cite{1996PASP..108.1051S}, which accurately computes atmospheric refraction for zenith distances below $65^{\circ}$ and wavelengths between 3000 and 10,000 \AA. This method is highly practical for most telescopes' needs as it requires only local meteorological data, such as ambient temperature, atmospheric pressure, and dew point or relative humidity.
The procedure has three stages: (1) compute local thermodynamical quantities including the dew point and water vapour pressure; (2) compute the refractive index of air $n(\lambda)$ and the auxiliary parameters $\beta$ and $\kappa$; (3) compute the refraction $R$ as an expansion in $\tan z$ and project $R$ onto equatorial components $(\Delta\alpha,\Delta\delta)$.

\begin{figure*}[ht!]
\centering
\gridline{%
  \fig{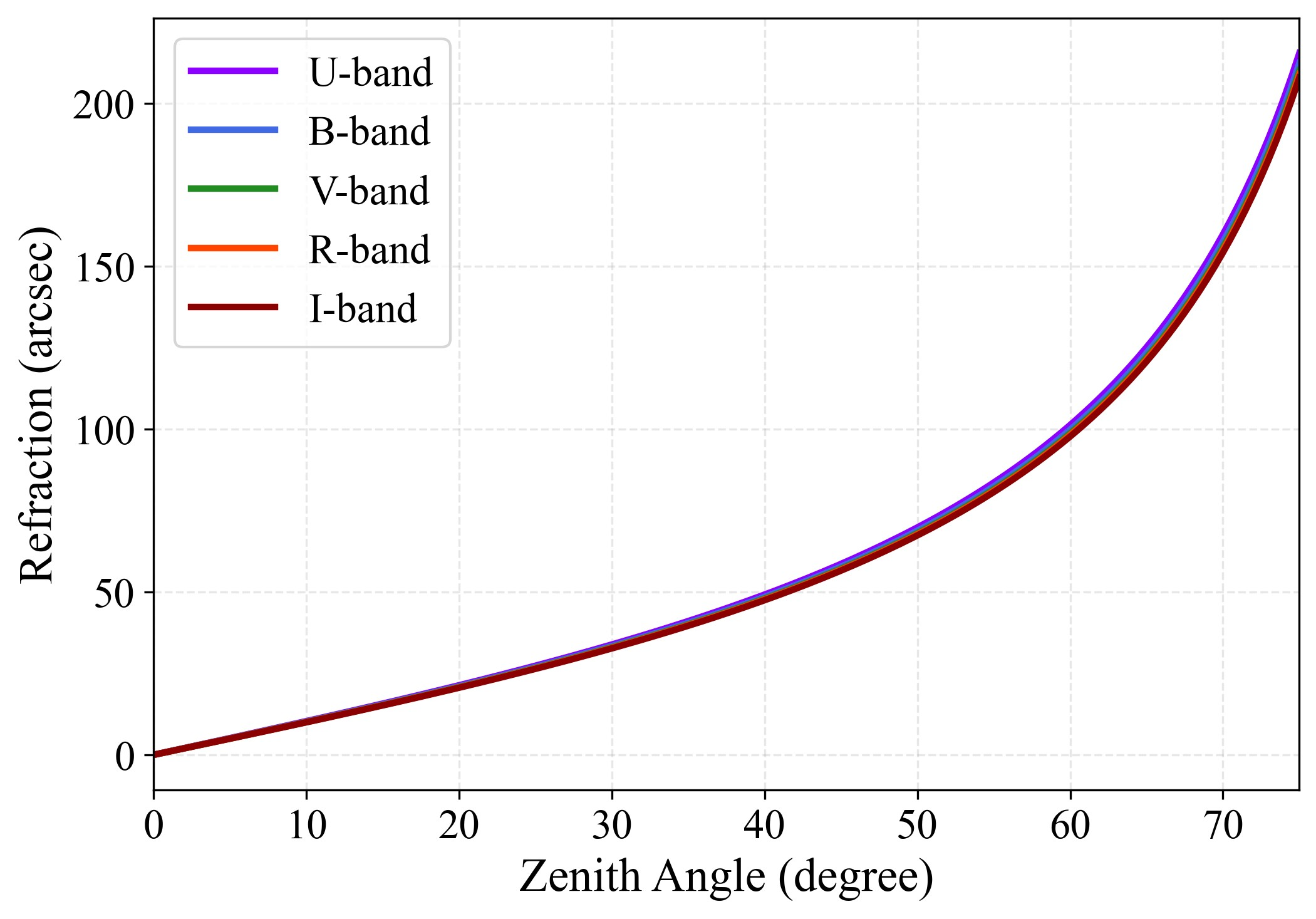}{0.43\textwidth}{(a) Atmospheric refraction at different zenith angle.}%
  \fig{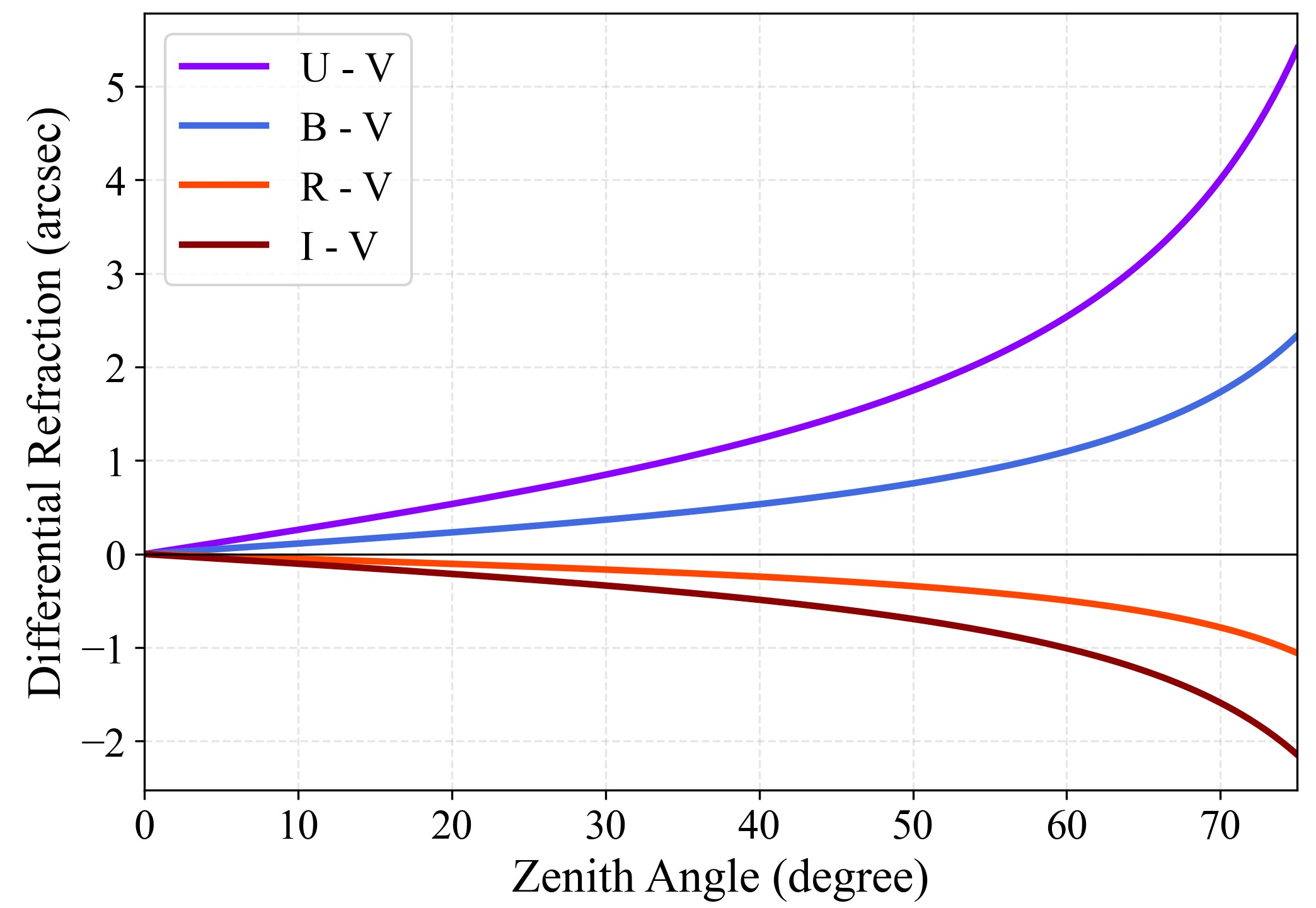}{0.43\textwidth}{(b) Differential refraction relative to V-band.}%
}
\caption{Differential atmospheric refraction at different zenith angle and band. The observation conditions were T=15$^\circ$C, $p_\mathrm s$=760mmHg,\ and RH=50$\%$. The left panel shows how atmospheric refraction varies with zenith angle under different band. The right panel shows differential refraction between other bands with V-band.}
\label{fig:ADR}
\end{figure*}

First, we can estimate the dew point $t_{\rm d}$ ($^\circ$C) with the empirical relation from temperature $t$ ($^\circ$C) and relative humidity RH (\%):
\begin{equation}\label{eq:scint}
\begin{split}
x = \ln\!\frac{\mathrm{RH}}{100}, \quad
t_{\mathrm d} = 238.3\;\frac{(t+238.3)\,x + 17.2694\,t}{(t+238.3)(17.2694-x)-17.2694\,t}\,.
\end{split}
\end{equation}
A polynomial approximation converts dew point to water vapour pressure (mmHg) by
\begin{equation}\label{eq:scint}
\begin{split}
\begin{array}{l}
p_w \;=\; 4.50874 + 0.341724t_d + 0.0106778t_d^2 + 0.000184889t_d^3 \\ \quad + 0.00000238294t_d^4 + 0.0000000203447t_d^5.
\end{array}
\end{split}
\end{equation}
Second, we calculate the dry-air and water-vapour density-like terms: 
\begin{equation}\label{eq:scint}
\begin{split}
T = 273.15 + t, \quad
P_\mathrm{s} = 1.333224(p_\mathrm{s} - p_\mathrm{w}),\quad
P_\mathrm{w} = 1.333224p_\mathrm{w},
\end{split}
\end{equation}
where $p_\mathrm s$ is atmospheric pressure. To obtain the excess refractivity, we calculate
\begin{equation}\label{eq:scint}
\begin{split}
\begin{array}{l}
D_\mathrm{s} = \frac{P_\mathrm{s}}{T}\left[1 + P_\mathrm{s}\,(57.90\times10^{-8} - 9.3250\times10^{-4}/T + 0.25844/T^2)\right],\\
D_\mathrm{w} = \frac{P_\mathrm{w}}{T}\left[1 + P_\mathrm{w}(1+3.7\times10^{-4}P_\mathrm{w})\,( -2.37321\times10^{-3} + \frac{2.23366}{T}  - \frac{710.792}{T^2} + \frac{7.75141\times10^4}{T^3})\right],
\end{array}
\end{split}
\end{equation}
where $P_{\rm s}$ and $P_{\rm w}$ are in millibars, and $T$ is in K. With the wave number $\sigma$ in reciprocal micrometres ($\sigma = 10^{4}/\lambda_{\text{[}\AA\text{]}}$ or $\sigma = 1/\lambda_{\mu\mathrm{m}}$), the excess refractivity $n-1$ is
\begin{equation}\label{eq:n}
\begin{split}
n(\lambda)-1 \;=\; 10^{-8}\Big[ (2371.34 + \frac{683939.7}{130-\sigma^2} + \frac{4547.3}{38.9-\sigma^2})\,D_\mathrm{s} \\
\;+\; (6487.31 + 58.058\sigma^2 - 0.71150\sigma^4 + 0.08851\sigma^6)\,D_\mathrm{w} \Big] .
\end{split}
\end{equation}
Subsequently, we calculate some auxiliary parameters: the ratio of scale heights $\beta$ and the gravity correction $\kappa$
\begin{equation}\label{eq:scint}
\begin{split}
\beta &= 0.001254\;\frac{T}{273.15}, \\
\kappa &= 1 + 0.005302\sin^2\phi - 5.83\times10^{-6}\sin^2(2\phi) - 3.15\times10^{-7}\,H_{\mathrm {obs}},
\end{split}
\end{equation}
where $\phi$ is the the astronomical latitude of the observing site, $H_{\rm obs}$ is the site elevation (m). Finally, the refraction (radian) is evaluated using the first two terms of the tangent expansion:
\begin{equation}\label{eq:scint}
R_\lambda \;=\; (n_0-1)\,\kappa\,(1-\beta)\,\tan z_0 \;-\; (n_0-1)\,\kappa\left(\beta-\frac{n_0-1}{2}\right)\tan^3 z_0 .
\end{equation}
$n_0$ can be calculated with the equation (\ref{eq:n}). For sufficiently large zenith distances ($z_0\gtrsim75^\circ$), higher-order corrections or full numerical integration along the ray path are recommended. Considering a band, it need to be recalculated based on weights to get total refraction $R$.
Then we project it onto equatorial components: to apply refraction as corrections in right ascension and declination, project $R$ along the parallactic angle $\psi$. Using the standard definitions:
\begin{equation}\label{eq:scint}
\begin{split}
\sin\psi &= \frac{\cos\phi\,\sin \mathrm {HA}}{\sin z_0},\\
\cos\psi \;&=\; \frac{\sin\phi - \sin\delta\cos z_0}{\cos\delta\sin z_0},
\end{split}
\end{equation}
where $\rm HA$ is hour angle, the refraction components (radian) in hour-angle and parallactic directions are
\begin{equation}\label{eq:scint}
\begin{split}
\Delta\alpha_{\mathrm {rad}} = -\,\frac{R \sin\psi}{\cos\delta},\\
\Delta\delta_{\mathrm {rad}} = -\,R \cos\psi.
\end{split}
\end{equation}
Differential atmospheric refraction at different zenith angle and band is shown in Fig. \ref{fig:ADR}. These offsets are added to the previously derived pixel coordinates, completing the correction for atmospheric differential refraction.

\subsubsection{Sky background}
The spatially varying sky background is a dominant source of photometric noise for faint objects and produces gradients that bias both source detection and background subtraction. The simulation must model the principal sky components. 
According to \cite{1991PASP..103.1033K}, the nighttime sky brightness in each pixel is computed as the sum of two magnitude components: the dark nighttime sky and the contribution from scattered moonlight:
\begin{equation}\label{eq:B0/moon}
\begin{array}{l}
B_0(Z) = B_{\mathrm {zen}}\;10^{-0.4\,k\,[X(Z)-1]},\\
B_{\mathrm {moon}}(Z, \rho) = f(\rho)\;I_{\mathrm {moon}}\;10^{-0.4\,K\,X(Z_\mathrm m)}\;\bigl(1-10^{-0.4\,K\,X(Z)}\bigr),
\end{array}
\end{equation}
where $B_0$ is the dark nighttime sky component and $B_{\rm moon}$ is the contribution from scattered moonlight. The symbol B represents the sky brightness in nanoLamberts (B), which can be converted to magnitude with equation (\ref{eq:B}) in \cite{1989PASP..101..306G}. $Z$ is the zenith distance and $Z_{\rm m}$ is the zenith distance of the Moon. $X(Z)$ is the airmass function calculated with equation (\ref{eq:X(z)}). $f(\rho)$ is the composite scattering function with angular distance $\rho$ from the Moon to the corresponding sky position for each pixel. Different pixel coordinates correspond to different zenith distance and angular distance with the Moon. $K$ is the atmospheric extinction coefficient.
\begin{equation}\label{eq:B}
B = 34.08 \times \exp(20.7233 - 0.92104M).
\end{equation}

\begin{equation}\label{eq:X(z)}
X(Z) = (1 - 0.96\sin^2 Z)^{-1/2}.
\end{equation}
Then we calculate the total magnitude of sky background based on the relationship between flux and magnitude:
\begin{equation}\label{eq:M}
M_{\mathrm {total\_sky}}(Z, \rho) = -2.5\log_{10}\bigl(10^{-0.4M_{\mathrm {0}}(Z)}+10^{-0.4M_{\mathrm {moon}}(Z, \rho)}\bigr).
\end{equation}
The differences in different pixels produced by dark nighttime sky component are not significant. To make it simple, we can take the dark nighttime sky component as a constant and using the raw dark-sky magnitude $M_0$ from input catalog. Therefore, the equation (\ref{eq:M}) becomes
\begin{equation}\label{eq:M_s}
M_{\mathrm {total\_sky}}(Z, \rho) = -2.5\log_{10}\bigl(10^{-0.4M_{\mathrm {0}}}+10^{-0.4M_{\mathrm {moon}}(Z, \rho)}\bigr).
\end{equation}
Then we calculate the $M_{\rm moon}(Z, \rho)$ based on equation (\ref{eq:B0/moon}) and (\ref{eq:B}):
\begin{equation}\label{eq:scint}
\begin{split}
\begin{aligned}
\begin{array}{l}
 M_{\mathrm {moon}}(Z, \rho) = \frac{20.7233 - \ln(B_{\mathrm {moon}}(Z, \rho)/34.08)}{0.92104},\\
 B_{\mathrm {moon}}(Z, \rho) = f(\rho)\,I_{\mathrm {moon}}\,10^{-0.4KX(Z_\mathrm m)}\,(1-10^{-0.4KX(Z)}).
\end{array}
\end{aligned}
\end{split}
\end{equation}
$f(\rho)$ can be obtained through:
\begin{equation}\label{eq:scint}
f(\rho) = P_\mathrm A\,f_M(\rho) + P_\mathrm B\,f_\mathrm R(\rho),
\end{equation}
where $P_\mathrm A = 1.5$, $P_\mathrm B = 0.9$ \citep{2013RAA....13.1255Y},
\begin{equation}\label{eq:scint}
\begin{array}{l}
f_\mathrm M(\rho) = \left\{
\begin{array}{l}
6.2 \times 10^7 \rho^{-2} \quad \text{if } \rho < 10^\circ, \\
10^{6.15-\rho/40}  \quad \text{if } \rho \geq 10^\circ ,
\end{array}
\right. \\
f_\mathrm R(\rho) = 10^{5.36}(1.06 + \cos^2(\rho)).
\end{array}
\end{equation}
And $I_{\mathrm {moon}} = 10^{-0.4(m + 16.57)}$ where the lunar magnitude is
\begin{equation}\label{eq:scint}
m = -12.73 + 0.026|\alpha| + 4 \times 10^{-9}\alpha^4,
\end{equation}
where $\alpha = 180^\circ - \epsilon$ is the lunar phase angle and $\epsilon$ is the Sun-Moon elongation. By organizing these formulas, the calculation of $M_{\mathrm {moon}}(Z, \rho)$ can be performed. Because pixel coordinates correspond to different zenith angles and the Moon distances, we can replace the independent variables $Z$ and $\rho$ with the pixel coordinate values $x$ and $y$. After getting $M_{\mathrm {moon}}(\text{x, y})$, we calculate
\begin{equation}\label{eq:M_xy}
M_{\mathrm {total\_sky}}(\text{x, y}) = -2.5\log_{10}\bigl(10^{-0.4M_{\mathrm 0}}+10^{-0.4M_{\mathrm {moon}}(x,y)}\bigr).
\end{equation}
Theoretically the above steps should be repeated to calculate the Moon sky brightness corresponding to each pixel one by one, but doing that requires coordinate conversion and celestial computation for each pixel, which has a high total complexity and is slow to compute. Therefore, we also provide a simple linear interpolation
\begin{equation}\label{eq:simp}
M_{\mathrm {moon}}(x,y) = M_{\mathrm {moon\_center}} + \frac{\partial M_{\mathrm {moon\_center}}}{\partial \text{RA}}(x - x_c)\Delta_{\mathrm {pix}} + \frac{\partial M_{\mathrm {moon\_center}}}{\partial \text{DEC}}(y - y_c)\Delta_{\mathrm {pix}},
\end{equation}
where $(x_\mathrm c, y_\mathrm c) = (n_\mathrm x/2, n_\mathrm y/2)$ is the field center, $\Delta_{\mathrm {pix}}$ is the pixel scale in arcsec/pixel, and the derivatives are computed numerically by evaluating the Moon model at offset positions. The pixel-by-pixel method significantly provides higher accuracy but slows down the operation. For most applications, the accuracy of the gradient approximation method using linear interpolation is sufficient, unless the Moon is very close to the field of view or an extremely high precision simulation is required.

In addition, we include a simulation for the Sun's contribution following the empirical formulations and fitting results of \cite{2014SPIE.9149E..2HW}. Although the solar background is negligible during most nighttime observations, modeling this component is necessary for realistic simulations of twilight flat-field exposures and observations conducted at small solar elongations, such as near-Sun asteroids. The solar term also prevents the inadvertent generation of daytime-like images. Accordingly, equation (\ref{eq:M_xy}) is expanded to
\begin{equation}\label{eq:M_total_three}
M_{\mathrm {total\_sky}}(\text{x, y}) = -2.5\log_{10}\bigl(10^{-0.4M_{\mathrm {0}}}+10^{-0.4M_{\mathrm {moon}}(x,y)} + 10^{-0.4M_{\mathrm {sun}}(x, y)}\bigr).
\end{equation}
We calculate the scattered sunlight contribution from the Sun position and the telescope pointing. The $M_{\mathrm {sun}}(x,y)$ are the scattered light contributions from the Sun at pixel coordinates $(x,y)$. 
The solar contribution creates a magnitude gradient \citep{2014SPIE.9149E..2HW}:
\begin{equation}\label{eq:G_sun}
G_{\mathrm {sun}} = -\frac{2.5}{\ln(10)} \times 10^{-0.005555\,\theta_{\mathrm {az}} - 1} / 3600 ,
\end{equation}
where $G_{\rm sun}$ is in unit of  \(\mathrm{mag\,arcsec^{-1}}\) and $\theta_{\rm az}$ is the azimuthal separation between Sun and pointing. 
The background solar sky magnitude is
\begin{equation}\label{eq:M_sun}
M_{\mathrm {sun\_background}} = \min\{30, \max\{1, 8 - 1.03 h_{\mathrm {sun}}\}\},
\end{equation}
where $h_{\rm sun}$ is the solar altitude in degrees. The spatial variation follows
\begin{equation}\label{eq:M_sun}
M_{\mathrm {sun}}(x,y) = M_{\mathrm {sun\_background}} + G_{\mathrm {sun}} \left[ u_{\mathrm {RA}}(x - x_c)\Delta_{\mathrm {pix}} + u_{\mathrm {DEC}}(y - y_c)\Delta_{\mathrm {pix}} \right],
\end{equation}
where $(u_{\rm RA}, u_{\rm DEC})$ are the unit direction components from pointing to the Sun, normalized on the celestial sphere.

\begin{figure*}[ht!]
  \gridline{\fig{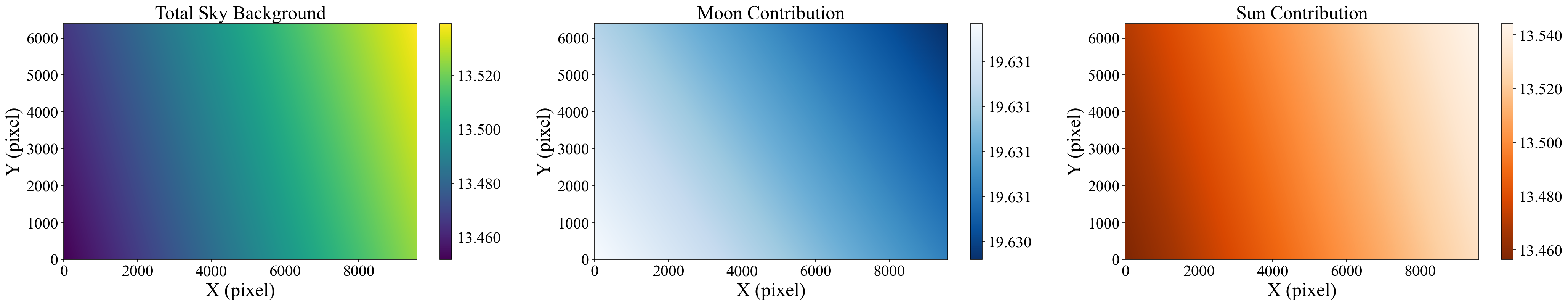}{0.95\textwidth}{(a) The spatial gradient of the total sky background dominated by the sunlight.}}
  \gridline{\fig{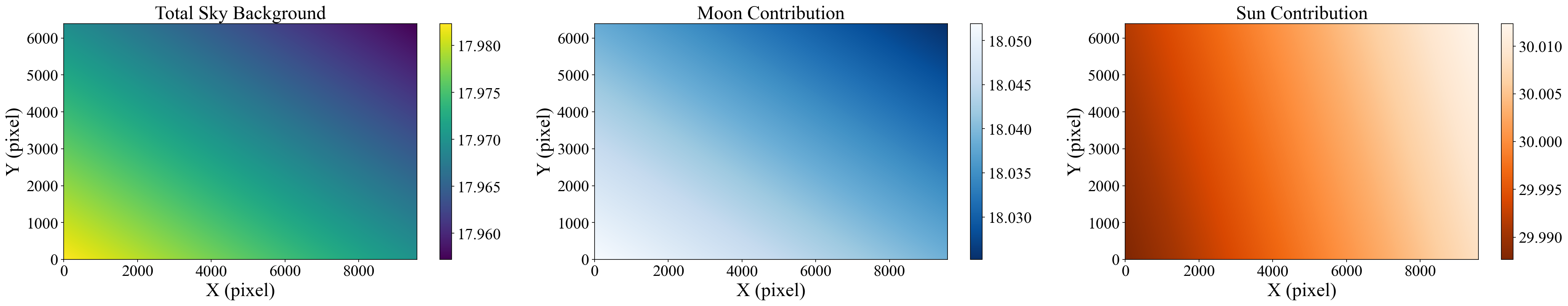}{0.95\textwidth}{(b) The spatial gradient of the total sky background dominated by the moonlight.}}
\caption{The spatial variation of the sky background magnitude across the field for the different times. Panel (a) describes the evening sky at the observation site, where the spatial gradient of the total sky background is dominated by sunlight. Panel (b) describes the midnight sky at the observation site, where the spatial gradient of the total sky background is dominated by moonlight.}
\label{fig:sky}
\end{figure*}

We present the final simulated sky-background results for Muguang Observatory, located at geographic coordinates $\mathrm{longitude} = 121.604^\circ$~E, $\mathrm{latitude} = 31.168^\circ$~N with an elevation of $100\ \mathrm{meter}$ above mean sea level. Key parameters of Muguang Observatory are listed in Table \ref{tab:obser_par}. We simulate a sky field center at R.A. = 47.369$^\circ$, Decl. = 30.674$^\circ$.
Fig. \ref{fig:sky} (a) shows the spatial variation of the sky background magnitude across the field for UTC 2025-02-12 10:00:00.000. At this time, the lunar zenith distance is 83.1$^\circ$, and the solar altitude is -5.3$^\circ$. The left, middle and right panels display the total sky background, the scattered moonlight contribution and the scattered sunlight contribution respectively. As the Sun is still near the horizon, the spatial gradient of the total sky background is predominantly shaped by the variation in scattered sunlight.
In contrast, Fig. \ref{fig:sky} (b) depicts the sky background for the same field at UTC 2025-02-12 19:00:00.000. The observational geometry has changed: the lunar zenith distance is 40.2$^\circ$, and the solar altitude is -46.7$^\circ$. With the Sun well below the horizon, the spatial structure of the total sky background is now clearly dominated by the gradient in scattered moonlight. 

\subsubsection{Atmospheric scintillation}
Atmospheric scintillation produces spatially correlated fluctuations in instantaneous flux. Simulating the atmospheric scintillation noise is a key step in mapping the temporal statistics of atmospheric turbulence onto the spatial domain of the image. In simulation, we synthesize a spatially correlated multiplicative field to represent scintillation-induced flux fluctuations which is often neglected in some simulators such as SkyMaker. Atmospheric scintillation primarily affects photometry and the PSF morphology, exerting a much smaller influence on the source's center position compared to ADR. Therefore, we make the following considerations here: we generate a real-valued white noise field $ w(x, y) \sim \mathcal{N}(0, 1) $ with the same dimensions as the image. White noise has a flat power spectrum in the frequency domain, making it a natural starting point for constructing a target power spectrum. This step yields a spatially uncorrelated random field that is ready for spectral shaping. We then compute its Fourier transform,
\begin{equation}\label{eq:scint}
\overline{W}(\mathbf{k}) = \mathcal{F}\{w(x, y)\} ,
\end{equation}
multiplication in frequency domain is computationally more efficient than convolution in spatial domain. Therefore, after forming the frequency-domain representation $\overline{W}(\mathbf{k})$, we apply the target spectrum directly in Fourier space. In practice we generate Kolmogorov style field by multiplying the amplitudes corresponding to the Kolmogorov spectrum derived by \cite{1976JOSA...66..207N}: 
\begin{equation}\label{eq:scint}
A(k) \propto (k^2 + k_0^2)^{-11/12}, \quad k = \sqrt{k_\mathrm x^2 + k_\mathrm y^2},
\end{equation}
where $ k_0 = 2\pi / L_0 $, $L_0$ is the outer scale. This item prevents divergence when $ k \to 0 $. Finally, we apply an inverse Fourier transform to return to the spatial domain:
\begin{equation}\label{eq:scint}
 H_\mathrm t(\mathbf{k}) = \overline{W}(\mathbf{k}) \cdot A(k), \quad 
 h(x, y) = \Re\{\mathcal{F}^{-1}[H_\mathrm t]\} .
\end{equation}
Up to now, we get zero-mean, real-valued random field $ h(x, y) $ with Kolmogorov-style correlation. In order to normalize the field and map it to a specified variance, we apply a log-normal transformation. First, we standardize:
\begin{equation}\label{eq:scint}
H(x, y) = \frac{h(x, y) - \overline{h}}{\sqrt{\text{Var}(h)}} \quad 
\end{equation}    
Let the relative RMS be $\sigma$. We do Log-normal mapping:
\begin{equation}\label{eq:scint}
a = \sqrt{\ln(1 + \sigma^2)}, \quad f(x, y) = \exp\left(aH(x, y) - \frac{a^2}{2}\right).
\end{equation}
This construction ensures $\mathbb{E}[f] = 1$ and $\text{Var}(f) = \sigma^2$. Log-normal transformation guarantees positive field values and allows precise control of mean and variance. It's more robust than the simple linear form $ 1 + \sigma H $, which may produce negative values for large $\sigma$. Multiplicative perturbation field $f(x, y)$ satisfying statistical objectives. We generate perturbed image:
\begin{equation}\label{eq:scint}
I(x, y) = I_{\mathrm{real}}(x, y) \cdot f(x, y). 
\end{equation}

We treated the atmospheric scintillation index $\sigma_Y^{2}$ using \cite{Young1967PhotometricEA} approximation, empirically adjusted according to \cite{2015MNRAS.452.1707O}. The adjustment introduces a site-dependent scaling coefficient \(C_\mathrm Y\), tuned to reproduce the realistic scintillation index measured at the specific observatory.
The scintillation index is given by
\begin{equation}\label{eq:scint}
\sigma_\mathrm{Y}^{2} \;=\; \frac{\langle I^2 \rangle - \langle I \rangle ^2}{\langle I \rangle ^2} \;= \; 10\times10^{-6}\,C_\mathrm{Y}^{2}\,D^{-4/3}\,t^{-1}\,(\cos\gamma)^{-3}\,
\exp\!\biggl(-\frac{2\,h_{\mathrm {obs}}}{H}\biggr),
\end{equation}
where $I$ is the measured intensity from the object as a function of time and $\langle \rangle$ denotes an average. \(D\) is the telescope diameter in units of meter, \(t\) is the exposure time in units of second, \(\gamma\) is the zenith angle, \(h_{\mathrm{obs}}\) is the observatory height and \(H\) is the atmospheric scale height (typically \(H\!\approx\!8000\) m). 

\begin{figure*}[ht!]
\centering
\includegraphics[width=0.4\textwidth]{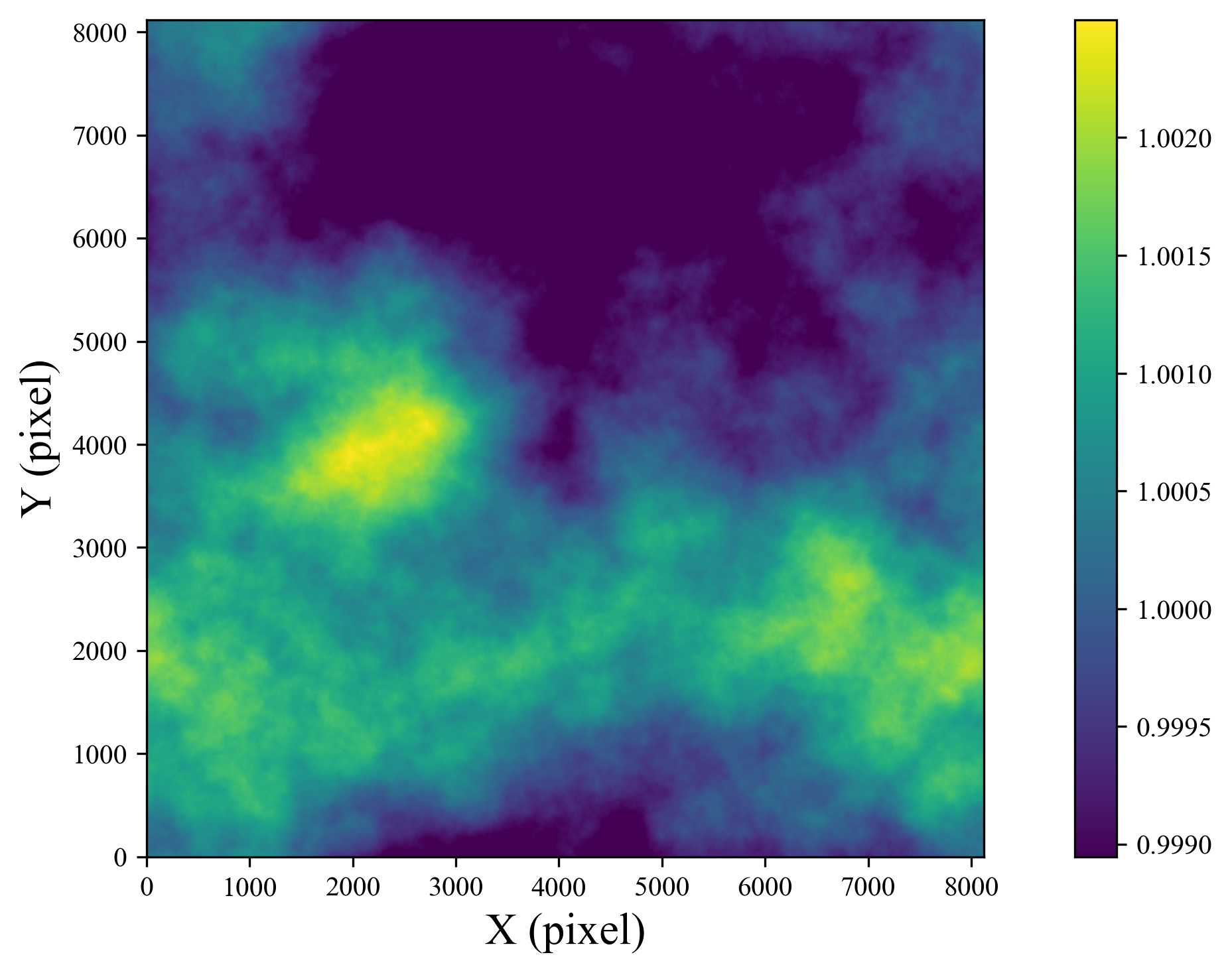}
\caption{Realizations of the atmospheric scintillation multiplicative field $f(x,y)$. The simulation is based on the Tianyu site parameters shown in Table \ref{tab:obser_par}. Exposure time is 5 second and outer scale is 25 meter.}
\label{fig:scint_L0}
\end{figure*}

While the empirical formulation above provides a robust, site-dependent estimate of the integrated scintillation variance over the time, a numerical simulation of scintillation effect on astronomical images requires a spatial realization of the underlying stochastic process as discussed earlier. The key link between the temporal variance $\sigma_\mathrm{Y}^2$ and the spatial simulation is established by setting the target variance parameter in the spatial domain, $\sigma^2$, equal to the temporally integrated variance, $\sigma_\mathrm{Y}^2$, under the assumption of Taylor's frozen flow hypothesis \citep{1938RSPSA.164..476T}. We can show realizations of the multiplicative atmospheric scintillation field $f(x,y)$ for the observing geometry based on Tianyu whose key parameters are in Table \ref{tab:obser_par}: image size $8120\times8120$ pixels, zenith angle $z=50^\circ$, telescope diameter $D=1.0\ \mathrm{m}$, exposure $t_{\rm exp}=5\ \mathrm{s}$, observatory altitude $h_{\mathrm{obs}}=4000\ \mathrm{m}$, they generate the same target relative RMS $\sigma = 0.001053$. We use the outer scale $L_0 = 25\ \rm meter$ and convert it into pixel units using Tianyu's pixel scale. Simulation result is in Fig. \ref{fig:scint_L0}.
The outer scale $L_0$ controls the characteristic transverse size of atmospheric scintillation structures in the focal plane while the target RMS $\sigma$ controls their amplitude. 

\subsection{Hardware simulator}
Following the generation of a photon distribution map at the focal plane by external scene photon simulator, hardware simulator first accounts for the effects of the optical system. The primary optical effect relevant to the imaging is vignetting, a radial fall-off in illumination caused by geometric obstructions within the light path. Accurately modeling this effect is essential because it modulates the incident photon flux before it reaches the sensor, directly impacting both photometric uniformity across the field and the signal-to-noise ratio at the image periphery.

\begin{figure*}[ht!]
\gridline{%
  \fig{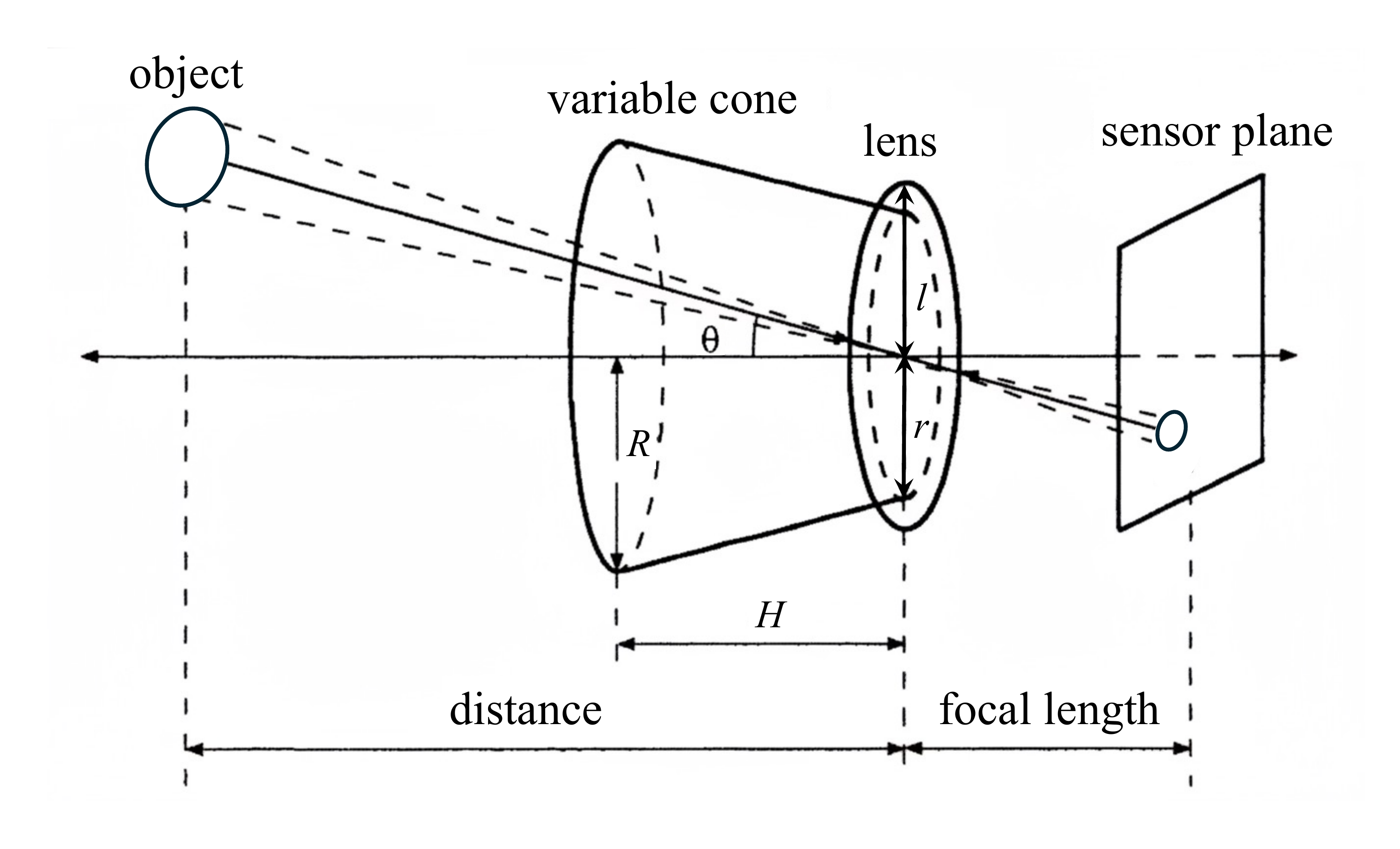}{0.36\textwidth}{(a) A variable cone in vignetting model.}%
  \fig{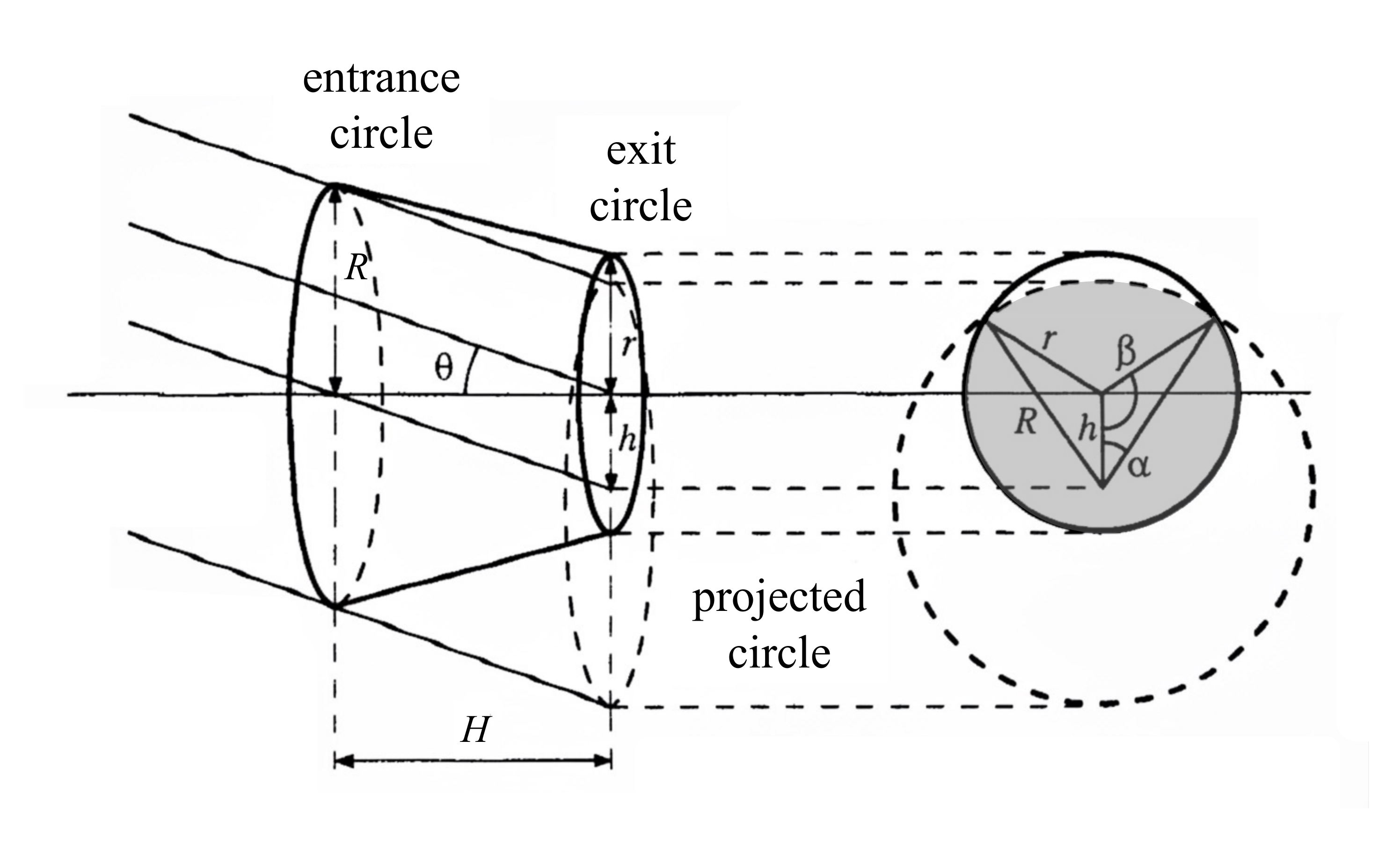}{0.36\textwidth}{(b) Effective opening area of the variable cone.}%
  \fig{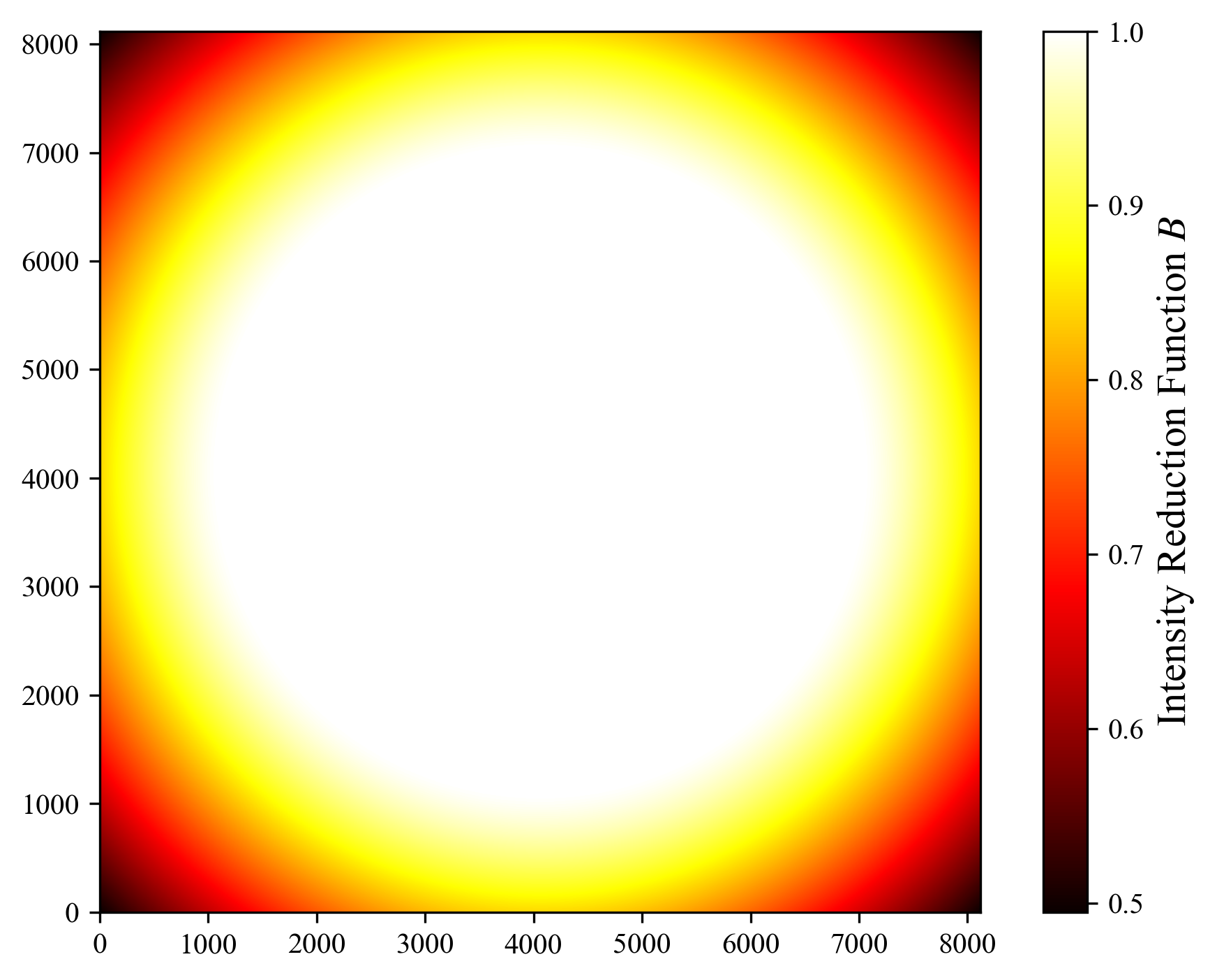}{0.26\textwidth}{(c) The simulation result.}%
}
  \caption{Vignetting model and one result example. The left and middle panels are adapted from \cite{546016} which consider the vignetting effect with a variable cone and calculate the attenuation factor by determining the overlapping area of the projections. The right panel shows the simulation result of the vignetting effect for a system with an image size of 8120 × 8120 pixel, a pixel size of 1e-5 meter and a focal length of 1.57 meter. The geometric parameters for the vignetting model were set to $H = 26$, $R = 0.99$, $r = l = 0.5$.}
  \label{fig:vig_model}
\end{figure*}

We simulate vignetting effect according to \cite{546016}. This model simulates the propagation and obstruction of light rays within the lens by introducing a variable cone, thereby explaining the phenomenon of reduced brightness at the edge of the image. The variable cone in the model is defined by three parameters: the radius of the entrance circle $R$, the radius of the exit circle $r$ and the height of the cone $H$. By considering the propagation paths and obstruction of light rays entering the lens at different angles, the model can calculate the attenuation of light intensity causing the vignetting effect, as shown in Fig. \ref{fig:vig_model}. 
We can calculate the angles $\alpha$ and $\beta$ by
\begin{equation}
\alpha = \cos^{-1}\!\left(\frac{R^{2}-r^{2}+h^{2}}{2 R h}\right),
\end{equation}
\begin{equation}
\beta = \cos^{-1}\!\left(\frac{r^{2}-R^{2}+h^{2}}{2 r h}\right), 
\end{equation}
where the displacement $h$ of the center of the projected circle from the optical axis is
\begin{equation}
h = H \tan\theta, \quad \theta = \frac{((x-x_{\mathrm {center}})^2+(y-y_{\mathrm {center}})^2)\times \mathrm{pixel\ size}}{\mathrm{focal\ length}}.
\end{equation}
The shaded area $A$ in Fig. \ref{fig:vig_model} is calculated by
\begin{equation}
A = R^{2}\big(\alpha - \sin\alpha\cos\alpha\big) + r^{2}\big(\beta - \sin\beta\cos\beta\big). 
\end{equation}
By normalizing $A$ with the area of the lens, we can get the intensity reduction function due to vignetting:
\begin{equation}
B = \frac{A}{\pi l^{2}}, 
\end{equation}
where the $l$ is the radius of lens. And the distortion-free range can be obtained by
\begin{equation}
|\theta| \le \tan^{-1}\!\left(\frac{R - r}{H}\right). 
\end{equation}
The choices of parameters $R$, $r$ and $H$ are telescope-specific and critical for an accurate simulation. Typically, $r$ is set equal to the radius of lens to ensure intensity reduction function is between 0 and 1. Parameters $R$ and $H$ can get analytical solution by matching some predictions and physical characteristics, such as the mean value of intensity reduction function or the fractional area of the field where the intensity reduction function exceeds a specified threshold. We simulate the vignetting effect for a system with an image size of 8120 × 8120 pixel, a pixel size of $1\times10^{-5}$ meter and a focal length of 1.57 meter based on Tianyu. The geometric parameters for the vignetting model are set to $H = 26$, $R = 0.99$, $r = l = 0.5$. The resulting vignetting pattern is presented in Fig. \ref{fig:vig_model} (c).

Once the light has passed through the optics, the sensor response must be simulated to transform the corrected photon flux into a digital readout. Simulating the complete signal chain—from photoelectric conversion to digitization and its associated noises is crucial for producing synthetic images that match the statistical properties and instrumental signatures of actual observational data from CCD or CMOS detectors.
We have done the high-level numerical simulations of noise in CCD or CMOS camera according to \cite{2014arXiv1412.4031K}. Moreover we have implemented a three-stage transformation: photon to electron, electron to voltage, and voltage to digital signal. We inject noise consistent with the input sensor parameters in the different phases following \cite{2014arXiv1412.4031K}. The modeled noise components include photon response non-uniformity, photon shot noise, dark current fixed pattern noise, dark current shot noise, offset fixed pattern noise, source follower noise and sense node reset noise. Saturation effects were also included in the simulations.

\section{Validation and example tests} \label{sec:cite}
\subsection{Comparison of real observation, AstroSkyFlow and SkyMaker}  \label{sec:comp}
AstroSkyFlow is flexible and can be adapted to different observation systems by adjusting parameters in the user configuration file. Since the Tianyu telescope has not yet commenced operations, we are unable to obtain its actual observational images. 
Therefore, we validate AstroSkyFlow using data from existing facilities. One validation dataset is obtained at the Muguang Observatory, located on the rooftop of the Tsung-Dao Lee Institute in Shanghai, China. Some key parameters of the Muguang Observatory are listed in Table \ref{tab:obser_par}. For this dataset, we use real observations acquired on January 11, 2024 with a 14-inch Corrected Dall-Kirkham (CDK14) telescope equipped with a QHY600 CMOS camera. During the transit of WASP-11 b, 75 consecutive images were obtained, each with an exposure time of 180 s. Another validation dataset is obtained with the 85 cm telescope at Xinglong Observatory equipped with an Andor iKon-M 936 CCD camera. Some key parameters of the Xinglong system are listed in Table \ref{tab:obser_par}. Specifically, we use 188 images of the W Ursae Majoris (W UMa)-type binary V0554 Dra obtained on May 23, 2025, each with an exposure time of 30 s. For both validation datasets, we reproduce the characteristics of the corresponding observatory, telescope, camera, and observing conditions through a set of input parameters. We generate AstroSkyFlow images for the same observing times and sky regions, and conduct a comprehensive comparison between AstroSkyFlow images and real observational images.
The evaluation focuses on three key aspects: recovery of injected signals, photometric precision and point spread function properties — including both individual PSF profiles and their FWHM distributions in the field of view. Unless otherwise specified, all AstroSkyFlow images appearing in the subsequent part are based on parameters from the Muguang Observatory.

To evaluate the performance and efficiency of our pipeline, it is essential to conduct a comparison with other simulators. 
While PhoSim provides a more physically complete, first-principles simulation framework, it is designed for computationally intensive end-to-end modeling of photon propagation through the atmosphere, telescope, and instrument, and such high-fidelity photon level simulations are typically much slower than lightweight image generation tools. Their computationally intensive nature makes them optimally suited for deep, end-to-end physical calculations of complex systems rather than rapid time-series generation. By contrast, AstroSkyFlow is intended for flexible, efficient production of time-domain image streams across epochs and sky fields, making it more appropriate to compare against lightweight image simulators with similar workflow characteristics. For these reasons, we adopt SkyMaker \citep{2010ascl.soft10066B} as the baseline for comparison.

SkyMaker is a widely used, user-friendly and  lightweight image simulation tool that operates based on a configuration file and an ASCII source list containing a catalog of objects to be injected into the simulated image. While efficient, it adopts a generalized approach that lacks detailed modeling of specific physical processes or instrument-specific noise characteristics. While SkyMaker's internal PSF model lacks realism, it allows users to supply an external PSF, though this typically results in a spatially invariant PSF across the entire image. 
For time-domain simulations, SkyMaker would require users to prepare a series object lists for each exposure in advance — making multi-epoch simulations complicated. AstroSkyFlow integrates these temporal variations internally. Furthermore, SkyMaker can not perform subsampled integration for rapidly moving targets, which show streak-like trails in single exposure. AstroSkyFlow supports temporal oversampling within a single exposure for modeling the trailing effects caused by fast-moving objects. In the comparative analysis, we pre-generate a series lists containing various signals for SkyMaker to simulate images in different times. And we update the SkyMaker configuration files based on the parameters of the Muguang and Xinglong Observatory. We use the Moffat profile with the same parameters of AstroSkyFlow as the input PSF in SkyMaker.

\begin{table}[!h]
    \centering
    \caption{Key parameters of the Muguang and Xinglong Observatory and Tianyu.}
    \label{tab:obser_par}
    \begin{tabular}{lllllll}
    \hline
         Name&Diameter (cm)& Focal length (mm)&Camera type & Pixel scale ($\mu$m)& Pixel number&FOV (arcmin)\\\hline
         Muguang 35cm& 35&2563&QHY 600M &3.76&9576$\times$6388&48$\times$32\\
         Xinglong 85cm& 85&2987&Andor iKon-M 936 &13.5&2048$\times$2048&31.84$\times$31.84\\
          Tianyu 100cm& 100&1570&COSMOS66 &10&8120$\times$8120&177.89$\times$177.89\\\hline
    \end{tabular}
\end{table}

    \subsubsection{Recovery of injected signals}
    \label{r and i}
    We can validate the pipeline by recovering photometric and motion signals. To assess photometric recovery, we simulate two cases: the transit of WASP-11 b observed on January 11, 2024, and the W UMa-type binary V0554 Dra observed on May 23, 2025. The orbital ephemerides and stellar parameters for WASP-11 b are taken from the NASA Exoplanet Archive\footnote{\url{https://exoplanetarchive.ipac.caltech.edu}}, while the simulation parameters for V0554 Dra are adopted from \cite{2026ApJ...997...39D}. Using these parameters, we inject the corresponding photometric signal into AstroSkyFlow and SkyMaker images. For calibration, we set the zero exposure to generate the bias and set uniform light to generate the flat. 
    The processing sequence includes calibration, point source detection via DAOStarFinder, alignment to a reference frame, fixed-radius aperture photometry using CircularAperture, and propagating flux uncertainties through calc\_total\_error. All of these steps are implemented using the photutils package \citep{larry_bradley_2019_3368647}.
    Finally, we use the differential photometry to extract the WASP-11 and V0554 Dra light curve. 

\begin{figure*}[ht!]
\gridline{%
  \fig{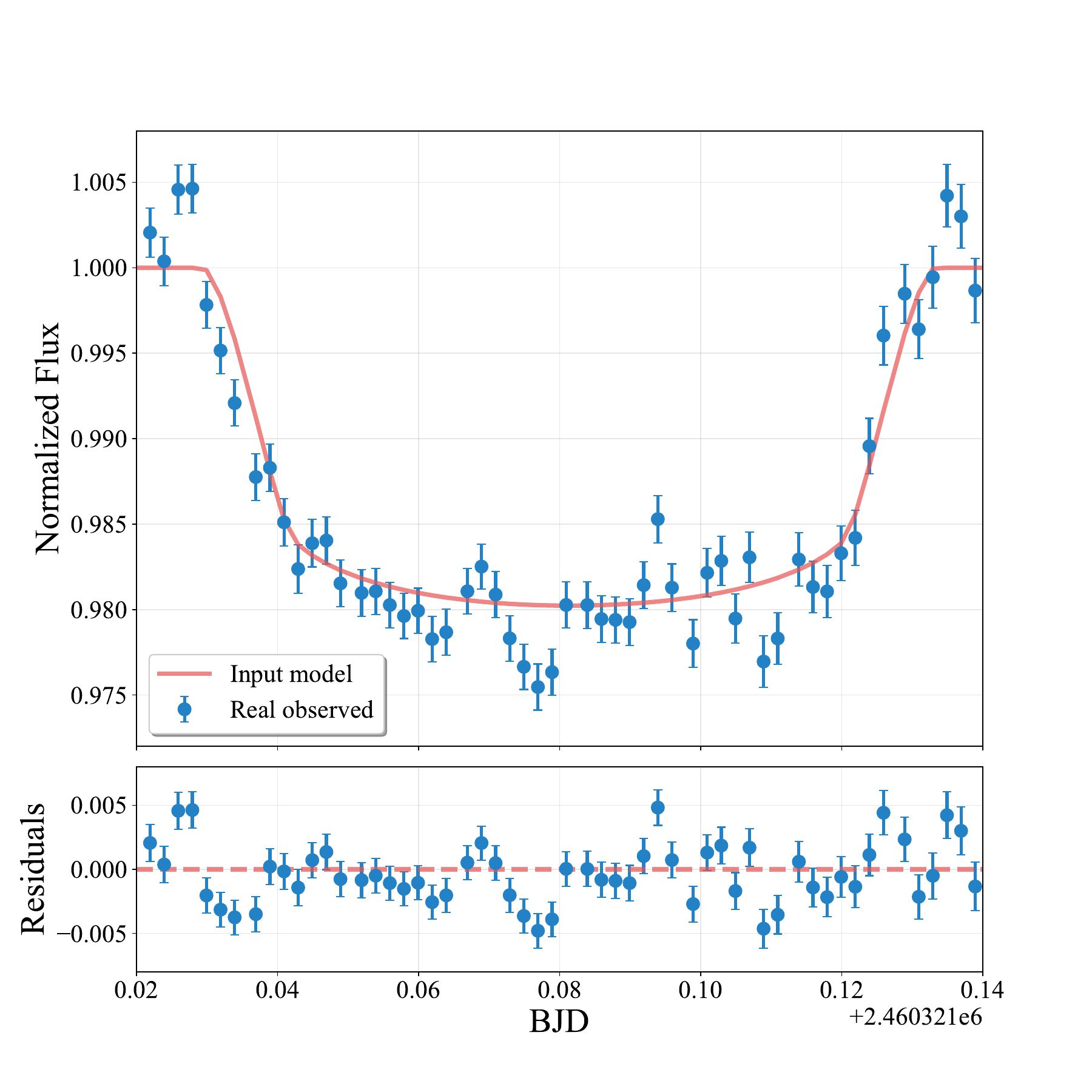}{0.33\textwidth}{(a) Real observation light curve.}%
  \fig{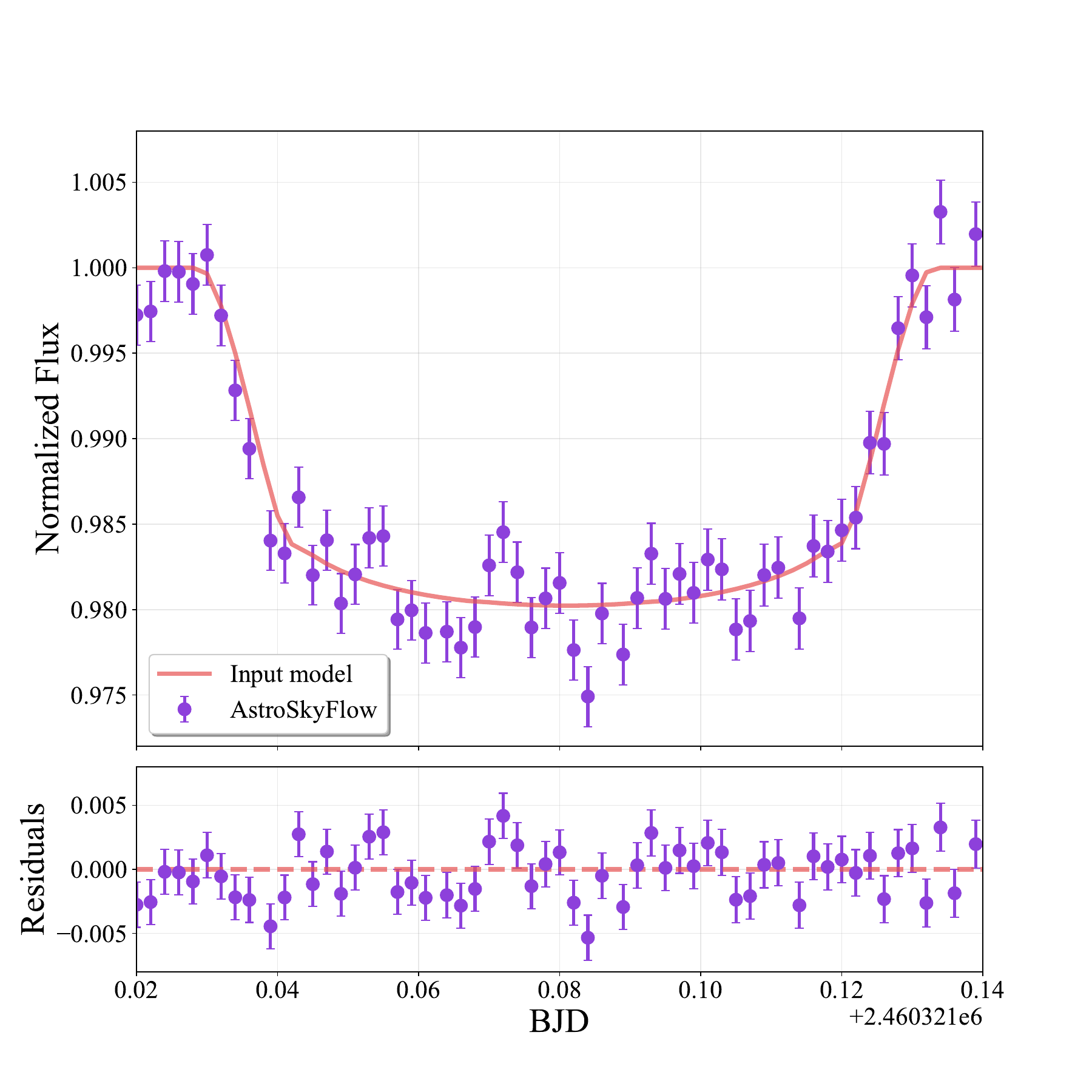}{0.33\textwidth}{(b) AstroSkyFlow light curve.}%
  \fig{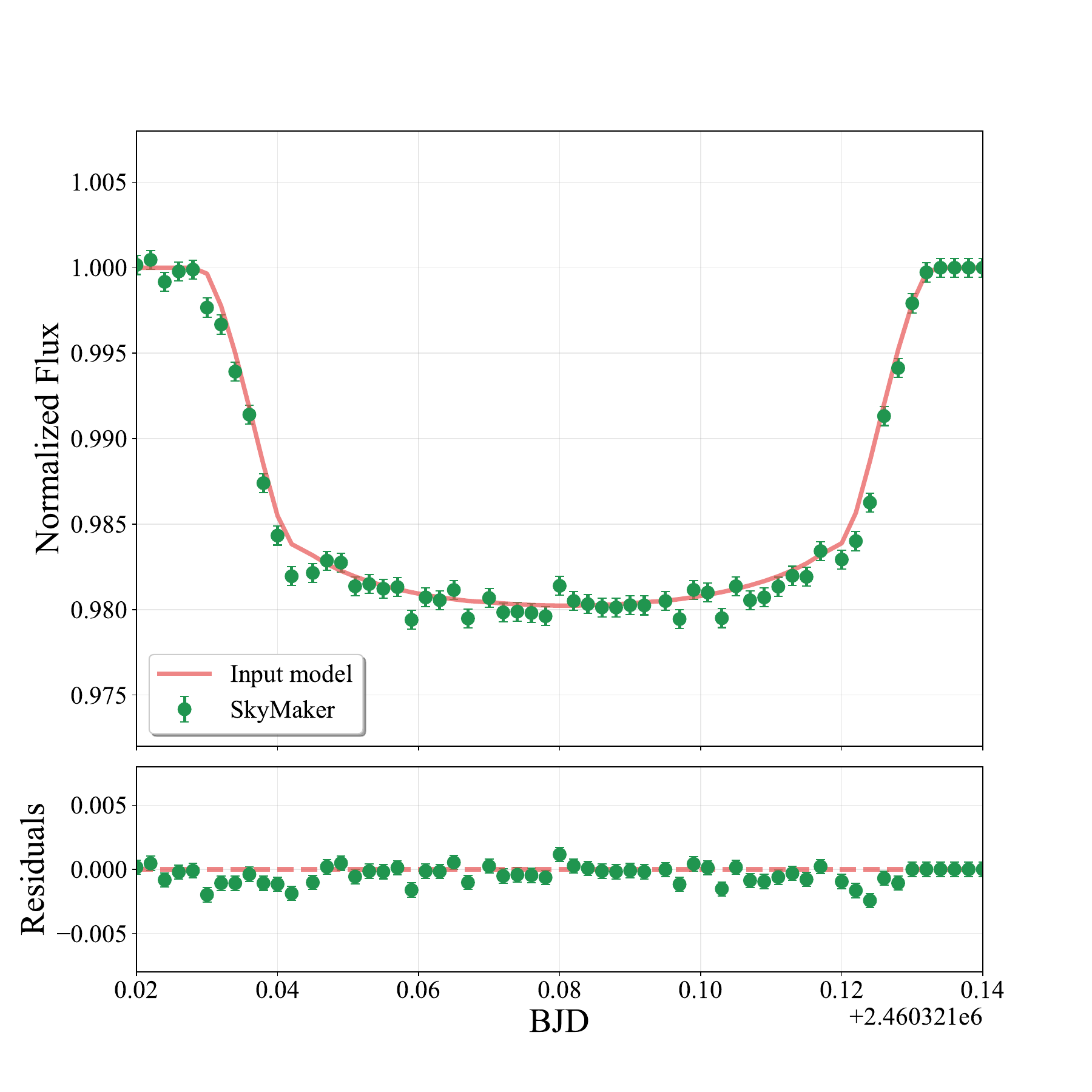}{0.33\textwidth}{(c) SkyMaker light curve.}%
}
\caption{The compared results between real observation, AstroSkyFlow, and SkyMaker WASP-11 b transit light curves. The red solid line is the theoretically injected binary light curve.}
\label{fig:v0554}
\end{figure*}

\begin{figure*}[ht!]
\gridline{%
  \fig{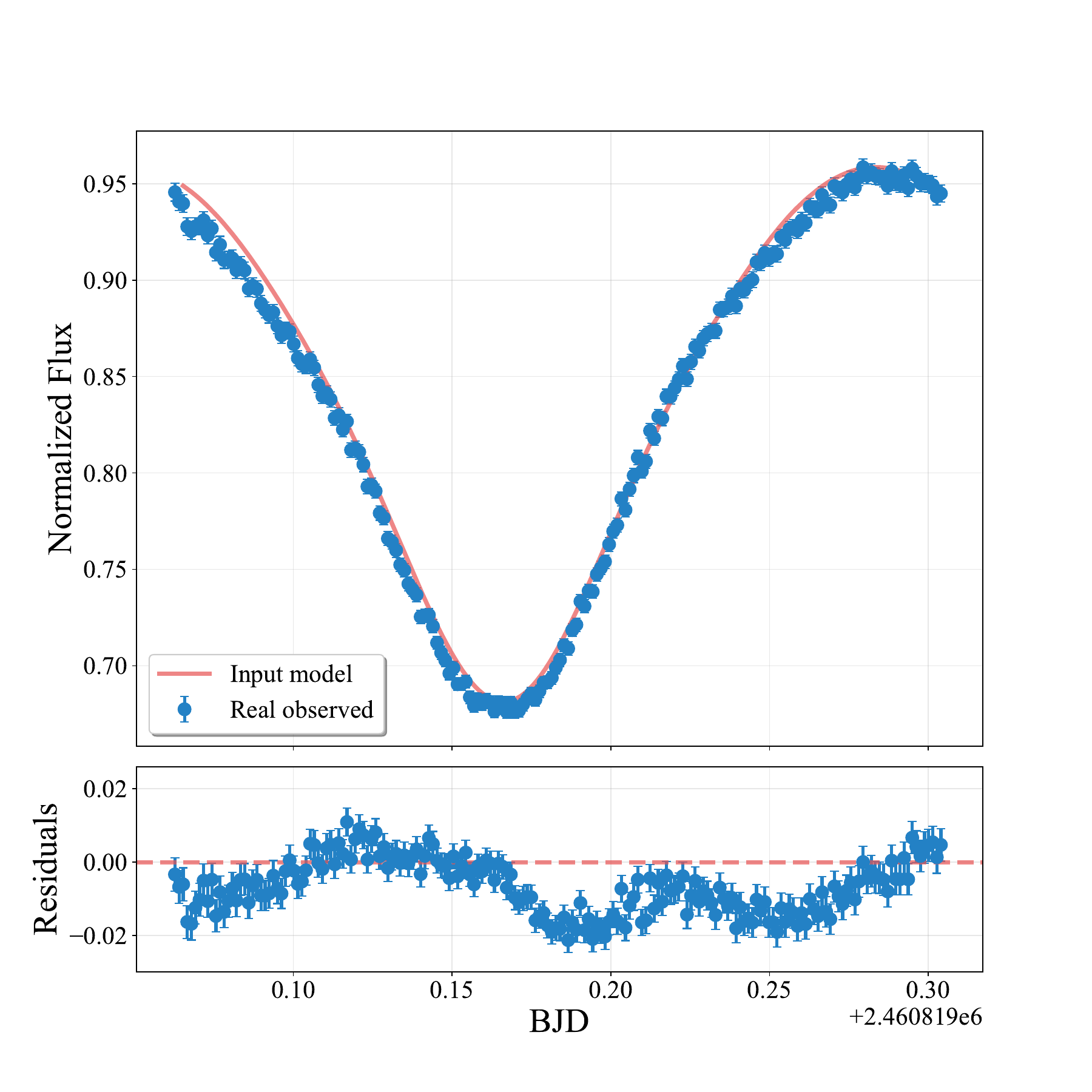}{0.33\textwidth}{(a) Real observation light curve.}%
  \fig{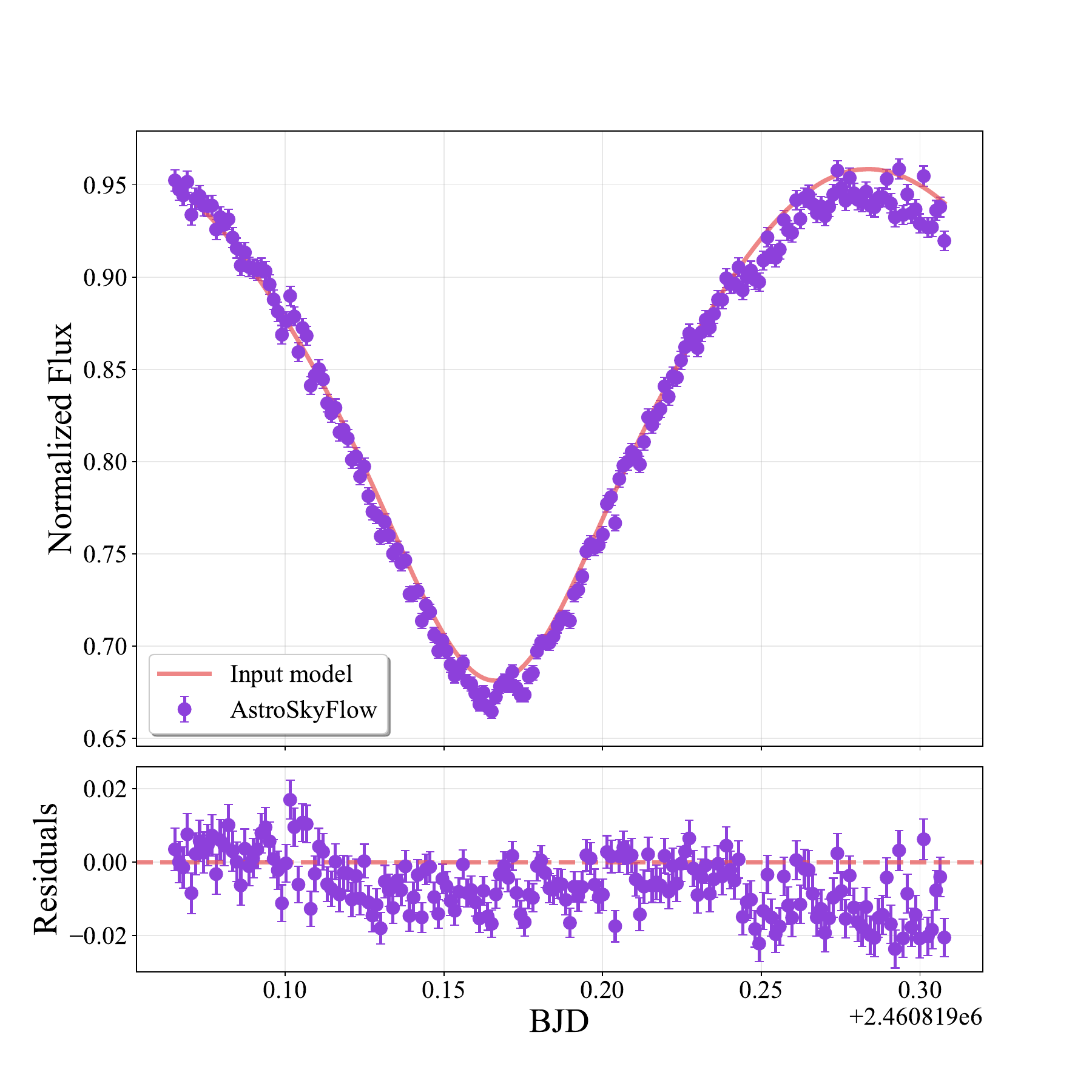}{0.33\textwidth}{(b) AstroSkyFlow light curve.}%
  \fig{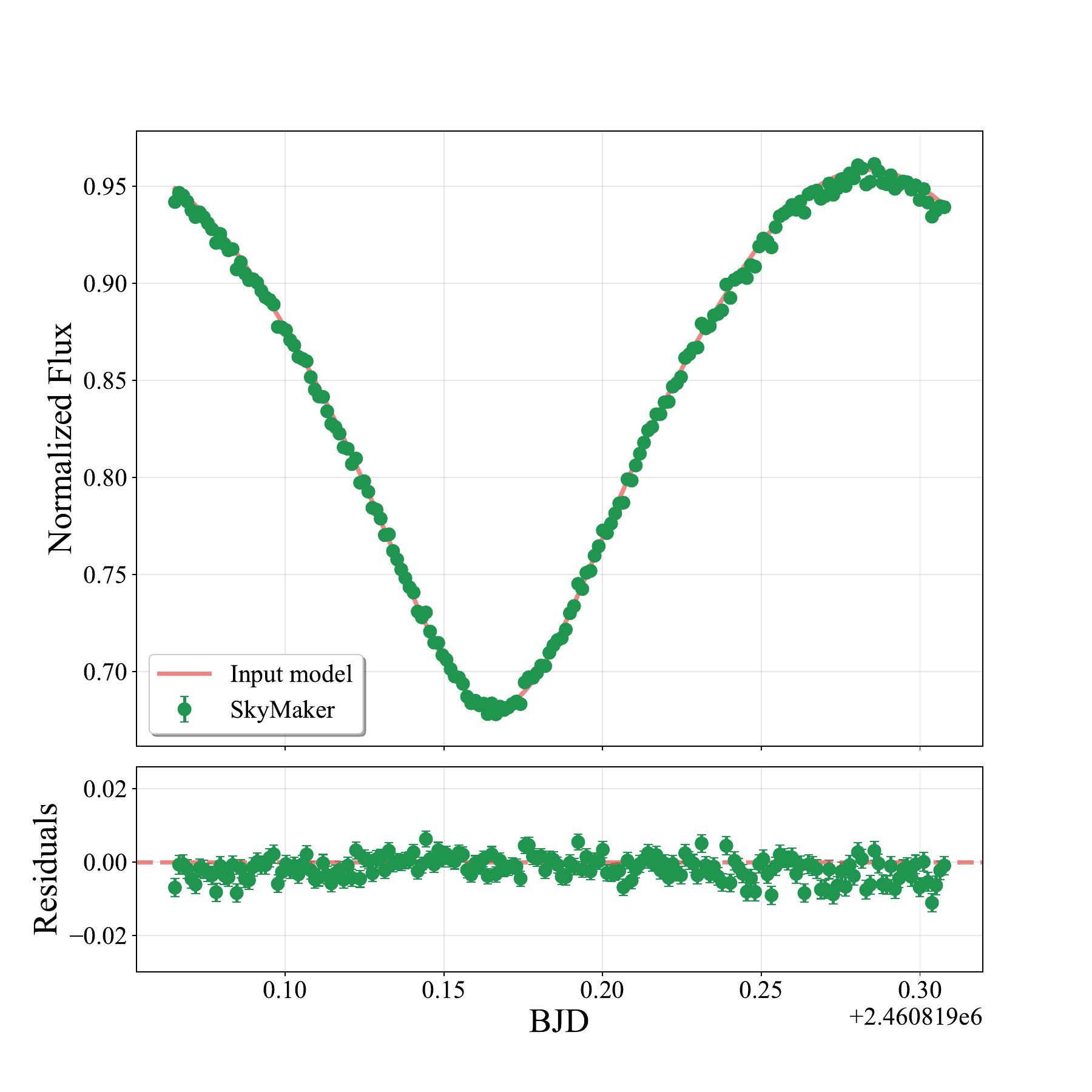}{0.33\textwidth}{(c) SkyMaker light curve.}%
}
\caption{The compared results between real observation, AstroSkyFlow, and SkyMaker V0554 Dra light curves. The red solid line is the theoretically injected transit light curve.}
\label{fig:transit}
\end{figure*}

    The photometric recovery results for WASP-11 b and V0554 Dra are shown in Fig. \ref{fig:transit} and \ref{fig:v0554}, respectively. In each figure, the left, middle and right panel display the injected light curve model and the normalized flux extracted from real observation, AstroSkyFlow and SkyMaker images individually. 
    For WASP-11 b, the residual RMS values are $2.466 \times10^{-3}$ for the real observation, $2.345 \times10^{-3}$ for AstroSkyFlow, and $7.678 \times10^{-4}$ for SkyMaker. AstroSkyFlow reproduces the noise amplitude of the real data to within approximately 5\%, whereas SkyMaker underestimates the noise level by about 69\%, indicating that several dominant noise contributions present in real observations are not included in its simulation framework. The left panel in Fig. \ref{fig:transit} shows that the simulated transit depth of injected light curve model is fundamentally consistent with the real depth. 
    For V0554 Dra, the residual RMS values are $1.001 \times10^{-2}$ for the real observations, $9.929 \times10^{-3}$ for AstroSkyFlow, and $3.592 \times10^{-3}$ for SkyMaker. Compared with the real data, the AstroSkyFlow results preserve morphology, amplitude and a comparable residual scatter of light curve, whereas the SkyMaker results recover the main periodic modulation but exhibit systematically lower residual scatter than real observed.
    These two cases show that AstroSkyFlow reproduces the noise behavior of real observations more realistically than SkyMaker, as reflected in the error bars, RMS values, and the amplitude and temporal structure of the residuals. This indicates that AstroSkyFlow captures the dominant observational uncertainties. SkyMaker results exhibit lower residual scatter, since it does not account for noise contributions from specific instrumental effects, atmospheric conditions, or sky background.
    
\begin{figure*}[ht!]
\gridline{%
  \fig{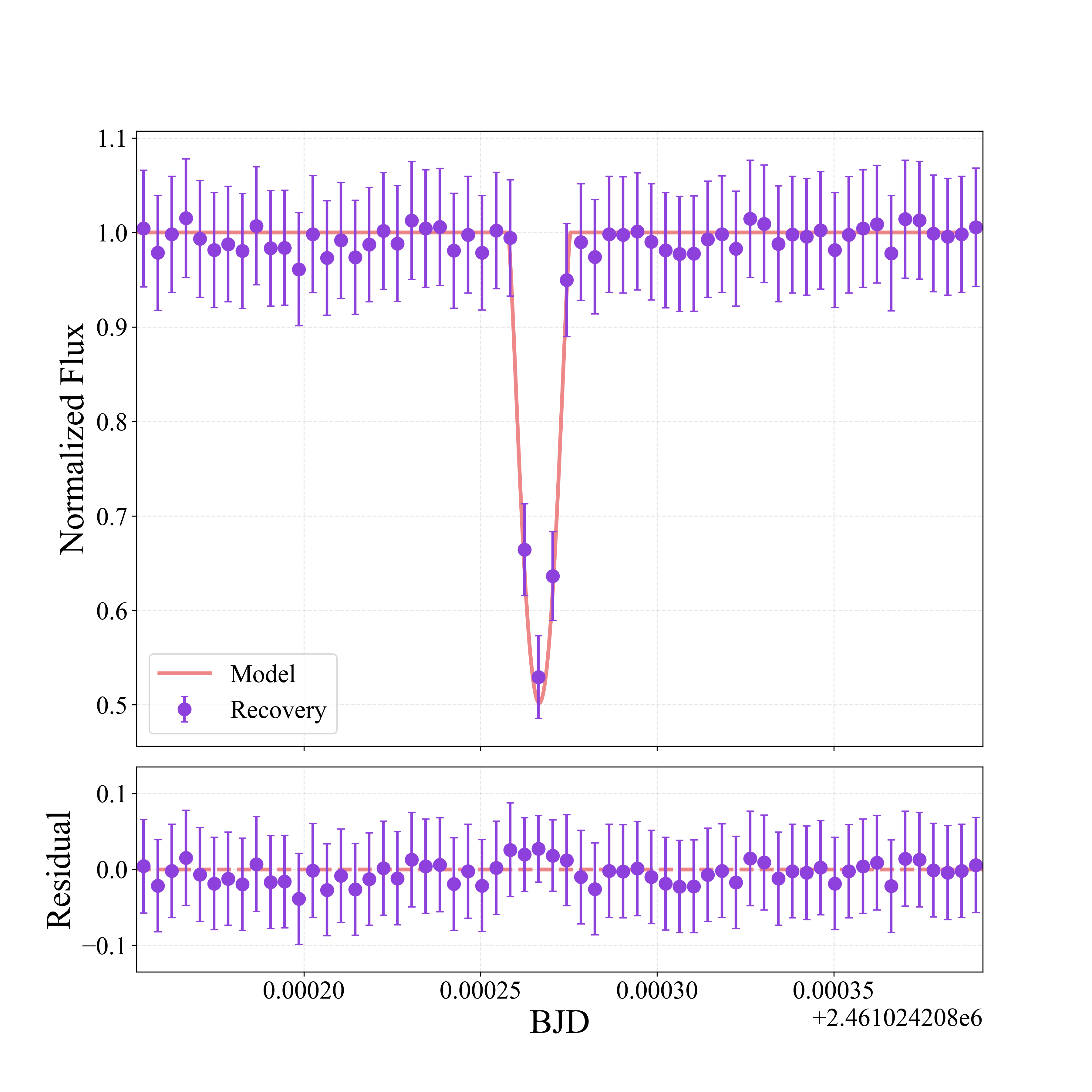}{0.33\textwidth}{(a) Occultation in geometric approximation.}%
  \fig{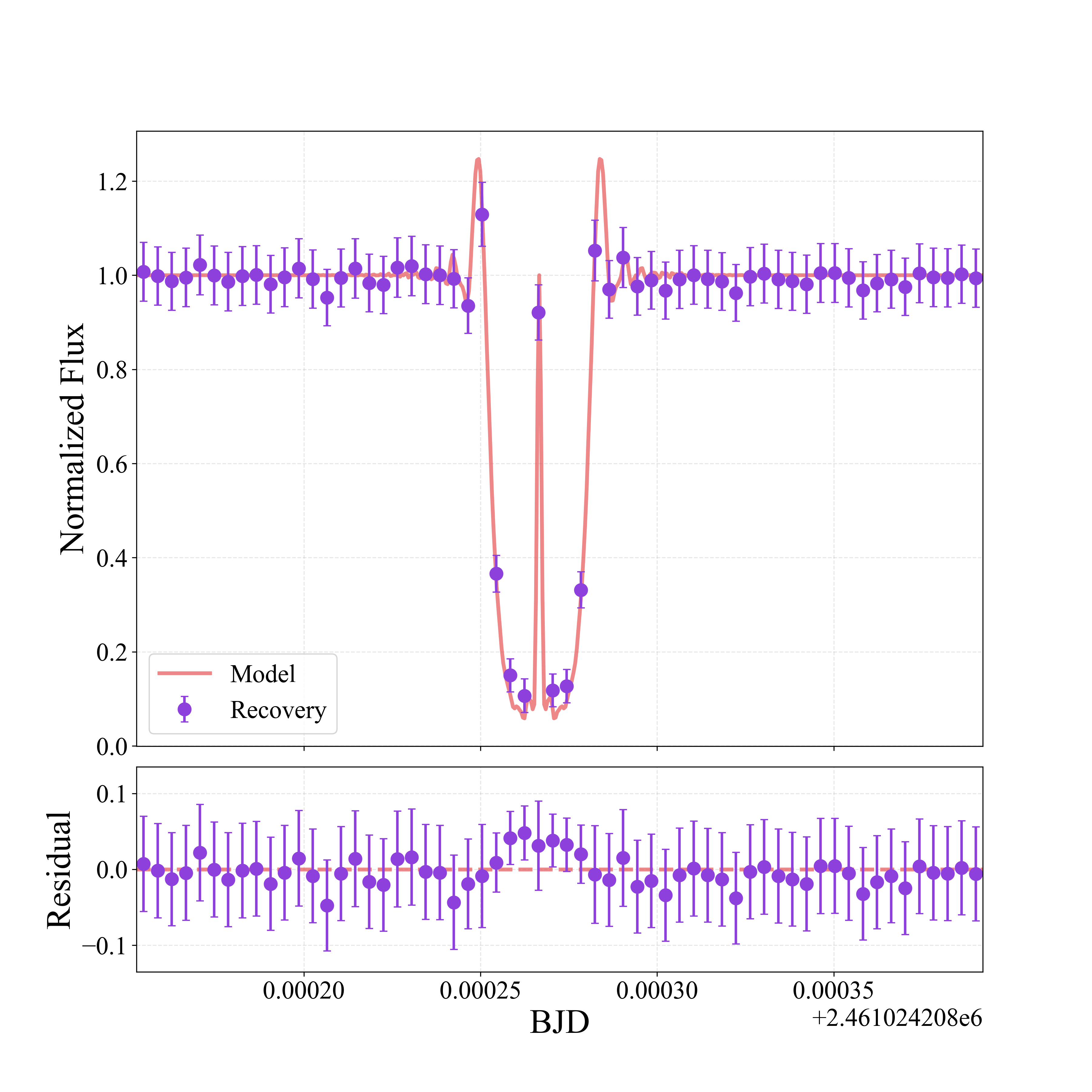}{0.33\textwidth}{(b) Occultation in Fresnel diffraction pattern.}%
  \fig{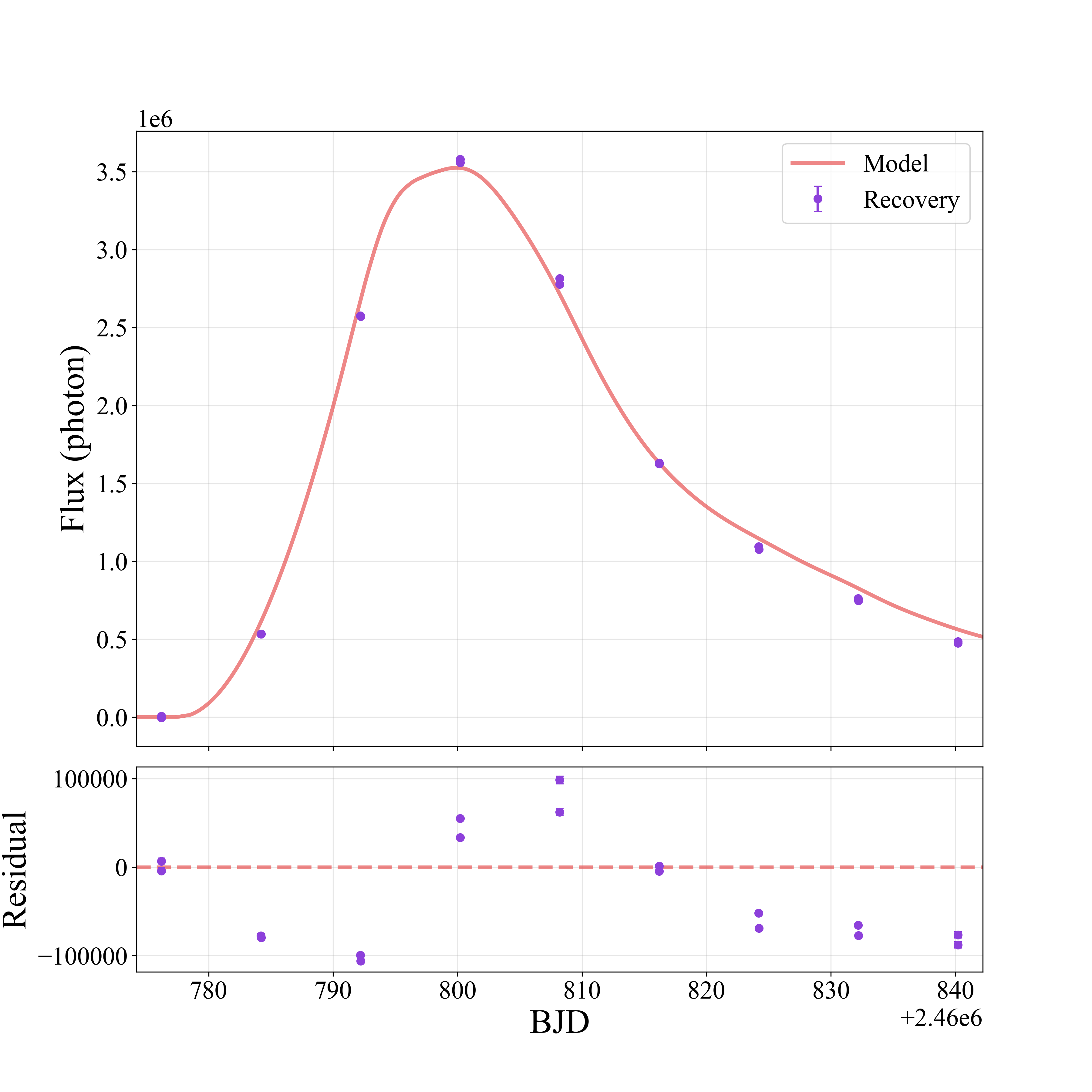}{0.33\textwidth}{(c) Supernova eruption.}%
}
\caption{The injection-recovery tests for time-domain phenomena. The figures show injected models and recovered signals, including occultation in both the geometric and Fresnel-diffraction patterns, as well as the supernova eruption. The red solid line is the theoretically injected transit light curve.}
\label{fig:transient}
\end{figure*}

In addition, we carry out injection-recovery tests for time-domain and transient phenomena to evaluate the suitability of the simulations for time-domain and transient-pipeline validation. These tests include occultation events in both the geometric and Fresnel-diffraction patterns, as well as astrophysical transients - supernova injected onto host galaxy. All experiments are performed using parameters representative of the Muguang Observatory. Because suitable real datasets are limited, we focus on injection-recovery analysis rather than direct comparisons with real observations or with SkyMaker outputs. All parameters employed in events models are listed in Table \ref{tab:transient_parameters}, and the recovery results are presented in Fig. \ref{fig:transient}. The residual RMS values are $1.486 \times10^{-2}$ for occultation in the geometric patterns, and $1.932 \times10^{-2}$ for occultation in the Fresnel-diffraction patterns. The relative residual RMS is $7.973 \times10^{-2}$ for the supernova eruption case. 
These results show that the recovery pipeline performs better for the geometric occultation case than for the Fresnel-diffraction case. The smaller residual RMS in the geometric pattern indicates that its light curve shape is more readily recovered, whereas the diffraction pattern yields a larger RMS because its finer oscillatory structure is more sensitive to photometric uncertainties and temporal sampling effects. For the supernova eruption case, the relative residual RMS of $7.973 \times 10^{-2}$ implies a typical fractional deviation of about 8\% between the recovered and input light curves. In addition, its temporal evolution and overall morphology also remain consistent. The experiments support that AstroSkyFlow is suitable for time-domain and transient-pipeline validation.

\begin{table}[htbp]
\centering
\caption{Parameters of Different Transient Events}
\label{tab:transient_parameters}
\resizebox{\textwidth}{!}{%
\begin{tabular}{lcclcll}
\hline
Variable type  & Parameter & Description & Value & Unit \\
\hline
occultation (geometric approximation) & $T_{\text{ref}}$ & reference time (center of the event) & 2461024.20826663 & BJD (TDB) \\
 & Mode & calculation mode: monochromatic or filters' name & Tianyu & -- \\
 & $\lambda$ & wavelength in \AA (if monochromatic) & -- & -- \\
 & $b$ & impact parameter & 2.0 & km \\
 & $R_{\text{occulter}}$ & occulter radius & 2000 & m \\
 & $D_{\text{obs}}$ & occulted object distance & $1.496 \times 10^{11}$ & m \\
 & $\theta_\star$ & stellar angular diameter & 0.1 & mas \\
 & $v_{\text{rel}}$ & relative speed of occulter& 0.5 & km/s \\
 & $f$ & brightness ratio between the occulter and the occulted object & 0.0 & -- \\
 & $\rho_{\text{limit}}$ & geometry or fresnel diffraction threshold & 5.0 & -- \\
\hline
occultation (Fresnel diffraction pattern) & $T_{\text{ref}}$ & reference time (center of the event) & 2461024.20826663 & BJD (TDB) \\
  & Mode & calculation mode: monochromatic or filters' name & Tianyu & -- \\
 & $\lambda$ & wavelength in \AA (if monochromatic) & -- & -- \\
 & $b$ & impact parameter & 2.0 & km \\
 & $R_{\text{occulter}}$ & occulter radius & 500 & m \\
 & $D_{\text{obs}}$ & occulted object distance & $1.496 \times 10^{11}$ & m \\
 & $\theta_\star$ & stellar angular diameter & 0.1 & mas \\
 & $v_{\text{rel}}$ & relative speed of occulter& 0.5 & km/s \\
 & $f$ & brightness ratio between the occulter and the occulted object & 0.0 & -- \\
 & $\rho_{\text{limit}}$ & geometry or fresnel diffraction threshold & 5.0 & -- \\
\hline
supernova &  $T_0$ & time of peak brightness & 2460800.2 & BJD (TDB)\\
 & $z$ & redshift & 0.10946385 & -- \\
 & $x_1$ & stretch parameter in SALT3 model & 0.1 & -- \\
 & $c$ & color parameter in SALT3 model & -0.1 & -- \\
 & filter & observation filter & Tianyu & -- \\
 & $M_\mathrm{V}$ & peak absolute magnitude & -19.3 & mag \\
\hline
\end{tabular}
}
\end{table}

\begin{figure*}[ht!]
\centering
\includegraphics[width=0.75\textwidth]{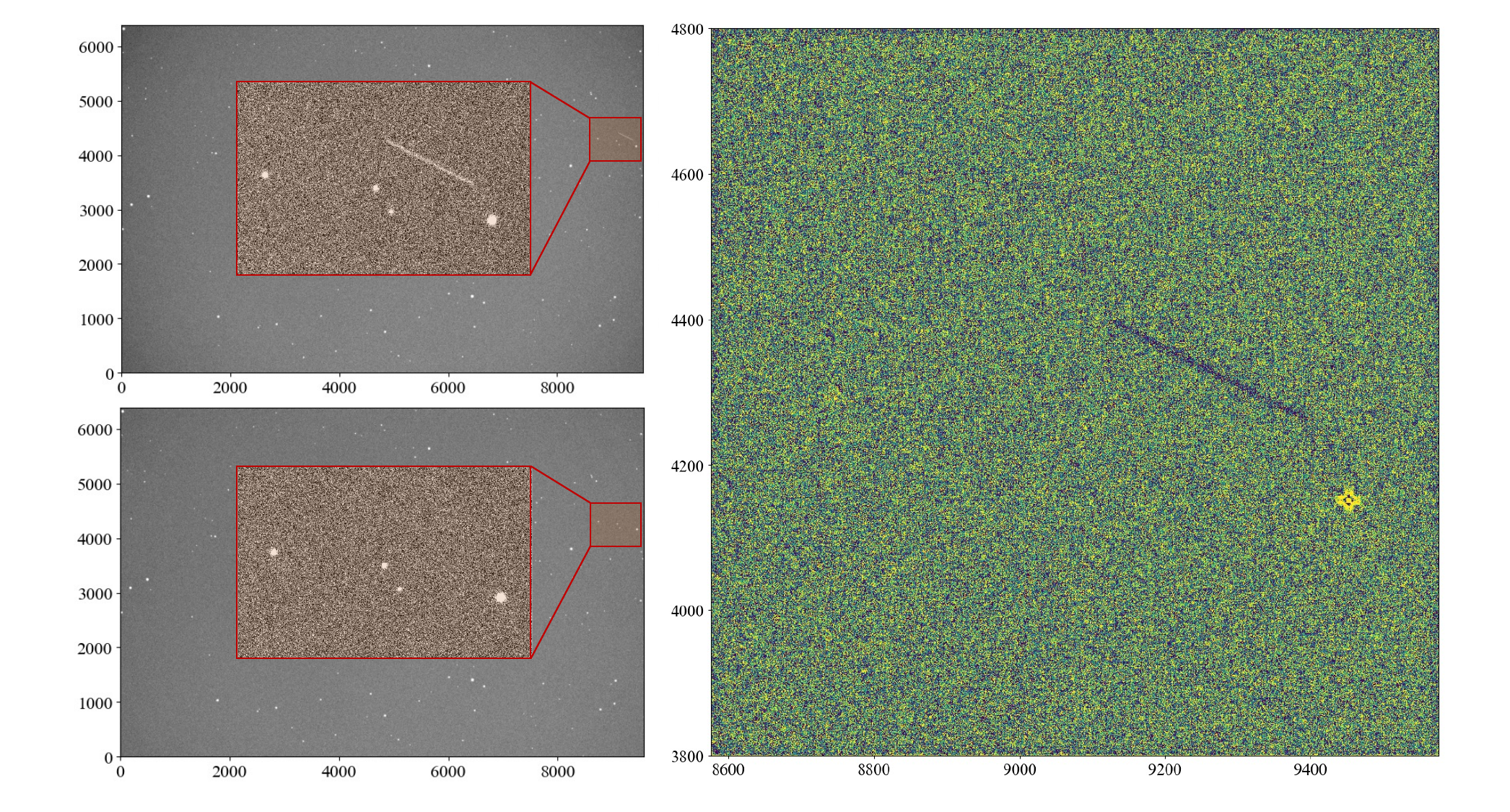}
\caption{Simulated image showing the streak of 2018 RC2 of Muguang Observatory. The left panel illustrates the AstroSkyFlow simulated frames during (top) and after (bottom) asteroid 2018 RC2 passage on Muguang Observatory. The right panel is the differential image between them.}
\label{fig:muguang_ios}
\end{figure*}

\begin{figure*}[ht!]
\centering
\includegraphics[width=0.95\textwidth]{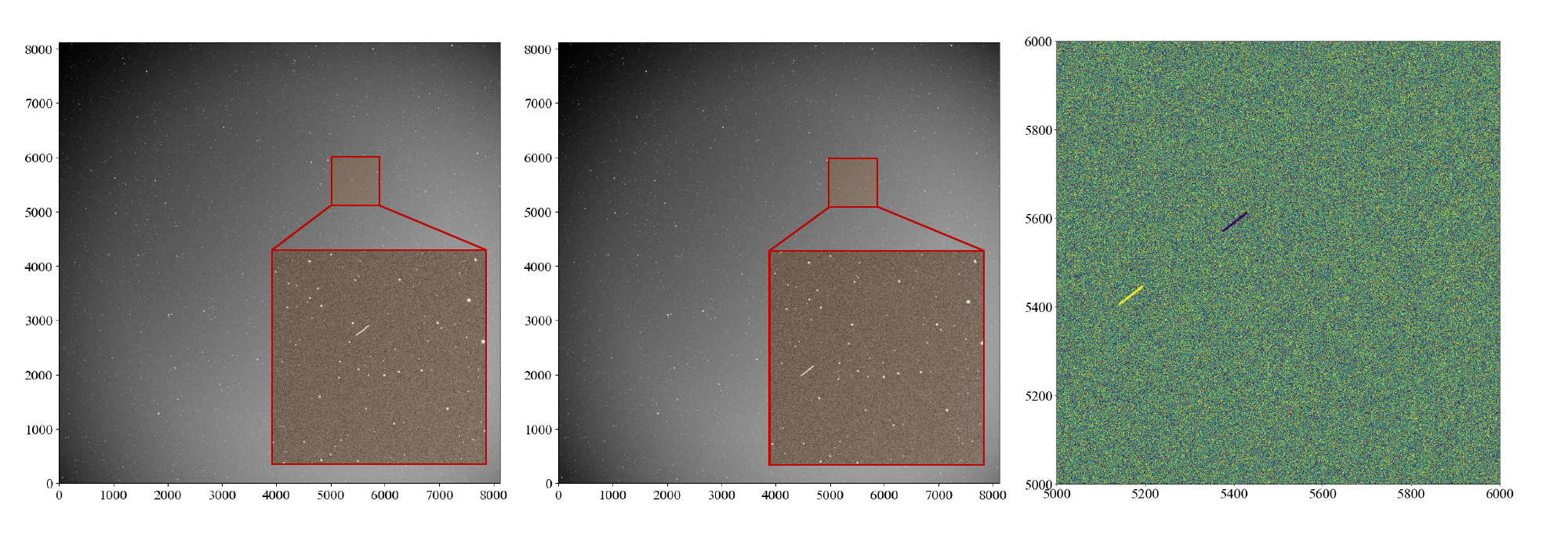}
\caption{Simulated image showing the streak of 2018 RC2 of Tianyu. The left and middle panels display the AstroSkyFlow simulated images captured at different epochs of Tianyu, where 2018 RC2 was in distinct positions. The right panel presents the corresponding differential image.}
\label{fig:tianyu_ios}
\end{figure*}

    To demonstrate the injection and recovery of motion signals, we employ asteroids as a test. Our simulation reproduces observations of the same field at different epochs, during which asteroids may traverse the field of view. Specifically, we simulate images capturing the passage of a fast-moving asteroid 2018 RC2 during March 8, 2025. As SkyMaker cannot simulate rapidly moving targets, we do not generate comparison images of SkyMaker. Asteroid detection is performed via differential imaging analysis using the Optimal Image Subtraction (OIS) \citep{2008MNRAS.386L..77B}. All images do calibration before differential image process. As illustrated in Fig. \ref{fig:muguang_ios}, the left panel shows the simulated frames during (top) and after (bottom) 2018 RC2 passage on Muguang Observatory. And the right panel is the differential image between them, which clearly demonstrates the streak caused by rapid movement, confirming successful injected motion signal recovery. We conduct similar tests in denser stellar fields, simulating images of 2018 RC2 passing by the Tianyu telescope. As shown in Fig. \ref{fig:tianyu_ios}, the left and middle panels display images captured at different times, where 2018 RC2 is in different positions. The right panel presents the corresponding differential image, revealing a clear moving streak that confirms the successful injection and recovery of the motion signal.
    
\subsubsection{Photometric precision}
To quantitatively assess the photometric precision of the AstroSkyFlow images, we use the Combined Differential Photometric Precision (CDPP) \citep{2010ApJ...713L..79K, 2012PASP..124.1279C} as a quantitative metric to compare the AstroSkyFlow simulated photometric image precision with the theoretical precision. We use lightkurve \citep{2018ascl.soft12013L} to calculate the measured CDPP of real observation, AstroSkyFlow and SkyMaker data separately. An 1-hour timescale detrending based on the Savitzky-Golay filter is applied to the relative flux, and the CDPP at a 0.5-hour timescale is computed from the detrended light curves obtained using photutils \citep{larry_bradley_2019_3368647}. \cite{2024AcASn..65...34F} provide a method for calculating the theoretical CDPP and explain its different components. 

\begin{figure*}[ht!]
\gridline{%
  \fig{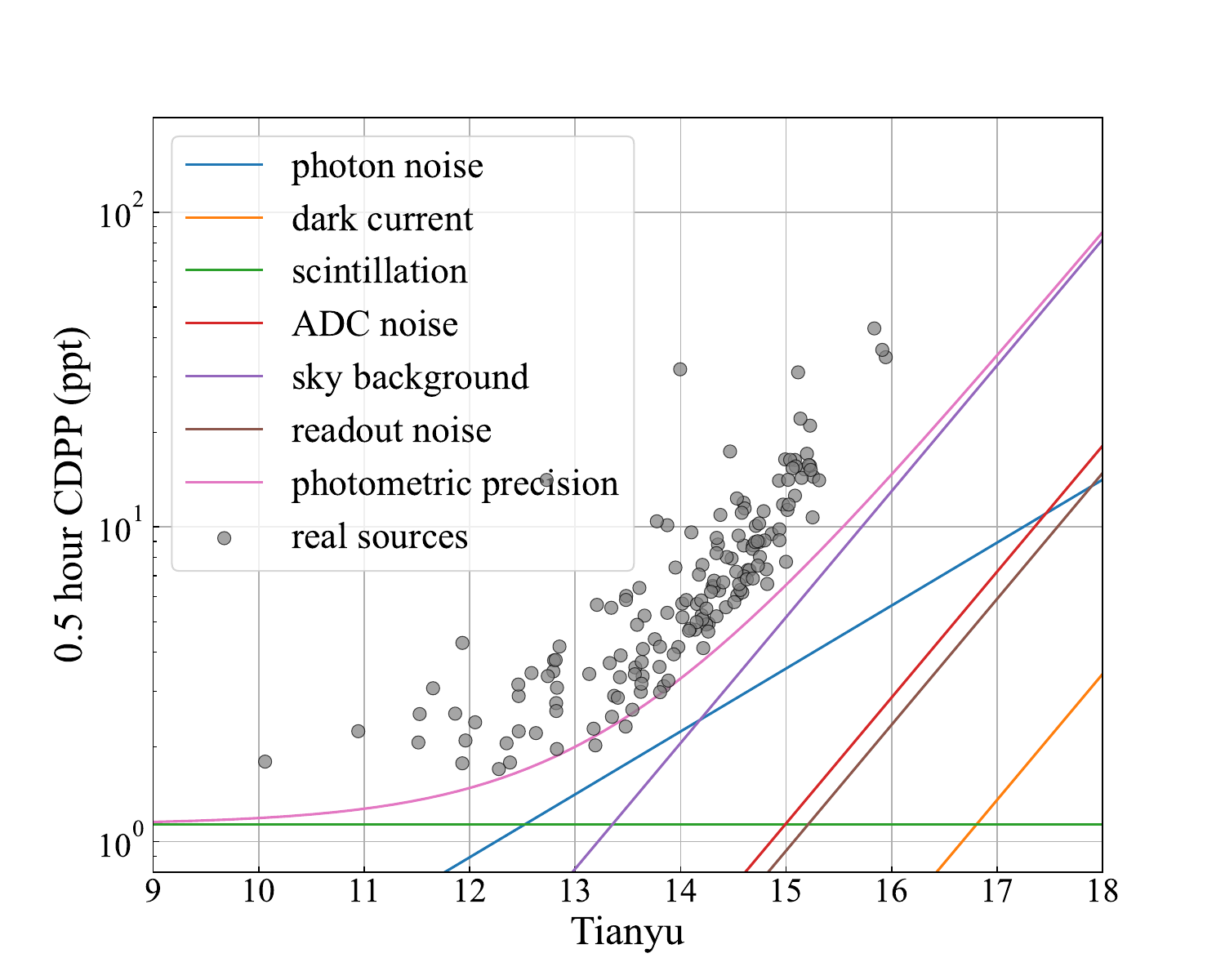}{0.33\textwidth}{(a) CDPP of real observation data.}%
  \fig{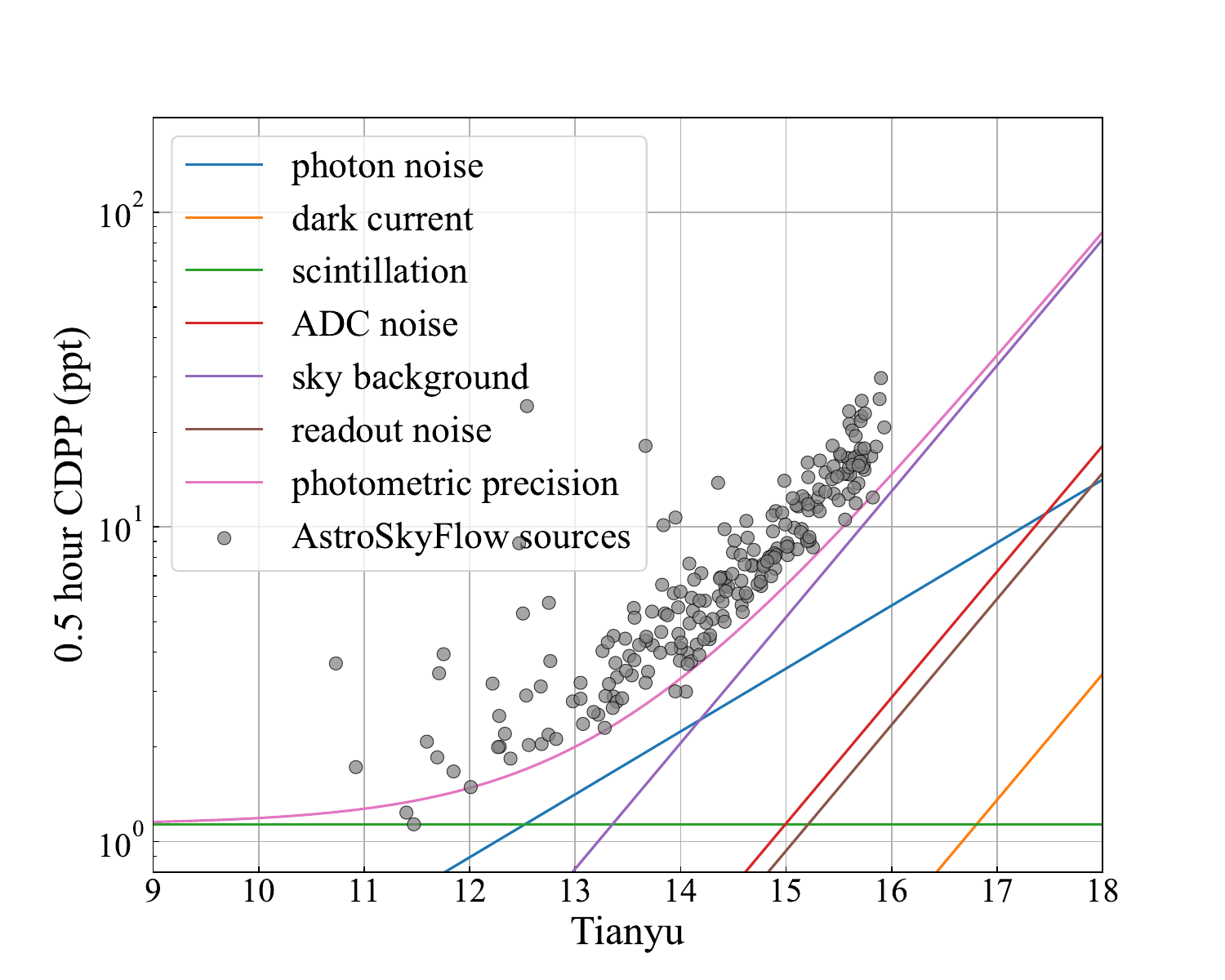}{0.33\textwidth}{(b) CDPP of AstroSkyFlow data.}%
  \fig{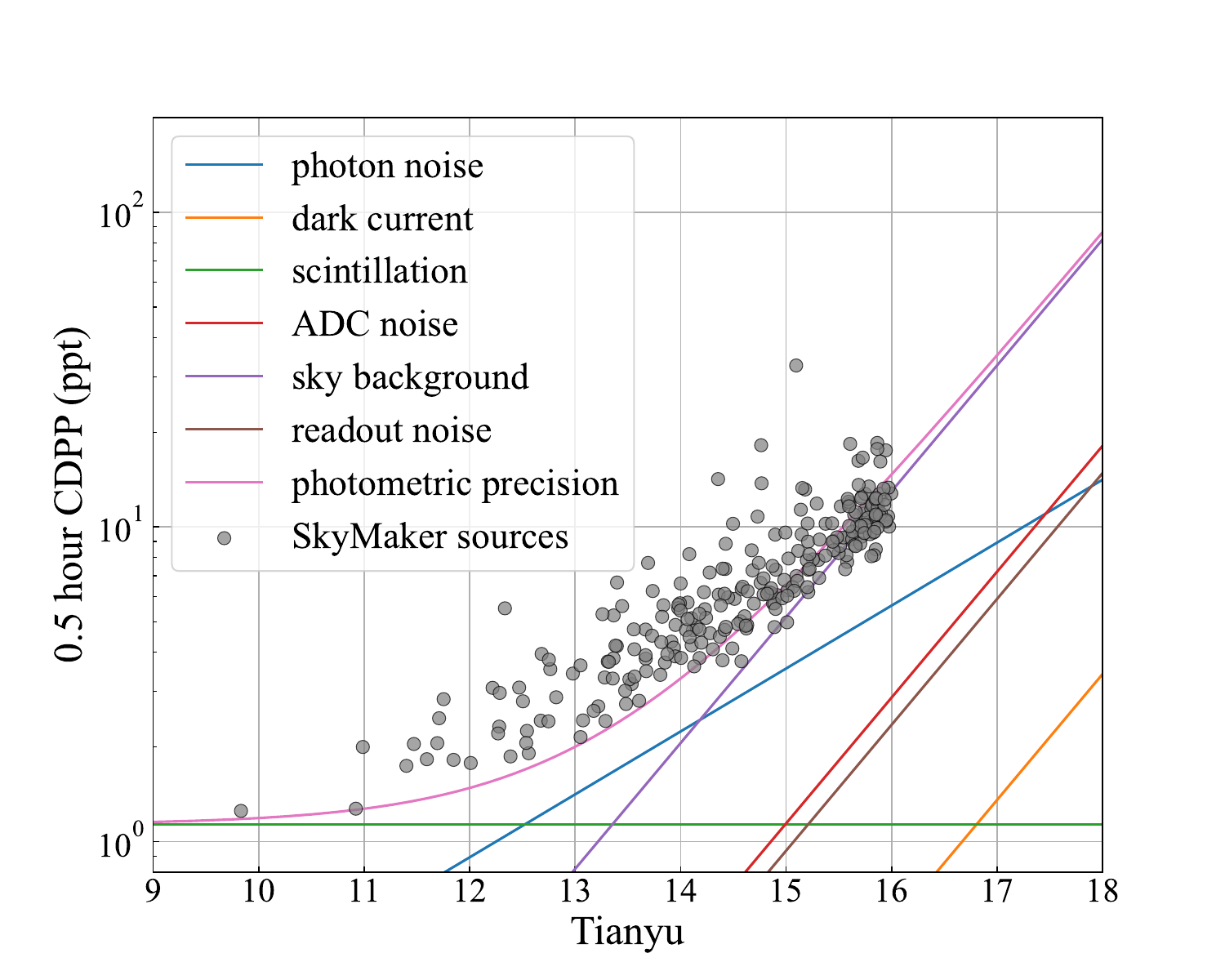}{0.33\textwidth}{(c) CDPP of SkyMaker data.}%
}
\caption{Comparison of the 0.5-hour CDPP from real observations, AstroSkyFlow, and SkyMaker for the Muguang-transit validation case. The measured 0.5-hour CDPP values are calculated from detrended light curves using \texttt{lightkurve} \citep{2018ascl.soft12013L}. The theoretical CDPP curve is computed following \citet{2024AcASn..65...34F}. Different straight lines indicate the contributions from different noise components. The horizontal axis shows the magnitude in the Tianyu band.}
\label{fig:CDPP}
\end{figure*}

\begin{figure*}[ht!]
\gridline{%
  \fig{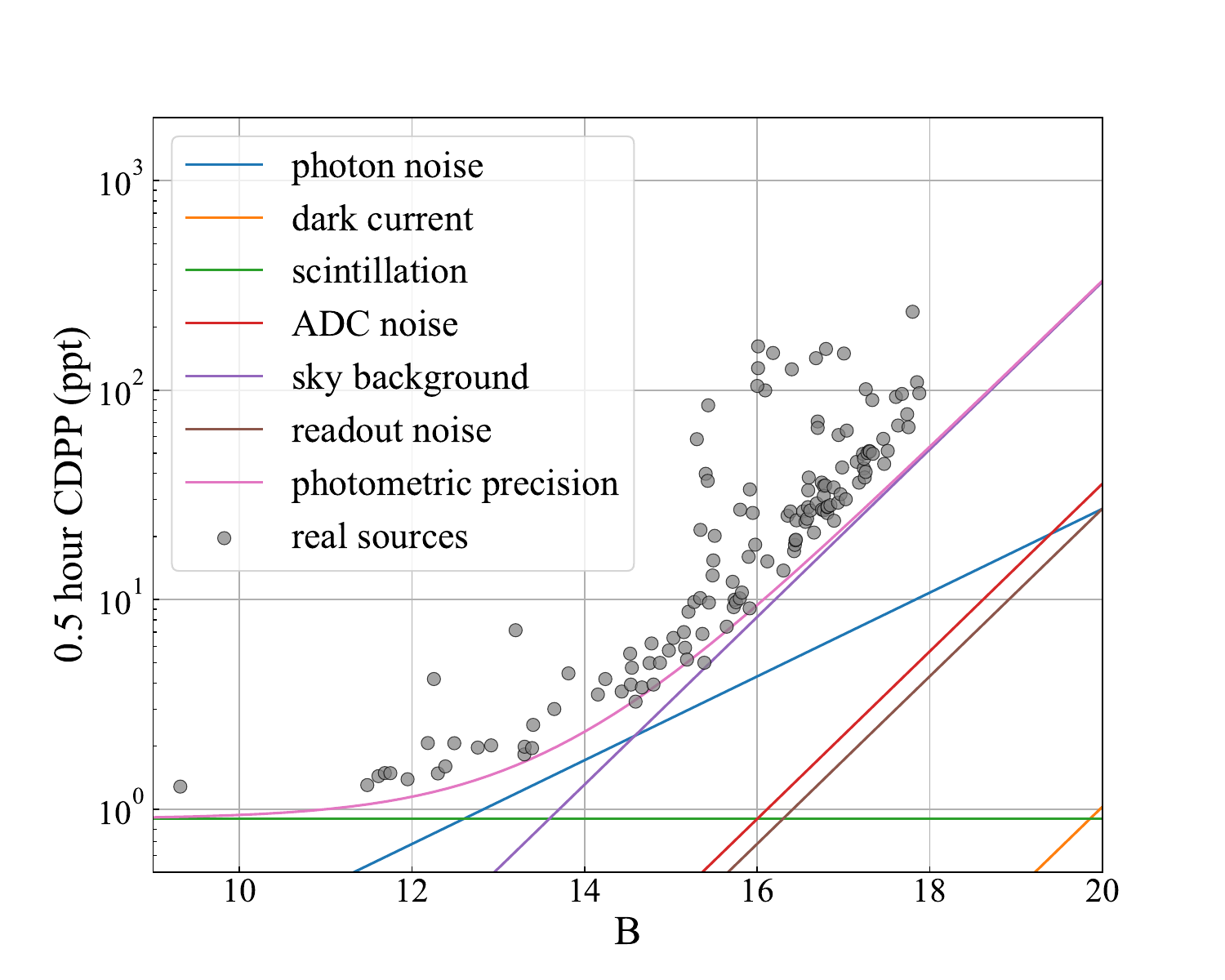}{0.33\textwidth}{(a) CDPP of real observation data.}%
  \fig{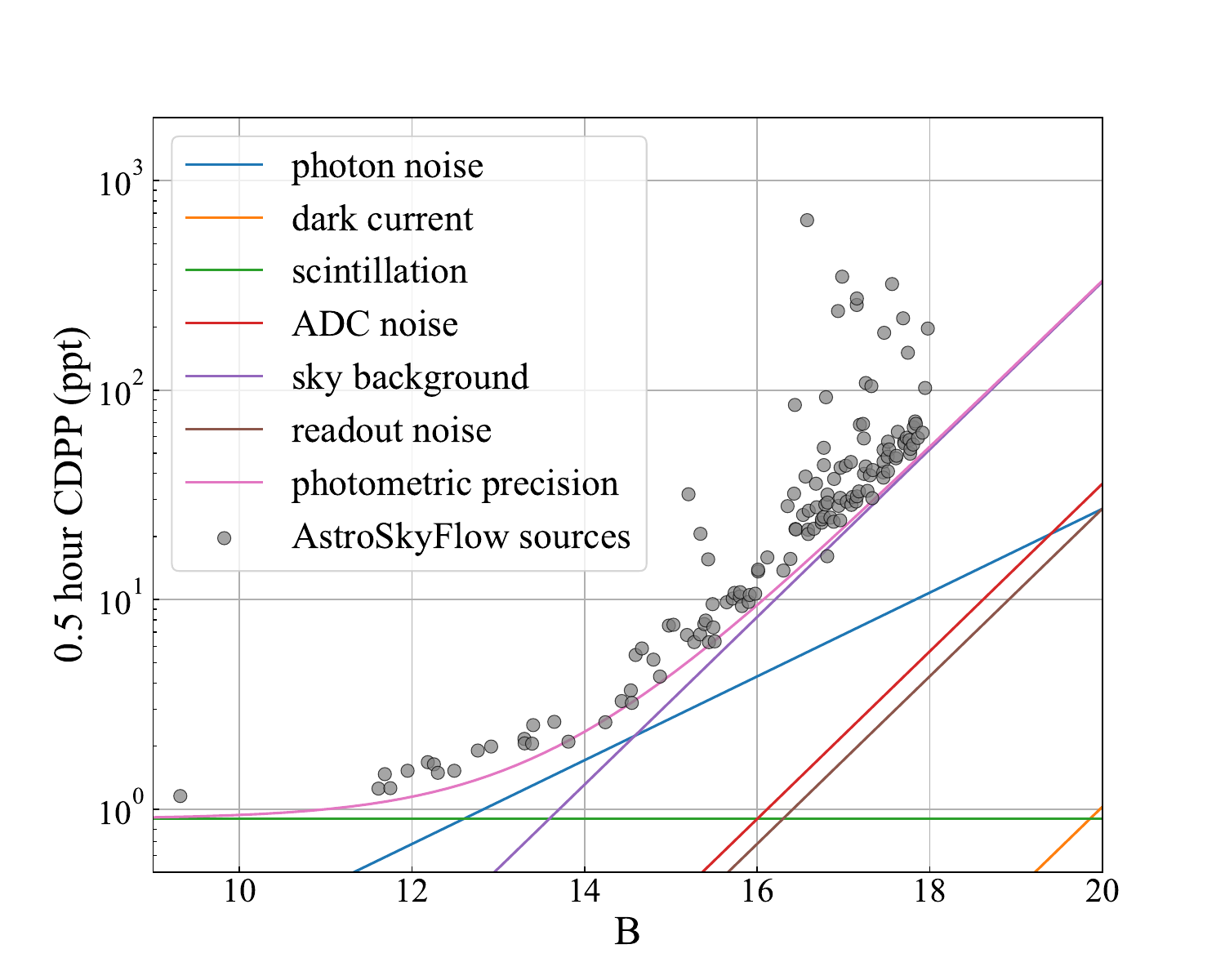}{0.33\textwidth}{(b) CDPP of AstroSkyFlow data.}%
  \fig{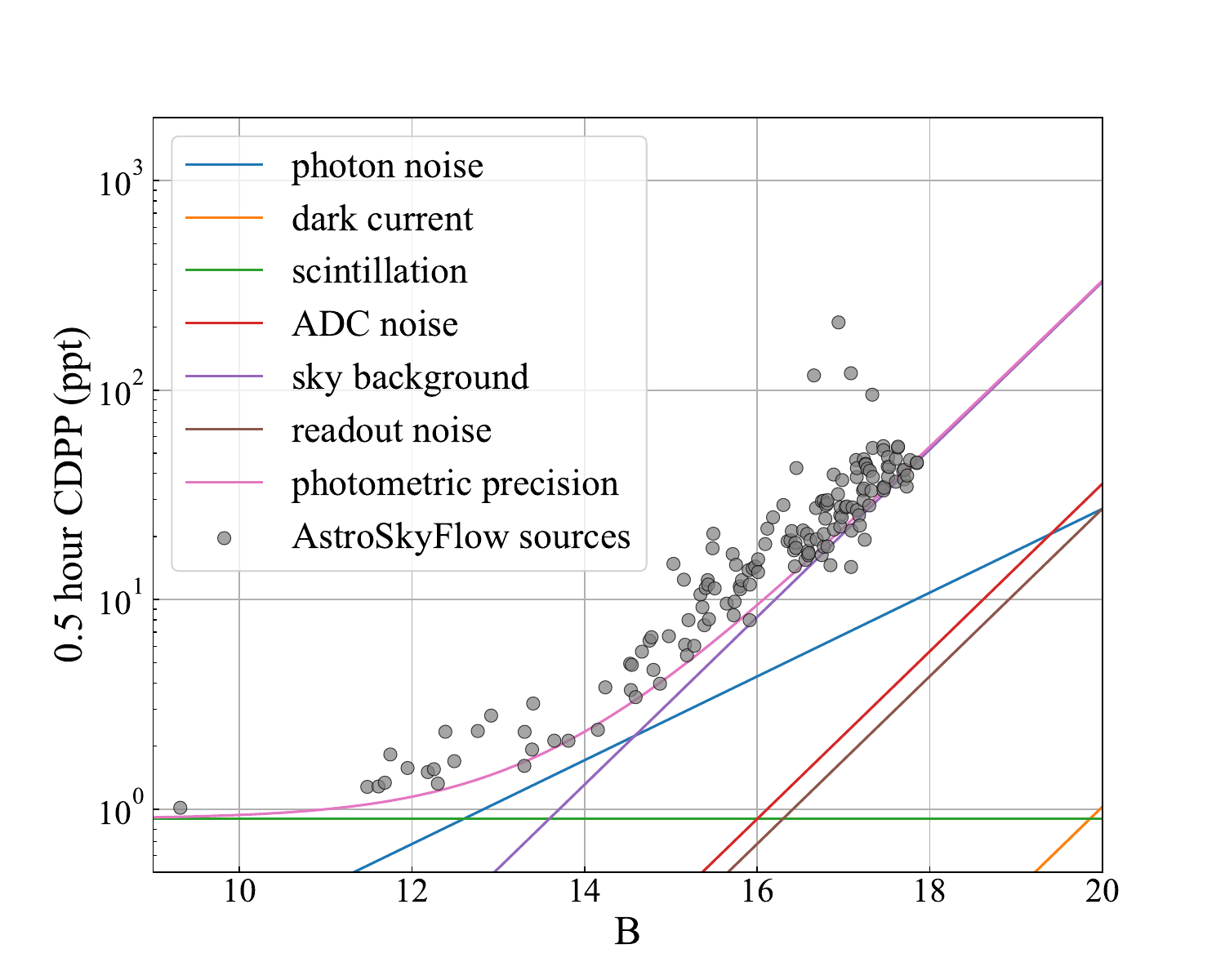}{0.33\textwidth}{(c) CDPP of SkyMaker data.}%
}
\caption{Comparison of the 0.5-hour CDPP from real observations, AstroSkyFlow, and SkyMaker for the Xinglong-binary validation case. The measured 0.5-hour CDPP values are calculated from detrended light curves using \texttt{lightkurve} \citep{2018ascl.soft12013L}. The theoretical CDPP curve is computed following \citet{2024AcASn..65...34F}. Different straight lines indicate the contributions from different noise components. The horizontal axis shows the magnitude in the B band.}
\label{fig:CDPP_binary}
\end{figure*}

The CDPP comparisons for the two validation cases are shown in Figures \ref{fig:CDPP} and \ref{fig:CDPP_binary}. Figure \ref{fig:CDPP} corresponds to the Muguang transit-validation case, while Figure \ref{fig:CDPP_binary} corresponds to the Xinglong binary-validation case. For the Muguang case, the horizontal axis is the synthetic Tianyu band magnitude rather than a calibrated band derived from real Tianyu observations. It denotes a synthetic instrumental bandpass defined from the expected total system response of Tianyu, including the telescope optics, detector quantum efficiency, filter configuration, and atmospheric transmission. 
Fig. \ref{fig:CDPP} exhibits a cutoff at 16 mag because the threshold magnitude in the AstroSkyFlow and SkyMaker input configuration files is set to 16. The lower boundary of the measured CDPP values for both the real observations and AstroSkyFlow is basically consistent with the theoretical curve. For targets with larger magnitudes, the theoretical CDPP curve nearly coincides with the sky background component, indicating that the photometric precision in this regime is primarily limited by the raw sky background. However, the theoretical prescription of \citet{2024AcASn..65...34F} includes only a single raw sky-background term. In practice, additional background contributions, such as scattered moonlight, scattered sunlight, and urban light pollution, further increase the effective background level. As a result, both real observation sources and AstroSkyFlow sources deviate upward from the theoretical CDPP curve at the larger magnitude end, with real observations showing a more pronounced deviation due to the influence of urban light pollution. This further demonstrates that for fainter stars with larger magnitudes, the CDPP should exhibit a steeper upward trend, as dimmer stars are more susceptible to various noise sources, harder to detect, and require higher measurement precision for reliable detection.
However, in the SkyMaker results, the CDPP for fainter stars does not exhibit this expected steep increase. Instead, it falls below the theoretical curve. This also indicates that some significant noises have been ignored in SkyMaker.
For the Muguang case, the median absolute deviation (MAD) of the residuals is 1.388 ppt for the real observations, 1.459 ppt for AstroSkyFlow, and 1.045 ppt for SkyMaker. 
For the Xinglong case, shown in Fig. \ref{fig:CDPP_binary}, the horizontal axis is the B band magnitude rather than the synthetic Tianyu-band magnitude. The same overall behavior is observed. The MAD of the residuals is 7.936 ppt for the real observations, 6.065 ppt for AstroSkyFlow, and 2.960 ppt for SkyMaker. Real observation and AstroSkyFlow sources exhibit greater dispersion than that of SkyMaker sources, indicating that SkyMaker remains more idealized than the other two.
Furthermore, the CDPP value is also influenced by the specific algorithms used for source extraction, light curve generation and the CDPP calculation. Different algorithms introduce different errors. 
Taken together, the two validation cases show that AstroSkyFlow reproduces the magnitude dependence of photometric precision more realistically than SkyMaker and AstroSkyFlow captures more observational noise sources.


\subsubsection{Point spread function properties}
\begin{figure*}[ht!]
\centering
  \gridline{
  \fig{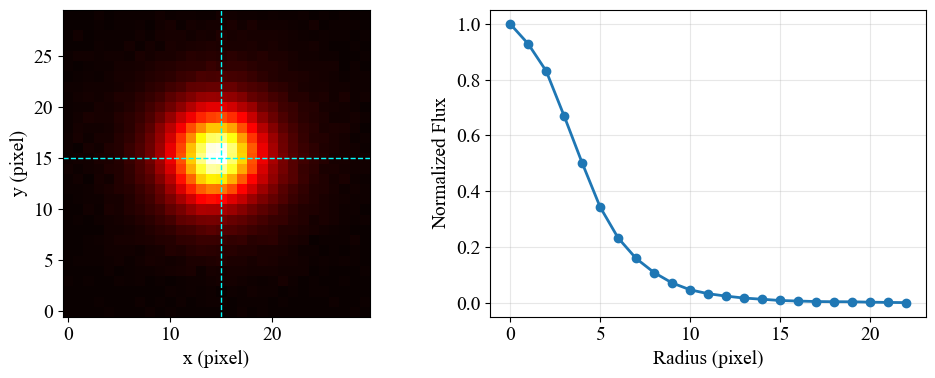}{0.5\textwidth}{(a) Real PSF of WASP-11 in the center of image.}
\fig{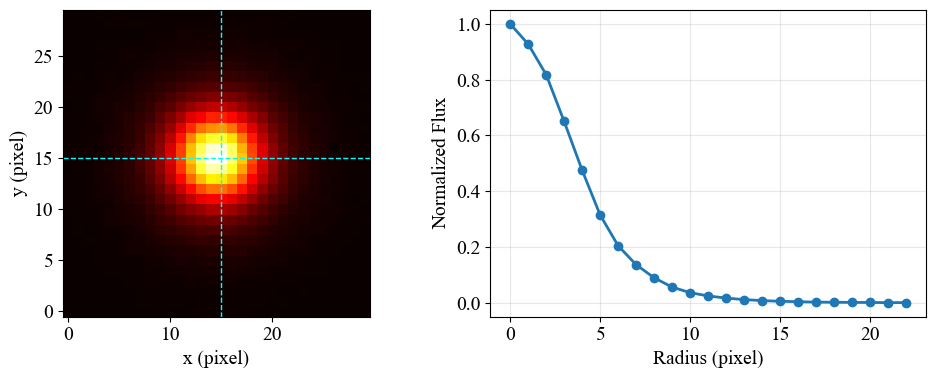}{0.5\textwidth}{(b) Simulated PSF of WASP-11 in the center of image.}
}
  \gridline{
  \fig{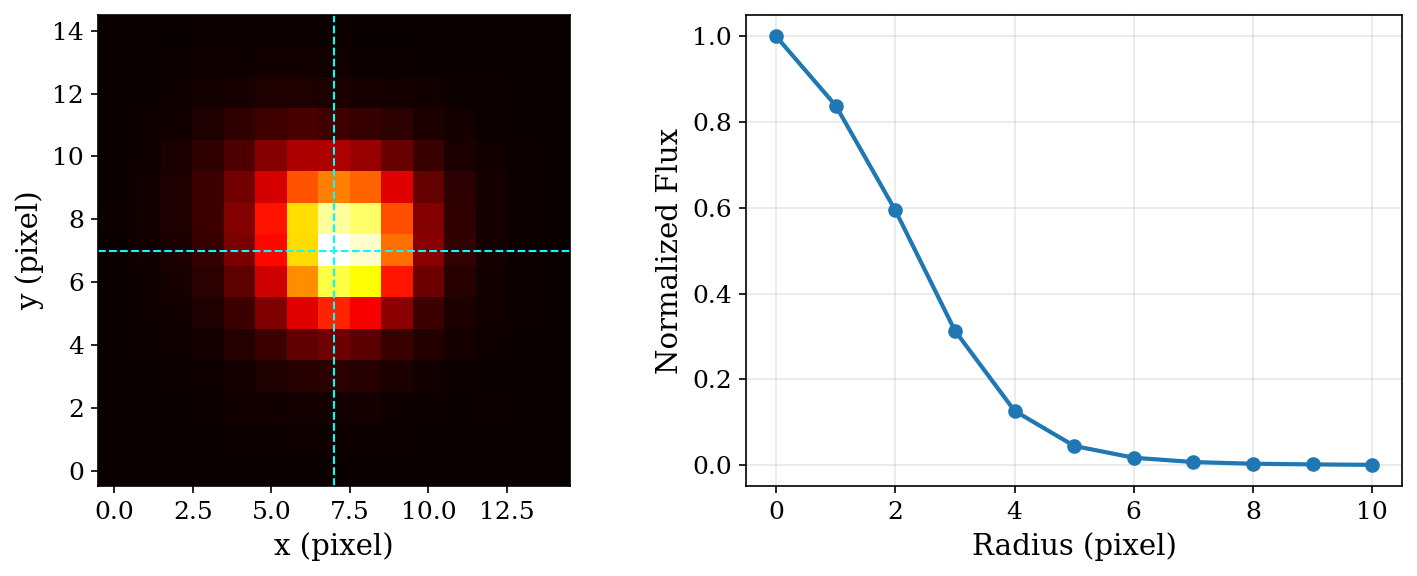}{0.5\textwidth}{(c) Real PSF of V0554 Dra in the center of image.}
\fig{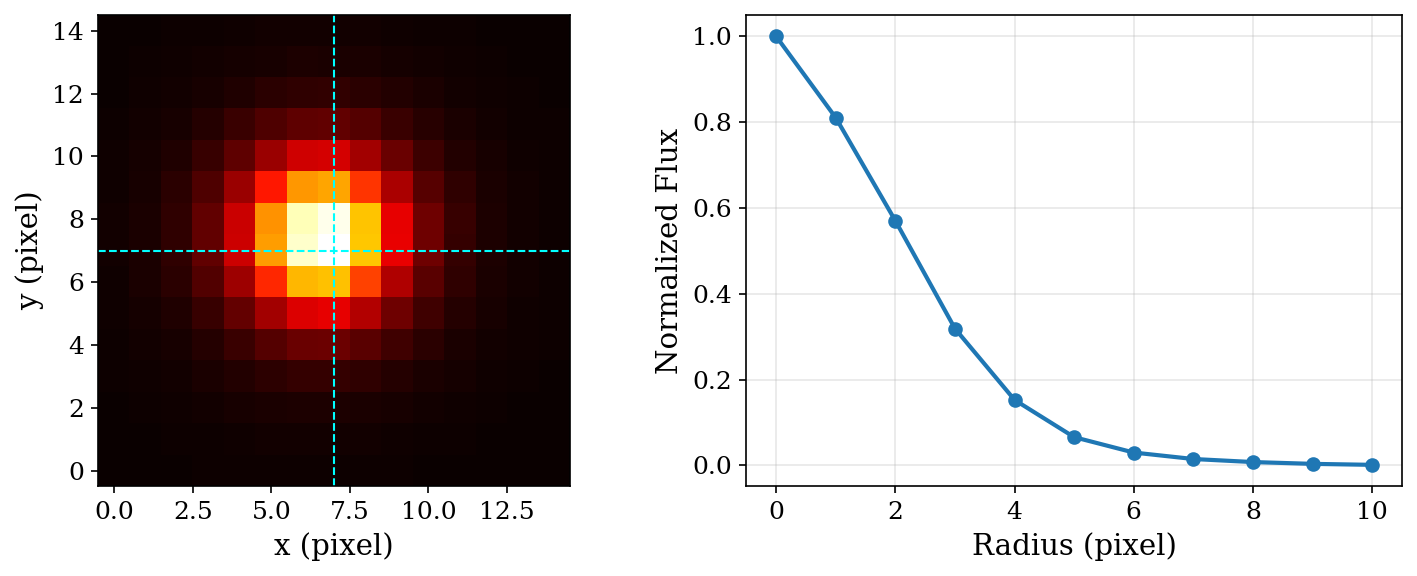}{0.5\textwidth}{(d) Simulated PSF of V0554 Dra in the center of image.}
  }
\caption{Real and simulated point spread function of WASP-11 and V0554 Dra. The brightest point is in the centre of the images.}
\label{fig:single_psf}
\end{figure*}

Accurate characterization of the PSF is critical for both photometric and astrometric measurements, as it directly impacts source detection, flux estimation and the deblending of overlapping sources. We conduct a comprehensive PSF analysis comparing real observations, AstroSkyFlow simulations, and SkyMaker simulations.
We begin with source WASP-11 and V0554 Dra which are in the center of images. As SkyMaker allows users to supply an external PSF, we use the Moffat profile with the same parameters of AstroSkyFlow as the input PSF in SkyMaker. As shown in Fig. \ref{fig:single_psf}, the brightest point is in the center of the images. Whether for WASP-11 or V0554 Dra, the real and simulated PSF exhibit nearly identical morphology and intensity distribution. 

\begin{figure*}[ht!]
  \gridline{\fig{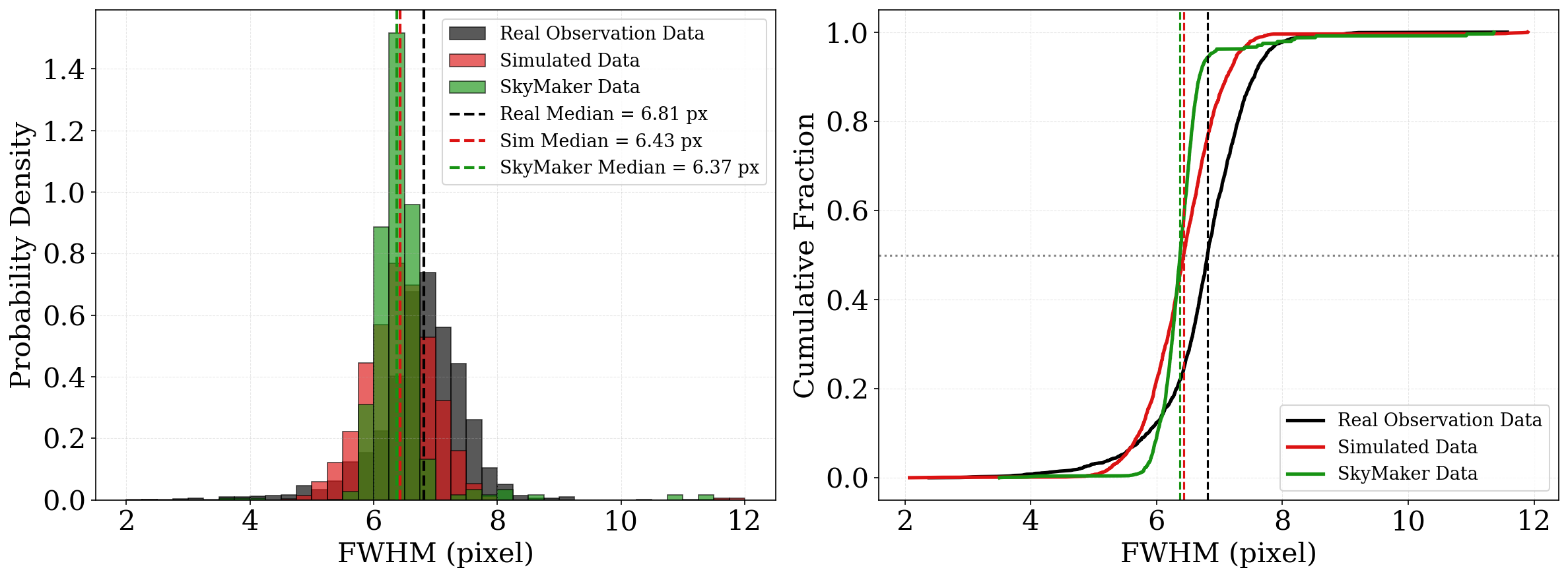}{0.75\textwidth}{(a) FWHM distribution in the real observation, AstroSkyFlow and SkyMaker images and their cumulative distribution for Muguang-transit validation case.}}
  \gridline{\fig{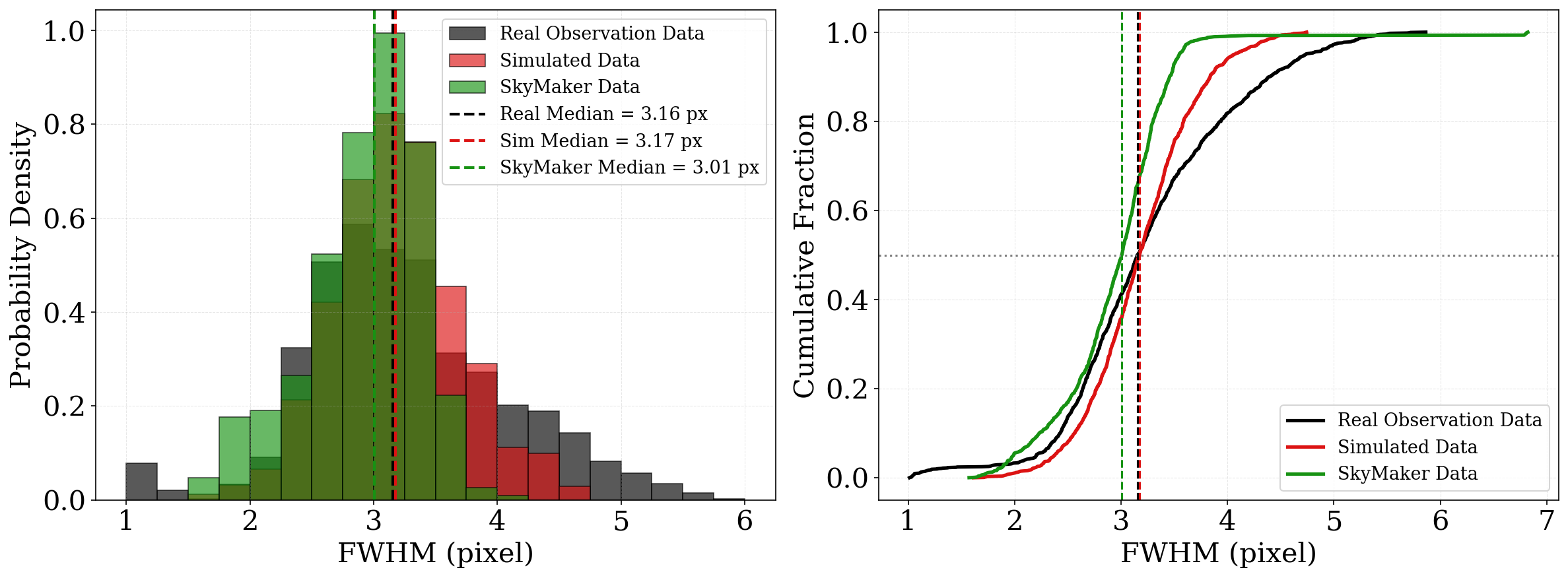}{0.75\textwidth}{(b) FWHM distribution in the real observation, AstroSkyFlow and SkyMaker images and their cumulative distribution for Xinglong-binary validation case.}}
  \caption{FWHM distribution in the real observation, AstroSkyFlow and SkyMaker images and their cumulative distribution comparison.}
  \label{fig:FWHM}
\end{figure*}

To quantify the variation in FWHM across the field of view, we perform a comparison of the PSF FWHM values extracted from the first ten 180 second exposure frames of real observations, AstroSkyFlow  and SkyMaker images, shown in Fig. \ref{fig:FWHM}. =For Muguang-transit validation case, the median FWHM is 6.81 pixel for the real observation data, 6.43 pixel for the AstroSkyFlow data and 6.37 pixel for the SkyMaker data, with the AstroSkyFlow data being smaller by approximately 5.58\% and SkyMaker data being smaller by approximately 6.46\%, as shown in Fig. \ref{fig:FWHM} (a). Although the three distributions overlap partially, both simulated distributions are more concentrated than the real one, as indicated by their steeper cumulative trends. To interpret this difference, we adopt a simplified model for the PSF width,
\begin{equation}
\sigma_{\mathrm{FWHM}} = \left( \sigma_\mathrm{s}^2 + \sigma_\mathrm{i}^2 \right)^{1/2},
\end{equation}
where $\sigma_\mathrm{s}$ denotes the contribution from atmospheric seeing and $\sigma_\mathrm{i}$ represents instrumental broadening. We infer that AstroSkyFlow likely underestimates the instrumental term $\sigma_i$, as it does not fully capture the complex optical aberrations, defocus, distortions, detector charge diffusion, pixelization effects and other random perturbations that contribute to PSF broadening in real observations. If we could integrate a spot diagram specific to the telescope-detector system, we can obtain a more accurate simulation of the PSF variations, thereby significantly improving its consistency with observational data. In Fig. \ref{fig:FWHM} (a), it is evident that the FWHM distribution from SkyMaker is highly concentrated, and its cumulative distribution curve is both less smooth and markedly steeper. This further highlights limitations in SkyMaker simulation.
For Xinglong-binary validation case, the median FWHM is 3.16 pixel for the real observation data, 3.17 pixel for the AstroSkyFlow data and 3.01 pixel for the SkyMaker data, with the AstroSkyFlow data being smaller by approximately 0.32\% and SkyMaker data being smaller by approximately 5.05\%\, as shown in Fig. \ref{fig:FWHM} (b). Similar to the Muguang-transit validation case, the simulated FWHM distributions are more concentrated than the real one, with the SkyMaker distribution showing the strongest concentration. 
Overall, although AstroSkyFlow does not perfectly reproduce the full complexity of the real observations, it provides a substantially more realistic representation of the PSF than SkyMaker while preserving the main characteristics observed in the data.
    


\subsection{Execution time}

\begin{figure*}[ht!]
\centering
  \gridline{
  \fig{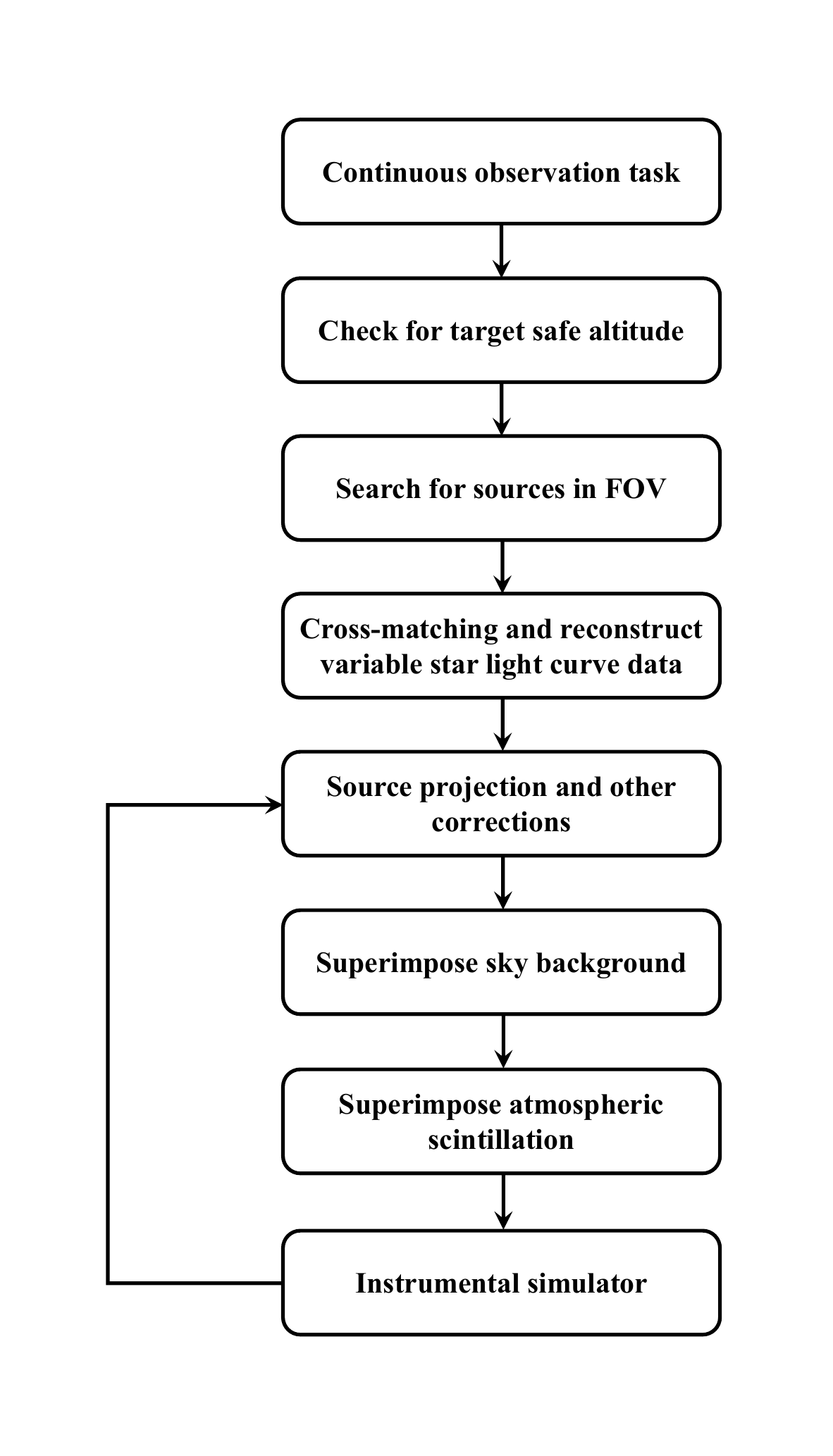}{0.25\textwidth}{(a) Operational sequence of AstroSkyFlow.}
\fig{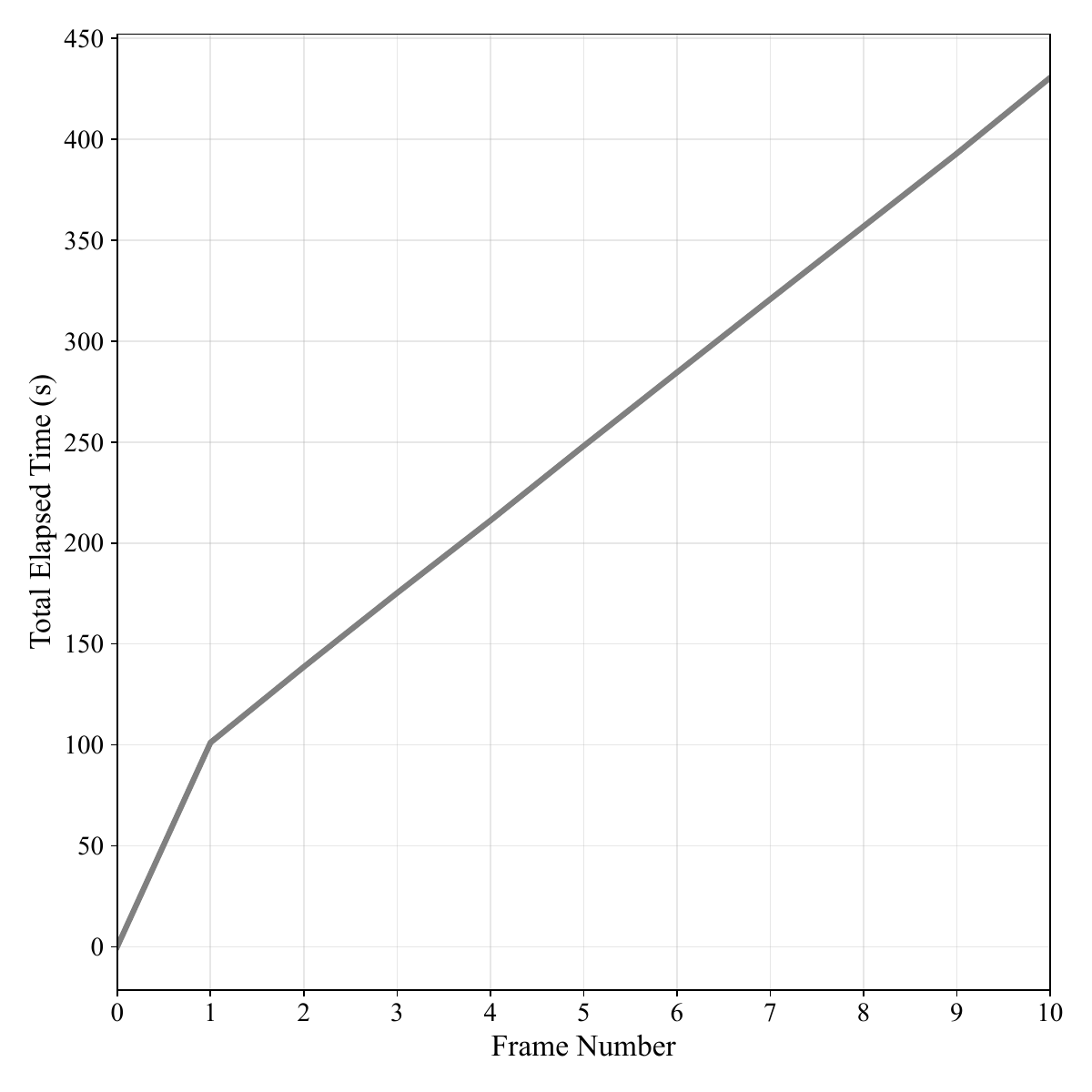}{0.425\textwidth}{(b) Total elapsed time as a function of the first 10 captured frames.}
}
\caption{The simulation loop and the time of first 10 captured frames. The left panel is operational sequence of AstroSkyFlow. When simulating continuous observations, it performs some one-time preprocessing and then enters a loop process to get continuous images. The right panel is the total elapsed time as a function of the first 10 captured frames when simulated the images during the transit of WASP-11 b in Section \ref{sec:comp}.}
\label{fig:time_flow}
\end{figure*}

We have tested the performance of AstroSkyFlow on both a laptop (MacBook Pro with Apple M3, 8 cores, and 16 GB of RAM) and a server node (dual Intel Xeon Ice Lake Platinum CPUs, 64 CPU cores in total, and 512 GB of memory). Importantly, the current version of AstroSkyFlow is executed as a serial, single-threaded pure-Python script on all systems. On the server node, we repeat with 16, 32, and 64 cores to assess the performance gains from increased core counts under the existing workflow. The operational sequence of AstroSkyFlow is illustrated in Fig. \ref{fig:time_flow} (a). And Fig. \ref{fig:time_flow} (b) shows the total elapsed time as a function of the first 10 captured frames when simulating the images during the transit of WASP-11 b in Section \ref{sec:comp}. When simulating continuous observations including multiple exposures of a fixed sky region, it performs some one-time preprocessing and then enters a loop process to get continuous images.

We break AstroSkyFlow into the following major time-consuming components: the time required to generate the first simulated image, denoted $T_{0}$, is significantly longer than the time required for subsequent image generation $T_{1}$. Because the former includes several one-time preprocessing steps. First, the master star catalog is filtered using a chunked search strategy to identify stellar sources within the FOV, with a time cost of $C_{\mathrm{chunk}}$. To accommodate different catalog sizes and prevent kernel crashes, we implement a chunk-based query strategy that reads and filters 5,000,000 rows per chunk and enforces a 2 GB memory cap per chunk. Using this approach, a catalog containing 127,498,636 rows can be filtered in approximately 30 seconds. Second, the galaxy catalog is filtered in a similar chunked manner to identify galaxies within the FOV, with a time cost of $C_\mathrm{galaxy}$. This step takes about 15 seconds to traverse the Gaia galaxy catalog once. Both of the above processes require traversing the entire catalog, since they are independent of the search radius. Third, time cost $C_{\mathrm{vsx}}$ of cross-matching variable stars in the field with the VSX database should be considered. Additionally, we account for $C_{\mathrm{lk}}$ during the first image generation process, which is the time required to search and download TESS/Kepler data for identified variable stars using the lightkurve package. After completing these one-time preprocessing steps, the additional time for each image mainly spent on the modeling process of the photon distribution of stars and galaxies, source corrections where the majority of the time spent on ADR, sky background, atmospheric scintillation, and hardware simulator including vignetting effects and sensor response. When considering asteroids, their rapid motion necessitates real-time position updates. Each image generation process must query the current asteroid positions via an API. The time cost of this step, denoted as $C_{\mathrm{asteroid}}$, depends on the specific website being queried and the network protocol used.

We characterize  how these simulation runtime scales with the image pixel count, considered effects, and the number of variable sources. And we report how the elapsed time is distributed across the pipeline stages. The specific relationship is described by equation (\ref{eq:time})
\begin{equation}\label{eq:time}
\begin{split}
\left\{
\begin{array}{l}
T_0(T_1,N_\mathrm v) = T_1 + \sum\limits_{i} C_j + N_\mathrm v \left( C_{lk} + A_{\mathrm{lk}}\right),\\
\\
T_1(P, P_\mathrm m, N_\mathrm s, N_\mathrm g, N_\mathrm v) = P\sum\limits_{i}A_\mathrm i + P \times log P \times A_{\mathrm{scint}} + N_{\mathrm g} \times A_{\mathrm g}\\
\quad+ N_{\mathrm s} \times P_{\mathrm m} \times log P_{\mathrm m} \times A_{\mathrm m} +  A_{\mathrm{ADR}}\times \sum\limits_{i}N_\mathrm i + C_{\mathrm{asteroid}},
\end{array}
\right.
\end{split}
\end{equation}
where the majority of computational operations in AstroSkyFlow exhibit linear time complexity, with processing times scaling linearly with input size. Only components that perform spatial convolutions or FFT-based operations exhibit the $S_{\mathrm{ize}}\log S_{\mathrm{ize}}$ complexity. In $T_0$, $N_\mathrm v$  represents the number of matched variables. $C_\mathrm i$ include $C_{\mathrm{chunk}}$, $C_\mathrm{galaxy}$,  $C_{\mathrm{vsx}}$ and $C_{\mathrm{lk}}$. $C_{\mathrm{chunk}}$ is determined primarily by local catalog size and chunk size; $C_{\mathrm{galaxy}}$ is determined mainly by galaxy catalog size; $C_{\mathrm{vsx}}$ is dominated  by the VSX website response time, the queried FOV, and the number of returned sources; $C_{\mathrm{lk}}$ is dominated by remote search/download latency in lightkurve package, the number of returned products, and network throughput. By construction, $A_{\mathrm{lk}}$ excludes these remote-query costs and denotes only local light curve processing, including quality filtering, normalization, and Lomb-Scargle modeling.
In $T_1$, $N_\mathrm s$, $N_\mathrm g$ and $N_\mathrm v$ represent the number of stars, galaxies and variables. $P$ is number of total pixels and $P_\mathrm m$ is the number of pixels covered by the Moffat calculation. $N_\mathrm i$ include $N_\mathrm s$, $N_\mathrm g$ and $N_\mathrm v$. The different coefficients A represent the temporal correlation of different parts. $A_\mathrm i$ include $A_{\mathrm{sky}}$, $A_{\mathrm{vignetting}}$ and $A_{\mathrm{sensor}}$. The detailed explanations for different coefficients A and the specific fitting results are presented in Table \ref{tab:A_detailed}. And $C_{\mathrm{asteroid}}$ is determined by the API response, network conditions, the requested FOV, magnitude limit, and the number of returned asteroids.
\begin{table}[!h]
    \centering
    \caption{Detailed descriptions and values of different coefficients $A$. The 16/32/64-core measurements were obtained by running the same serial, single-threaded pure-Python script on a single 64c512g node (2 $\times$ Intel Xeon ICX Platinum, 64 CPU cores total, 512 GB RAM).}
    \label{tab:A_detailed}
    \resizebox{\textwidth}{!}{%
    \begin{tabular}{llccccc}
    \hline
    Coefficient & Description & MacBook Pro (1 core) & 16 cores & 32 cores & 64 cores & Unit \\ \hline
    $A_{\mathrm{lk}}$ & Local light curve processing after lightkurve search. & $6.589 \times 10^{-2}$ & $6.488 \times 10^{-2}$ & $6.562 \times 10^{-2}$ & $6.486 \times 10^{-2}$ & $\mathrm{sec\,target^{-1}}$\\
    $A_{\mathrm{sky}}$ & Modeling the sky background level. & $1.571 \times 10^{-7}$ & $6.342 \times 10^{-8}$ & $6.454 \times 10^{-8}$ & $6.500 \times 10^{-8}$ & $\mathrm{sec\,pixel^{-1}}$ \\
    $A_{\mathrm{vignetting}}$ & Calculating the vignetting effect. & $1.590 \times 10^{-7}$ & $8.680 \times 10^{-8}$ & $8.906 \times 10^{-8}$ & $8.977 \times 10^{-8}$ & $\mathrm{sec\,pixel^{-1}}$ \\
    $A_{\mathrm{sensor}}$ & Simulating the sensor response and noise chain. & $1.830 \times 10^{-7}$ & $2.264 \times 10^{-7}$ & $2.255 \times 10^{-7}$ & $2.258 \times 10^{-7}$ & $\mathrm{sec\,pixel^{-1}}$ \\
    $A_{\mathrm{scint}}$ & Modeling atmospheric scintillation. & $9.853 \times 10^{-9}$ & $ 6.952\times 10^{-9}$ & $7.053 \times 10^{-9}$ & $7.096 \times 10^{-9}$ & $\mathrm{sec\,pixel^{-1}}$ \\
    $A_{\mathrm g}$ & Modeling galaxy morphology and photon distribution. & $7.617 \times 10^{-2}$ & $1.411 \times 10^{-1}$ & $1.465 \times 10^{-1}$ & $1.459 \times 10^{-1}$ & $\mathrm{sec\,target^{-1}}$ \\
    $A_{\mathrm m}$ & Modeling the stellar PSF with a Moffat profile. & $2.543 \times 10^{-9}$ & $1.563 \times 10^{-9}$ & $1.607 \times 10^{-9}$ & $1.636 \times 10^{-9}$ & $\mathrm{sec\,pixel^{-1}\,target^{-1}}$ \\
    $A_{\mathrm{ADR}}$ & Correcting atmospheric differential refraction. & $2.881 \times 10^{-3}$ & $4.136 \times 10^{-3}$ & $4.302 \times 10^{-3}$ & $4.344 \times 10^{-3}$ & $\mathrm{sec\,target^{-1}}$ \\ \hline
    \end{tabular}}
\end{table}

\begin{figure*}[ht!]
\centering
\includegraphics[width=0.8\textwidth]{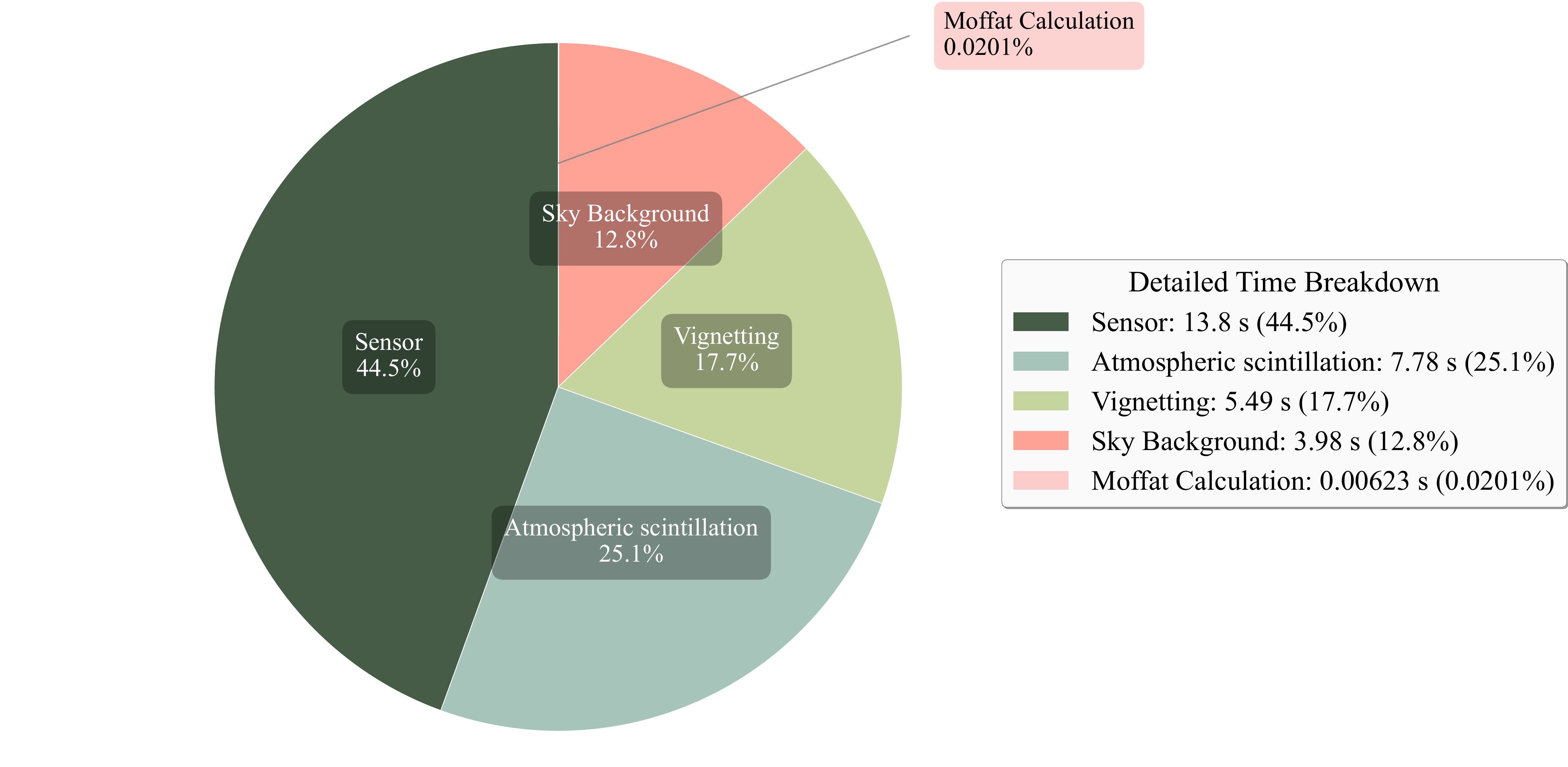}
\caption{The proportion of elapsed time for the computation of each effect.}
\label{fig:time_pro}
\end{figure*}

In AstroSkyFlow, each physical effect is controlled by an independent flag, allowing for modular testing and performance analysis. 
Table \ref{tab:A_detailed} shows that simply allocating more CPU cores to the current script does not lead to a systematic speed-up. The measured coefficients on the 16, 32, and 64 cores allocations
differ only at the few-percent level, which is consistent with the fact that the present version of AstroSkyFlow is still executed as a serial, single-thread Python program. In the present serial implementation, the relatively large coefficients are $A_{\mathrm{ADR}}$ and $A_\mathrm g$, so the per-frame cost rises rapidly as the number of corrected targets and galaxies increases.
Most modern observatories employ guiding systems that track the field center in real time, compensating for ADR offsets as field center zenith angle and observing conditions change. If guiding is effective, differential ADR across the field caused by variations in zenith angle is neglected, and when an ADC is used to minimize band-dependent dispersion, the ADR term can be negligible. Furthermore, if the scientific goal does not focus on galaxies or asteroids, the corresponding terms can also be bypassed. Under these assumptions, $T_1$ can be simplified in a relatively simple case:
\begin{equation}
\label{eq:simple}
\begin{array}{l}
   T_1(P, P_{\mathrm m}, N_{\mathrm s}) =  P(A_\mathrm{sky} + A_\mathrm{vignetting} + A_\mathrm{sensor}) + P\times log P \times A_\mathrm{scint} \\
   \quad + N_{\mathrm s} \times P_{\mathrm m} \times log P_{\mathrm m} \times A_\mathrm{m}.
\end{array}
\end{equation}
Based on the equation (\ref{eq:simple}), the total elapsed time when simulated a simplified case with 9576 $\times$ 6388 pixels, 500 stars, and a 33 $\times$ 33 Moffat footprint is approximately 31.07 s per image. The proportion of elapsed time for each effect calculation is shown in Fig. \ref{fig:time_pro}. For a 9576 $\times$ 6388 float32 image, the data volume per frame is about 244.7 MB. And the corresponding image-generation throughput is about $\sim 8$ MB s$^{-1}$. This is significantly below the processing requirement of the Tianyu, whose software system is intended to support raw-data rates exceeding 500 MB s$^{-1}$ for transiting-exoplanet, variable-star, and transient surveys \citep{2024AcASn..65...34F, 2025PASP..137f4501R}. Therefore, the current implementation of AstroSkyFlow is not suitable for real-time or near-real-time use. Instead, AstroSkyFlow is better regarded as an offline or ahead-of-time image generator for pipeline validation, injection--recovery tests, and machine-learning training. Substantially higher throughput will require explicit parallelization in future versions.

\section{Discussion and Conclusion} \label{sec: con}
We have developed AstroSkyFlow, a modular, photometric image simulator designed to generate high-fidelity, multi-epoch image sequences for pipeline validation and machine-learning training. AstroSkyFlow incorporates an external scene photon simulator that accounts for celestial sources, intrinsic variability, atmospheric effects including extinction, differential refraction, sky background, scintillation, and other user-added scatter light 
and a hardware simulator that models optics and sensor response. AstroSkyFlow is driven by an observing schedule and can inject a variety of time-domain signals while remaining highly configurable via a user configuration file.

To validate the performance of AstroSkyFlow, we conducted a comparative analysis between AstroSkyFlow, real observations and SkyMaker. Our validations against real observations from the Muguang and Xinglong Observatories demonstrate that AstroSkyFlow produces images with noise characteristics and PSF properties that are significantly more consistent with real data than those generated by the popular SkyMaker simulator. Specifically, the CDPP of AstroSkyFlow images follows the theoretical curve and real data more closely, while SkyMaker underestimates the noise, especially for fainter sources. The PSF FWHM distribution in AstroSkyFlow images is in better agreement with real observations than in SkyMaker. Moreover, AstroSkyFlow successfully recovers injected photometric and motion signals, which are challenging for normal simulators. 

AstroSkyFlow offers superior flexibility in signal injection and workflow automation. It is capable of batch-generating multi-epoch images across distinct sky fields from an input comprehensive observation schedule file. And AstroSkyFlow is highly configurable and can be adapted to different telescopes and sites by adjusting input parameters, making it a versatile tool for survey strategy optimization, pipeline testing, and data-challenge generation. Its ability to generate realistic, time-series images with accurate noise and variability makes it particularly valuable for the development of machine-learning methods in astronomy. As more and more time-domain surveys become operational, tools like AstroSkyFlow will be crucial for preparing and validating the analysis pipelines that will handle the massive data streams.

Our current simulator assumes instrumental characteristics and typical atmospheric profiles. We currently neglect some second-order effects. For instance, the PSF is approximated with a Moffat profile, without incorporating corrections for optical aberrations. Future developments will consider more second-order effects. At present, galaxies are modeled as simple Sérsic profiles using GalSim. Future work will implement more realistic galaxy morphology. Additionally, we plan to incorporate corrections for proper motion and parallax of static sources, simulate more categories of variable stars, and extend the simulation to include additional classes of fast-moving objects such as comets. Critically, to address current throughput limitations and facilitate large-scale data production, future versions will focus on the parallelization of the core computational modules.
The current version of the AstroSkyFlow code is archived on Zenodo \citep{li_2026_19733852}, and the corresponding GitHub repository is available at \url{https://github.com/laomuliubingbing/AstroSkyFlow}, and we encourage the community to use and extend it for their own applications.

\begin{acknowledgments}
We would like to thank Peng Jia and Ming Yang for helpful discussions. We would like to thank Min Fang for help with the Tianyu star catalog, and Jiarui Sun for assistance with galaxy modeling.
This work is supported by the National Key R\&D Program of China, No. 2024YFA1611801 and No. 2024YFC2207700, by
the National Natural Science Foundation of China (NSFC) under Grant No. 12473066, by the Shanghai Jiao Tong
University 2030 Initiative, and by the China-Chile Joint Research
Fund (CCJRF No. 2205). CCJRF is provided by Chinese Academy of Sciences South America Center for Astronomy
(CASSACA) and established by National Astronomical Observatories, Chinese Academy of Sciences (NAOC) and
Chilean Astronomy Society (SOCHIAS) to support China-Chile collaborations in astronomy. This project is supported in part by Office of Science and Technology, Shanghai Municipal Government (grant Nos. 24DX1400100,ZJ2023-ZD-001). This paper includes data collected by the Kepler mission and obtained from the MAST data archive at the Space Telescope Science Institute (STScI). Funding for the Kepler mission is provided by the NASA Science Mission Directorate. STScI is operated by the Association of Universities for Research in Astronomy, Inc., under NASA contract NAS 5–26555. This paper includes data collected with the TESS mission, obtained from the MAST data archive at the Space Telescope Science Institute (STScI). Funding for the TESS mission is provided by the NASA Explorer Program. STScI is operated by the Association of Universities for Research in Astronomy, Inc., under NASA contract NAS 5–26555.

\end{acknowledgments}




%

\facilities{Kepler, TESS, Tianyu, Xinglong 85cm, Muguang Observatory}

\software{numpy \citep{Harris20}, pandas \citep{reback2020pandas}, lightkurve \citep{2018ascl.soft12013L}, astropy \citep{astropy13}, batman \citep{2015PASP..127.1161K}, phoebe \citep{2016ApJS..227...29P}, sncosmo \citep{barbary_2025_15019859}, scipy \citep{2020NatMe..17..261V}, astroquery \citep{Ginsburg2019}, skyfield \citep{2019ascl.soft07024R}, matplotlib \citep{Hunter2007}, GalSim \citep{2015A&C....10..121R}, tqdm \citep{2021zndo...5517697D}, opencv \citep{opencv_library}.}


\appendix

\section{AstroSkyFlow implementation and user guide} \label{sec:guide}

\begin{figure*}[ht!]
\centering
\includegraphics[width=0.99\textwidth]{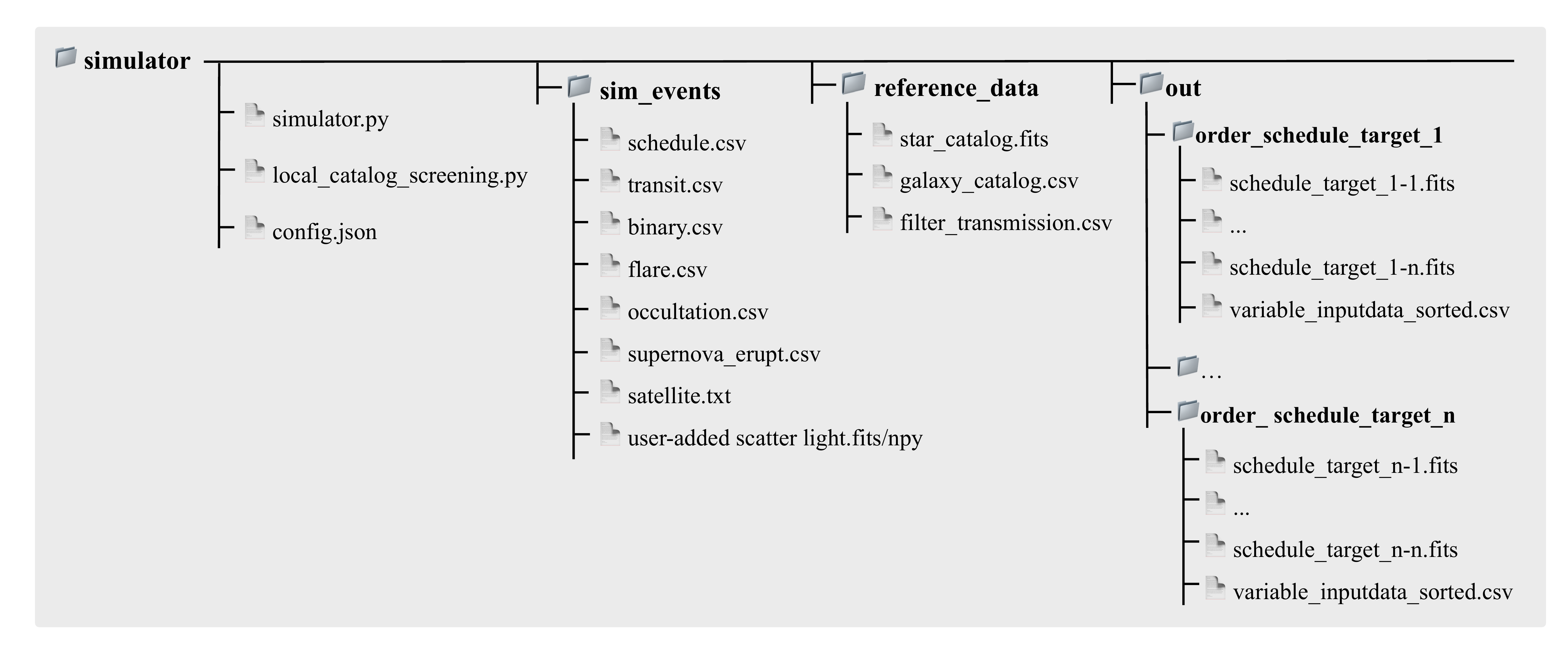}
\caption{Modular directory structure of AstroSkyFlow.}
\label{fig:folder}
\end{figure*}

AstroSkyFlow is organized into a modular directory structure to separate input configurations, reference data and simulation outputs, as shown in Fig. \ref{fig:folder}. The main components are core scripts, configuration files, events directory, reference photometric data directory and output directory. The main execution logic is implemented in simulator.py which comprises several cooperating classes, while local\_catalog\_screening.py handles local catalog filtering for specific FOV in different observation tasks and is called by simulator.py during execution. The simulatior parameters of different classes are specified in config.json. We have listed all parameters in Table \ref{Table:parameter} and give some brief description. The events directory contains schedule file and other variable files. The schedule.csv includes the observation observation order, targets name, celestial coordinates, number of frames, exposure, start and end time. The other variable files include transit, binary, flare, occultation, supernova eruptions and satellite files. Each event type is described by various of parameters as shown in Table \ref{tab:target_parameters} and the latest satellite files come from the celestrak website. Users can also provide the scatter light file. The reference photometric data directory stores the stellar catalog, galaxy catalog and filter transmission curves for different observation systems. In output directory, each scheduled target has a dedicated subdirectory which contains the simulated FITS images and some corresponding input variable files containing injected information for subsequent comparison.

AstroSkyFlow requires users to supply reference photometric data specifying the observational system: a stellar catalog, a galaxy catalog and filter transmission curves. These files should be placed in the reference\_data directory. If the simulated observation system lacks a proprietary stellar catalog, users can set the corresponding parameter star\_catalog in config.json to ``online". AstroSkyFlow will query the Gaia archive over the network and use the Gaia DR3 catalog and its photometry as the reference stellar catalog. Note that this mode requires an internet connection and may be subject to archive query limits and latency. Then modify the observation tasks in schedule.csv to get observation sequence, where individual observations must be arranged in strict chronological order. If users require to inject the specific variable sources, users can modify the parameters within the corresponding variable files. To successfully inject and simulate the variable star, it is essential that this variable exists within the star catalog employed for the present observation. If users do not require the source truly existent, users may assign the desired optical variable properties to any star in the fixed star catalog: simply enter the Gaia DR3 ID of this fixed star to the corresponding column within the variable files. Final setup is completed by adjusting parameters in the config.json. Users can modify the only key parameters and all others can remove; the code will automatically apply their default values. Two essential operational constraints are that the initial time of the simulator set in config.json must be earlier than the start time of the first scheduled observation and all scheduled targets must be over an altitude greater than 25 degrees at their respective observation times. This constraint mirrors real-world telescope safety protocols, and AstroSkyFlow will generate a warning if it is violated. The current version of the AstroSkyFlow code is archived on Zenodo \citep{li_2026_19733852}, and the corresponding GitHub repository is available at \url{https://github.com/laomuliubingbing/AstroSkyFlow}.



\startlongtable
\begin{deluxetable*}{l l l}
\tabletypesize{\scriptsize}    
\tablecaption{Descriptions of the input parameters for each simulation class.\label{tab:sim_params}}
\tablewidth{0pt}          
\tablehead{
  \colhead{Class} & \colhead{Parameter} & \colhead{Description}
}
\startdata
    Camera & \texttt{camera\_name} (optional) & The name of camera. Default name is simulator\_camera. \\
    Camera & \texttt{pixel$\_$size$\_$m} & Pixel size in meters. \\
    Camera & \texttt{pixel$\_$number\_x}, \texttt{pixel\_number\_y} & Number of pixels in x and y dimensions. \\
    Camera & \texttt{readout\_time\_s} (optional) & Camera readout time per frame (s). Default value is  0.3448. \\
    Camera & \texttt{gain} & Conversion gain (e$^-$/ADU). \\
    Camera & \texttt{dark\_current\_e\_s\_1} & Dark current rate (e$^-$/s). \\
    Camera & \texttt{full\_well\_capacity\_e} (optional)& Pixel full well capacity (e$^-$). Default value is 200000. \\
    Camera & \texttt{read\_noise\_e} & Readout noise (e$^-$). \\
    Camera & \texttt{QE} & Quantum efficiency. \\
    Camera & \texttt{bit\_per\_pixel} & Bit depth per pixel. \\
    Camera & \texttt{bias\_level} (optional)& Bias level in photons (zero-exposure measurement is unattainable in practice). Default value is 400.\\
    Camera & \texttt{CCD/CMOS\_temperature} (optional)& Temperature of the detector. Default value is 265.15\\
    \hline
    Mount & \texttt{tracking\_mode} (optional) & Tracking mode (e.g. alt-az). Default mode is alt-az.\\
    Mount & \texttt{tracking\_speed\_deg\_s\_1} (optional)& Tracking speed (deg/s). Default value is 5\\
    Mount & \texttt{stable\_time\_s} (optional)& Time required to stabilize after a slew (s). Default value is 5.\\
    Mount & \texttt{goto\_error\_arcsec} (optional)& Pointing error of the mount (arcsec). Default value is 2.\\
    Mount & \texttt{tracking\_error\_arcsec\_min\_1} (optional)& Tracking drift per minute (arcsec/min). Default value is 0.1.\\
    \hline
    Telescope & \texttt{telescope\_name} (optional)& The name of telescope. Default name is simulator\_telescope.\\
    Telescope & \texttt{rot\_deg} & Rotation angle relative to celestial north (degree). \\
    Telescope & \texttt{latlonalt} & Observatory latitude, longitude, altitude. \\
    Telescope & \texttt{seeing\_arcsec} & Atmospheric seeing (arcsec). \\
    Telescope & \texttt{focal\_length\_m} & Telescope focal length (m). \\
    Telescope & \texttt{diameter\_m} & Telescope aperture diameter (m). \\
    \hline
    Photons & \texttt{sky\_raw\_mag} (optional) & Background sky brightness. Optional value is 21.\\
    Photons & \texttt{sky\_background\_mode} (optional) & The mode for generating the sky background. Default mode is fast.\\
    Photons & \texttt{$C_Y$} (optional) & Empirical site-dependent scaling factor for atmospheric scintillation.Default value is 2.\\
    Photons & \texttt{scintillation\_$L_0$} (optional) & Outer scale of scintillation. Default value is None.\\
    Photons & \texttt{scintillation\_method} (optional)& The method used to generate scintillation. Default method is lognormal.\\
    Photons & \texttt{scintillation\_seed} & Random seed for reproducible generation of the scintillation. \\
    Photons & \texttt{temp\_C} & Ambient temperature in degrees Celsius. \\
    Photons & \texttt{pressure\_mm} & Atmospheric pressure in millimeters of mercury. \\
    Photons & \texttt{relative\_humidity} & Relative humidity (percent). \\
    Photons & \texttt{star\_catalog} & Reference star catalog (FITS file). \\
    Photons & \texttt{delta\_lambda\_min} & Minimum Bandpass (m). \\
    Photons & \texttt{delta\_lambda\_max} & Maximum Bandpass (m). \\
    Photons & \texttt{zero\_mag} & Zero magnitude. \\
    Photons & \texttt{Mag\_limit} & Maximum magnitude limitation. \\
    Photons & \texttt{alpha} & Moffat distribution parameter. \\
    Photons & \texttt{FWHM\_error} & Fractional standard deviation applied when sampling the PSF FWHM. \\
    Photons & \texttt{moffat\_scale\_FWHM} & Pixel size of the Moffat distribution convolution (scale relative to FWHM). \\
    Photons & \texttt{ADR\_flag} & Flag to simulate atmospheric differential refraction. \\
    Photons & \texttt{historical\_variable\_star\_flag} & Flag to search and simulate variable stars outside the input user-specified catalog. \\
    Photons & \texttt{galaxy\_flag} & Flag to simulate galaxy. \\
    Photons & \texttt{galaxy\_local\_size} &  Pixel size using for the convolution in galaxy simulation. \\
    Photons & \texttt{supernova\_erupt} & Flag to simulate supernova eruptions. \\
    Photons & \texttt{satellite\_flag}, \texttt{asteroid\_flag} & Flags to include satellites/asteroids. \\
    Photons & \texttt{asteroid\_move\_limit\_pixel} & Lower limit of the asteroid's motion simulated as a streak (pixel/exposure). \\
    Photons & \texttt{move\_oversample\_factor} & Subsampling factor for pixels during simulated motion. \\
    Photons & \texttt{extinction\_coeffcient} & Atmospheric extinction coefficient. \\
    Photons & \texttt{supernova\_seed} & Random seed for supernova simulation. \\
    Photons & \texttt{filter} & Transmission filter file. \\
    Photons & \texttt{band\_name} & Band label used when registering your bands for Phoebe and sncosmo\\
    Photons & \texttt{custom} & Custom defined or native Phoebe band\\
    Photons & \texttt{mag\_system} & magnitude system (AB/Vega). \\ 
    Photons & \texttt{transit\_catalog} (optional) & Event catalogs including their key parameters. Default directory structure is shown in Fig. \ref{fig:folder}.\\
    & \texttt{binary\_catalog} (optional)&\\
    & \texttt{supernova\_erupt\_catalog} (optional)&\\
    &\texttt{flare\_catalog} (optional)&\\
    &\texttt{occultation\_catalog} (optional)&\\
    Photons & \texttt{user\_stray\_field} &  Scatter light field by user adding; set to null if none.\\
    Photons & \texttt{satellite\_catalog} (optional) &  Catalogs for satellites. Default directory structure is shown in Fig. \ref{fig:folder}.\\
    Photons & \texttt{gal\_catalog} & Catalogs for galaxies. \\
    Photons & \texttt{satellite\_mag} (optional) & Brightness of satellite (mag). Default value is 5.\\
    \hline
    Sensor & \texttt{vignetting\_flag}, \texttt{cone\_height} & Vignetting model parameters. \\
    &\texttt{entrance\_radius}, \texttt{exit\_radius}& \\
    Sensor & \texttt{PRNU\_flag}, \texttt{sigma\_prnu}, \texttt{seed\_prnu} & Photo response non-uniformity model. \\
    Sensor & \texttt{dark\_rate\_model} (optional)& Dark current model parameters. Default values are False, 0.001, 1, 248 and 1.1557 respectively.\\
    & \texttt{pixel\_area} (optional)&\\
    &\texttt{dark\_figure\_merit} (optional) &\\
    &\texttt{temperature} (optional) &\\
    & \texttt{energy\_gap} (optional)&\\
    Sensor & \texttt{dark\_current\_rate} & Dark current rate (e$^-$/s). \\
    Sensor & \texttt{dark\_current\_fpn\_flag} & Dark current fixed-pattern noise. \\
    &\texttt{dark\_noise\_factor}, \texttt{seed\_dark\_fpn}&\\
    Sensor & \texttt{SF\_flag}, \texttt{f\_clock} (optional)& Source follower noise model parameters. Default values are 2e7, 1.5e-8, 1000, 1e-6, 1e-12, 1e-6,  \\
    &\texttt{W} (optional), \texttt{f\_c} (optional)& and 5e-7.\\
    &\texttt{tau\_RTN} (optional), \texttt{delta\_I} (optional)& \\
    &\texttt{t\_s} (optional), \texttt{tau\_D} (optional)&\\
    Sensor &\texttt{A\_SN} & Three-stage gain components; their multiplication should equal the camera's gain.\\
    & \texttt{A\_SF} & \\
    & \texttt{A\_adc\_linear}  & \\
    Sensor & \texttt{reset\_noise\_flag} & Reset noise parameters. Default value of capacitance is 1e-15. \\
    &\texttt{sensing\_capacitance} (optional)&\\
    Sensor & \texttt{offset\_fpn\_flag}& Offset fixed-pattern noise. \\
    & \texttt{offset\_correlation\_factor}&\\
    &\texttt{offset\_fpn\_sigma\_U} &\\
    &\texttt{seed\_offset\_fpn}&\\
    Sensor & \texttt{V\_e\_nonlinear}, \texttt{alpha} (optional & Electron-to-voltage nonlinearity. Default value is 1e-15 respectively.\\
    Sensor & \texttt{V\_v\_nonlinear} & Voltage-to-voltage nonlinearity. Default values are 1.04 and 2.0 respectively.\\
    &\texttt{gamma\_nlr} (optional) &\\
    & \texttt{v\_full\_well} (optional) & \\
    Sensor & \texttt{A\_adc\_nonlinearity\_flag}& ADC response model. Default values are false, 1.01 and 2 respectively. \\
    &\texttt{gamma\_adc\_nonlin} (optional) & \\
    & \texttt{V\_adc\_ref} (optional) & \\
    \hline
    World & \texttt{output\_dir} (optional)& Path to output directory. Default directory structure is shown in Fig. \ref{fig:folder}. \\
    World & \texttt{julian\_date} & The initial time of the simulator. \\
    World & \texttt{input\_schedule} (optional) & Input observing schedule file. Default directory structure is shown in Fig. \ref{fig:folder}.\\
\enddata
\label{Table:parameter}
\end{deluxetable*}

\bibliography{main}{}
\bibliographystyle{aasjournalv7}



\end{document}